\documentclass[fleqn,usenatbib]{mnras}

\usepackage[T1]{fontenc}

\DeclareRobustCommand{\VAN}[3]{#2}
\let\VANthebibliography\thebibliography
\def\thebibliography{\DeclareRobustCommand{\VAN}[3]{##3}\VANthebibliography}

\usepackage{graphicx}	
\usepackage{amsmath}	
\usepackage{amssymb}	
\usepackage{newtxtext,newtxmath}
\usepackage{threeparttable}

\title[Mk\,33Na: a very massive colliding wind binary]{Melnick 33Na: a very massive colliding wind binary system in 30 Doradus\thanks{Based on observations collected at the European Southern Observatory under program ID 5106.D-0835(A).}}

\author[J. M. Bestenlehner et al.]{
Joachim\,M.\,Bestenlehner$^{1}$,\thanks{E-mail: j.m.bestenlehner@sheffield.ac.uk}
Paul\,A.\,Crowther$^{1}$,
Patrick S. Broos$^{2}$,
Andrew\,M.\,T.\,Pollock$^{1}$,
Leisa K. Townsley$^{2}$
\\
$^{1}$Department of Physics \& Astronomy, Hounsfield Road, University of Sheffield, Sheffield, S3 7RH, UK\\
$^{2}$Department of Astronomy \& Astrophysics, 525 Davey Laboratory, Pennsylvania State University, University Park, PA 16802, USA\\
}

\date{Accepted XXX. Received YYY; in original form ZZZ}

\pubyear{2021}

\begin{document}
\label{firstpage}
\pagerange{\pageref{firstpage}--\pageref{lastpage}}
\maketitle

\begin{abstract}
We present spectroscopic analysis of the luminous X-ray source Melnick 33Na (Mk\,33Na, HSH95 16) in the LMC 30 Doradus region (Tarantula Nebula), utilising new time-series VLT/UVES spectroscopy. We confirm Mk\,33Na as a double-lined O-type spectroscopic binary with a mass ratio $q = 0.63 \pm 0.02$, $e = 0.33 \pm 0.01$ and orbital period of $18.3 \pm 0.1$ days, supporting the favoured period from X-ray observations obtained via the Tarantula -- Revealed by X-rays (T-ReX) survey. Disentangled spectra of each component provide spectral types of OC2.5\,If* and O4\,V for the primary and secondary respectively - unusually for an O supergiant the primary exhibits strong \ion{C}{iv}\,4658 emission and weak \ion{N}{v} 4603-20, justifying the OC classification. Spectroscopic analysis favours extreme physical properties for the primary ($T_{\rm eff} = 50$ kK, $\log L/L_{\odot} = 6.15$) with system components of $M_{1} = 83 \pm 19 M_{\odot}$ and $M_{2} = 48 \pm 11 M_{\odot}$ obtained from evolutionary models, which can be reconciled with results from our orbital analysis (e.g. $M_{1} \sin^3 i = 20.0 \pm 1.2 M_{\odot}$) if the system inclination is $\sim 38^{\circ}$ and it has an age of 0.9 to 1.6 Myr. This establishes Mk\,33Na as one of the highest mass binary systems in the LMC, alongside other X-ray luminous early-type binaries Mk34 (WN5h+WN5h), R144 (WN5/6h+WN6/7h) and especially R139 (O6.5\,Iafc+O6\,Iaf).
\end{abstract}




\section{Introduction}

\begin{figure}
\begin{center}
		\includegraphics[width=0.8\columnwidth]{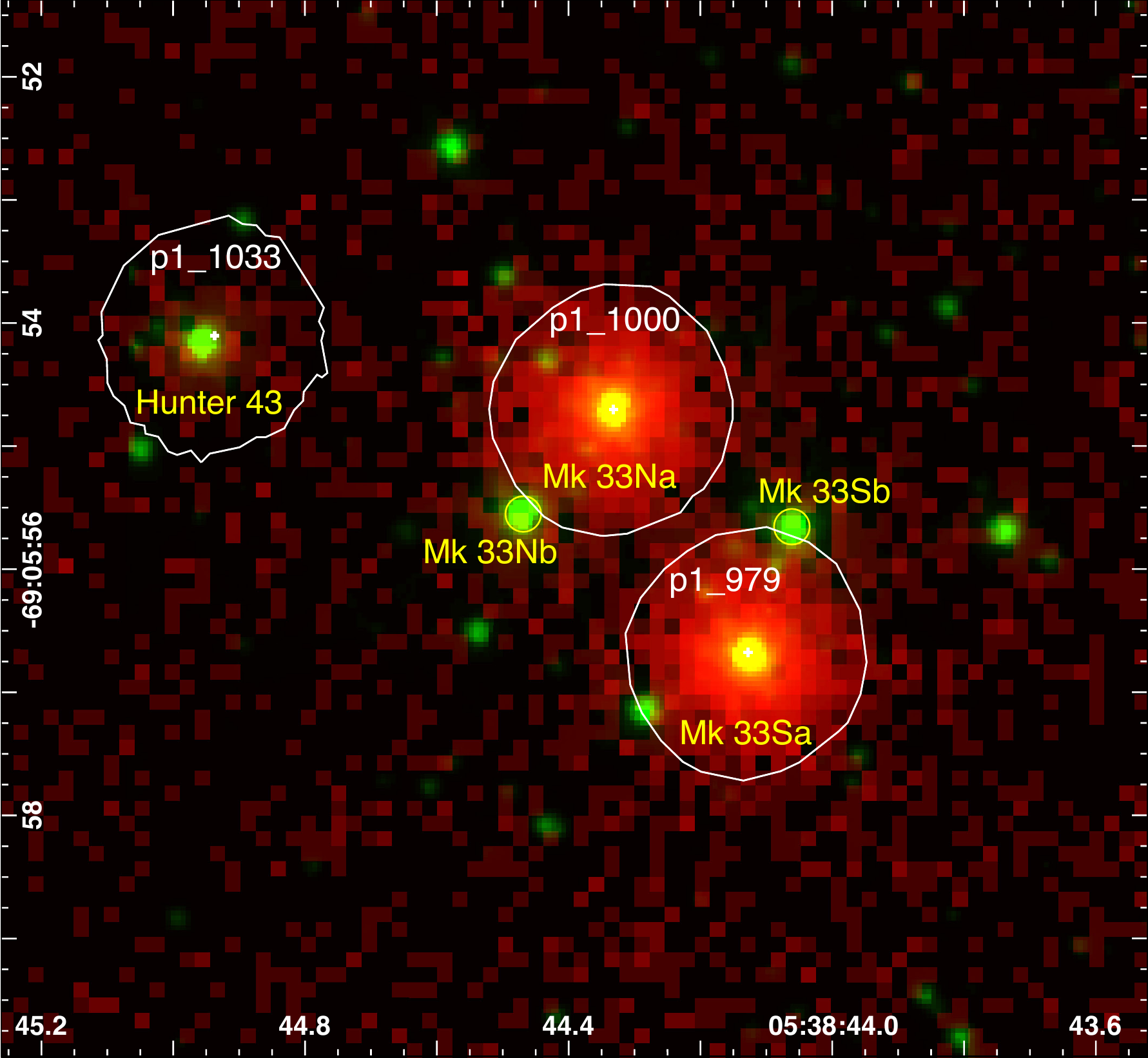}
		\end{center}
    \caption{2 colour composite image of Melnick 33 HST/WFC3 F555W \citep[green][]{deMarchi2011} and Chandra ACIS (red, Townsley+ in prep) indicating the target of the current study Mk\,33Na and its close visual companions Mk\,33Nb (O6.5\,V), Mk\,33Sa \citep[O3\,III(f*),][]{massey2005} and Mk\,33Sb \citep[WC4+OB,][]{smith1990}. 
    Mk\,33Na and Mk\,33Sa are bright X-ray sources, while Mk\,33Nb and Mk\,33Sb remain undetected in >2Ms of Chandra data.
    The field of view is 8$\times$8 arcsec (2$\times$2 parsec at the distance of the LMC), with North up and East to the left. The rich star cluster R136 lies 3 parsec to the SW.}
    \label{f:Mk33}
\end{figure}

The mass of a star is its fundamental defining property, but direct measurements remain elusive with the  exception of visual binaries or eclipsing spectroscopic binaries \citep{andersen1991, serenelli2021}.
Consequently mass estimates usually follow from spectroscopic results or comparison with evolutionary predictions. 
Uncertainties of evolutionary models increase with stellar mass \citep[e.g][]{martins2013} due to uncertainties in nuclear reaction rates, stellar structure, internal mixing processes and mass-loss properties \citep{langer2012}, so stellar mass estimates of high mass stars remain poorly constrained. Eclipsing double-lined spectroscopic binaries (SB2s) are the ideal gold standard calibrators to test single-star evolutionary models, because they permit accurate determinations of the dynamical mass, stellar parameters and chemical abundances for both components. Most studies of such systems have  focused on stars up to 15\,$M_{\odot}$ which only consider convective core-overshooting and rotational mixing \citep[e.g.][]{southworth2004, tkachenko2014, pavlovski2018}. \cite{higgins2019} also included the effect of mass loss, which strips the stellar envelope and reveals CNO-processed material. They calibrated their grid of stellar models with the 40 + 34 $M_{\odot}$ eclipsing massive supergiant binary HD 166734 (O7.5If + O9I(f)) using the results for this system from \cite{mahy2017}.

Despite the high binary frequency  amongst massive stars \citep{sana2012Sci, sana2013} eclipsing systems are rare.  Within the Milky Way there are very few massive eclipsing binaries whose component masses exceed a few tens of solar masses, including NGC~3603 A1 with system components of 116+89 $M_{\odot}$ \citep{schnurr2008}, F2 in the Arches cluster with 82+60 $M_{\odot}$ \citep{lohr2018} and WR20a with 71+69 $M_{\odot}$ \citep{bonanos2004}. Such systems are also scarce in the Magellanic Clouds, although a few massive systems have been identified via MACHO, OGLE or targeted photometric surveys \citep[e.g.][]{massey2002, ostrov2003, bonanos2009}.  By way of example, as part of the TMBM \citep{almeida2017} follow up to the VFTS survey of OB stars in the Tarantula Nebula \citep{evans2011}, only 4 of $\sim 100$ O-type binaries were identified as detached eclipsing binaries \citep{mahy2020}.

Double-lined spectroscopic binaries (SB2) are relatively common amongst massive stars and provide accurate orbital parameters, although the inclination must be inferred in order to derive the dynamical masses of the individual components. If the orbital solution is known, SB2 spectra can be disentangled and stellar parameters and chemical compositions obtained for the primary and secondary components. In the LMC there are a few very massive SB2s, the majority of which are located in the Tarantula Nebula \citep{crowther2019}, including Melnick 34 whose primary component has been estimated to be 139 $M_{\odot}$ \citep{tehrani2019}, R144 with a primary mass of 74 $M_{\odot}$ \citep{shenar2021} and  R139 whose primary exceeds 69 $M_{\odot}$ \citep{mahy2020}. In principle, inclinations  and in turn dynamical masses can be obtained for binaries lacking photometric eclipses via optical  \citep[e.g.][]{shenar2021} or X-ray light curve modelling. Indeed, Melnick 34 \citep{pollock2018} was first established as a colliding wind binary via the Chandra X-ray imaging survey of the Tarantula Nebula, T-ReX (Townsley et al., Broos \& Townsley, in prep.).

Mk\,33Na is located at a projected distance of $\sim$3.4 pc from the rich star cluster R136 in the Tarantula Nebula \citep{massey1998, crowther1998}. \cite{melnick1985} first identified "Mk\,33N" as an early O-type star (O4), with "Mk\,33S" $\sim$2 arcsec to the south west. \cite{massey1998} refined its spectral type to O3\,If* from HST/FOS spectroscopy \citep[alias \# 16 from][]{hunter1995} with "a" added by \cite{crowther1998} to distinguish it from "b", an O6.5\,V star \citep[alias \# 32 from][]{hunter1995} 1.2 arcsec away as shown in Fig.~\ref{f:Mk33}. Other aliases for Mk\,33Na include \# 1140 in \cite{parker1993}, \# 33 in \cite{selman1999a} and \# 1943 in \cite{castro2018}. 

\begin{figure}
	\includegraphics[width=\columnwidth]{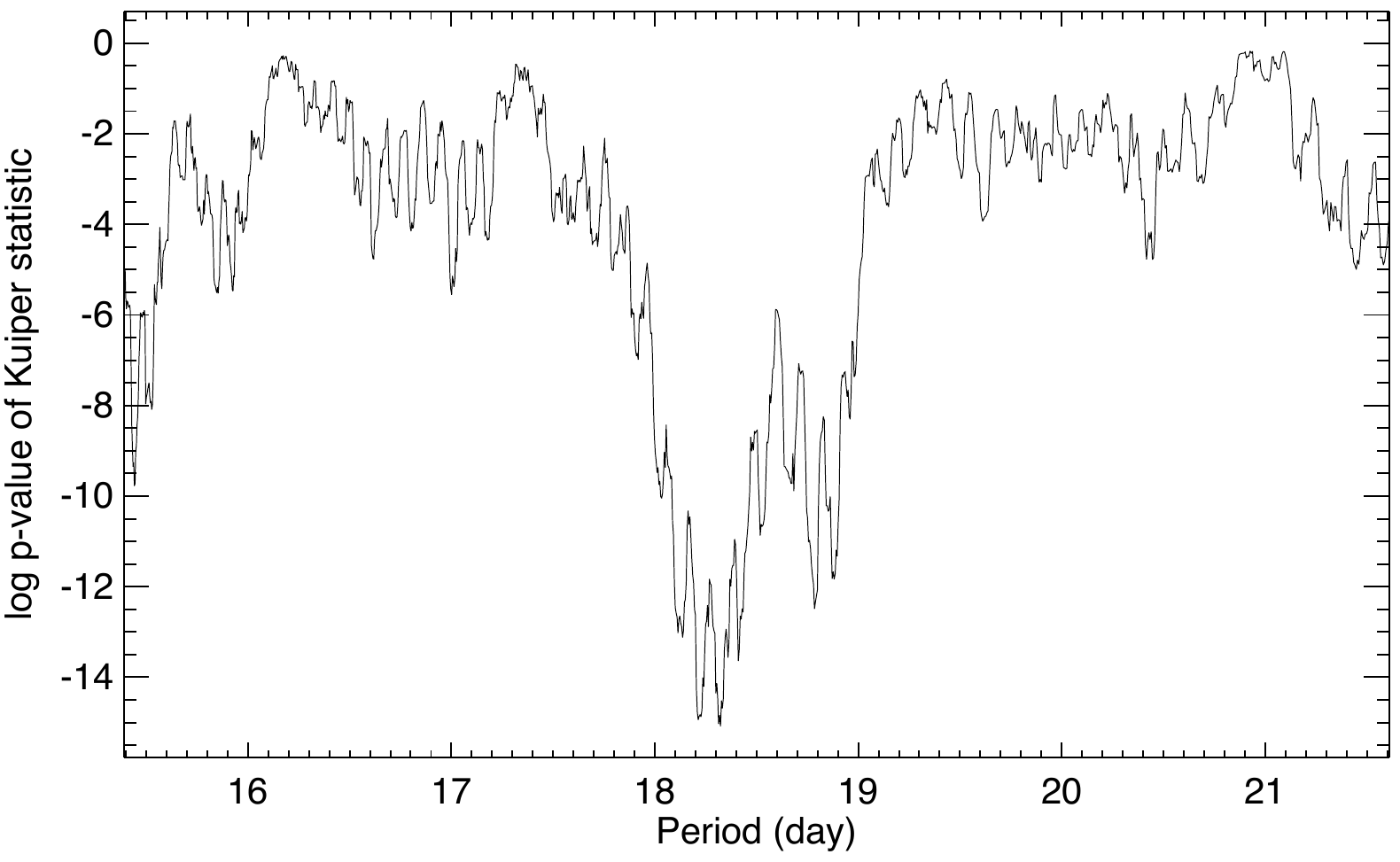}
    \caption{A period search for Mk\,33Na from  X-ray observations with the Chandra T-ReX survey of the Tarantula Nebula (Townsley et al.; Broos \& Townsley), favouring a period of $\sim$18.3 days from the observed Kuiper statistic (Paltani 2004).}
    \label{f:spec_dens_period}
\end{figure}

Shallow Chandra X-ray observations revealed Mk33\,Na as a  bright X-ray source (CX9 in \citealp{portegies2002}; \# 133 in \citealp{townsley2006}), with a similar X-ray luminosity ($L_{\rm X}$) to R139, suggesting the presence of a colliding-wind binary (CWB). Deep X-ray imaging via the T-ReX survey (Townsley et al. in prep.) confirmed Mk33Na as a bright X-ray source (Fig.~\ref{f:Mk33}) corresponding to an attenuated X-ray luminosity of $2.8 \times 10^{33}$ erg\,s$^{-1}$ at the distance of the LMC (Crowther et al. in prep.). 
On the basis of its Kuiper statistic,  a variant of the Kolmogorov-Smirnov (KS) test \citep{paltani2004} shown in Figure~\ref{f:spec_dens_period}, Broos \& Townsley (in prep.) identified significant X-ray variability suggesting a potential period of 18.3 days for Mk\,33Na most likely due to orbital modulation in a binary system. Figure~\ref{f:Mk33} shows that Mk33Sa is similarly bright in X-rays suggestive perhaps of another binary, although it is less variable than its neighbour.

Here we undertake an analysis of VLT/UVES time-series spectroscopy of Mk\,33Na motivated by the T-ReX observations, confirm that it is a double-lined spectroscopic binary with a period of $\sim$18.3 days, estimate its physical and wind properties, and reveal that it comprises a colliding wind binary with a high $L_{\rm X}/L_{\rm bol} = 1.2 \times 10^{-6}$ (Crowther et al. in prep.). In Section~\ref{UVES} we present our orbital analysis of new VLT/UVES spectroscopy and X-ray variability of Mk\,33Na, Section~\ref{s:spt_sa} presents the spectral types of each disentangled component plus a spectroscopic analysis from which physical, wind and chemical properties are obtained. Stellar masses, ages and orbital inclination are presented in Section~\ref{s:mass_age_i}. We discuss our results in the context of other massive binaries in the LMC in Section~\ref{s:disc} and brief conclusions are drawn in Section~\ref{conclusions}.




\section{Observations}\label{UVES}
Using optical multi-epoch observations (Sect.~\ref{s:opt_obs}) we determine the radial velocity variations of both components of the Mk\,33Na system (Sect.~\ref{s:rv}). The orbital parameters are calculated in Sect.~\ref{s:orbit_par} and the X-ray observations are presented in Sect.~\ref{X-ray}.

\subsection{Optical spectroscopy\label{s:opt_obs}}

To determine the orbital parameters we spectroscopically followed up Mk\,33Na \citep[RA\,05:38:44.34, DEC\,-69:05:54.6, J2015.5, GaiaDR2][]{gaia2018} with the Ultraviolet and Visual Echelle Spectrograph \citep[UVES,][]{dekker2000} mounted at the Very Large Telescope (VLT) in Paranal, Chile. 15 epochs of VLT/UVES spectra were obtained in service mode over a period of 17 days in October 2020 (Program ID 106.21MH.001 aka 5106.D-0835(A), PI Bestenlehner). The observations were initially planned to sample the orbital period over 2.5 months with preferences on dates where service mode was available. Each epoch or orbital phase could be potentially observed on 2 or 3 dates. According to the Phase 2 guidelines 1 to 2 epochs per orbital phase were scheduled on days when service mode was not available. In the unlikely event that all epochs are obtained in one go including dates where service mode was not available, possible coverage periods were between 17 and 19 days.

Due to Covid-19, regular observations with the VLT paused earlier in 2020. VLT/UVES was one of the first instruments to be brought back in operation for service mode only, from the beginning of October 2020. All 15 epochs were obtained on the nights of 8th through 25th October or in Modified Julian Date (MJD) from 59131.25 to 59148.18\,d, even though some nights were originally flagged as no service mode available, as indicated in the observations log (Table~\ref{t:rv}). The slit was placed north to south with a slight angle avoiding stars directly in the slit (Fig.~\ref{f:Mk33}). Observations were taken in Dichroic 2 mode with central wavelengths of 437 and 760nm, providing spectral ranges of 3730 to 4990 \AA\ (blue-arm, EEV 2k$\times$4k detector), 5670 to 7570 \AA\ (lower red-arm, EEV 2k$\times$4k detector) and 7670-9450 \AA\ (upper red-arm, MIT/LL 2k$\times$4k detector) with an integration time of 593 seconds. A 0.8 arcsec slit was used for all epochs with a spectral resolution between 50,000 and 60,000. A signal to noise ratio (S/N) between 30 to 50 was achieved for the blue and lower red-arm.

For the data reduction we used the ESO Reflex pipeline \citep{freudling2013} with standard calibration data for service mode. Mk\,33Na is located in a crowded region within the Tarantula nebula (Fig.~\ref{f:Mk33}). The default pipeline setting for flux calibrated spectra took the entire slit into account. Even though we placed the slit in a way that minimised contamination from nearby sources, some of the spectra were still contaminated owing to high air mass (up to 2.0) and variable seeing conditions (DIMM\footnote{Averaged full-width half-maximum seeing conditions calculated over the exposure time as measured by the Differential Image Motion Monitor.} 0.5 to 1.3 arcsec). The WC star Mk33Sb was the most prominent source which affected \ion{He}{ii}\,4686 and the \ion{C}{iv}\,5801/5812 spectral lines (Fig.~\ref{f:Mk33}). Therefore, we chose the 2D slit extraction and co-added the regions which were the least contaminated by nearby sources. However, this led to a reduction in the S/N that was achieved (35 and 25 in the blue and lower red-arm, respectively).

In the final step we applied to each epoch the barycentric correction to the wavelength array. To bring the wavelength array of the individual epoch to the same scale we used the  interstellar band \ion{Ca}{ii}\,3934 for the blue-arm and \ion{Na}{i}\,5890 for the lower red-arm. As a reference value we chose the average radial velocity of the interstellar lines.



\subsection{Radial velocities\label{s:rv}}

\begin{table*}
	\centering
	\caption{Observing log of VLT/UVES spectroscopy of Mk\,33Na and measured radial velocities for the primary OC2.5\,If* and secondary O4\,V components.}
	\label{t:rv}
	\begin{tabular}{lcclcccc|cc}
		\hline
		UT Date &  sec $z$ & DIMM & MJD & \ion{N}{iv}\,4058  & \ion{C}{iv}\,5801 & \ion{He}{ii}\,4542 & H$_{\eta}$
	 & \ion{He}{ii}\,4542 & H$_{\eta}$ \\
		         &               & arcsec &             & Primary           & Primary              & Primary               & Primary                                     & Secondary & Secondary \\
		\hline
		09 Oct 2020 & 1.62 & 0.71 & 59131.25 & $175.4^{+5.9 }_{-6.1 }$& $199.1 \pm 4.5        $& $203.7\pm 9.4       $& $210.4^{+11.6}_{-12.6}$& $318.3^{+11.9}_{-12.0}$& $330.3^{+25.4}_{-22.3}$\\
		10 Oct 2020 & 1.61 & 0.67 & 59132.25 & $233.4 \pm 6.9        $& $225.0^{+4.6 }_{-4.7 }$& $238.6\pm 8.4       $& $219.8^{+12.5}_{-13.1}$& $317.3\pm 10.8        $& $382.0^{+21.2}_{-19.1}$\\
		11 Oct 2020 & 1.60 & 0.49 & 59133.25 & $241.8^{+5.8 }_{-6.3 }$& $252.6^{+6.5 }_{-6.2 }$& $251.7\pm 8.9       $& $257.2^{+11.9}_{-12.2}$& $269.4^{+10.9}_{-10.8}$& $283.9^{+28.1}_{-27.1}$\\
		12 Oct 2020 & 1.78 & 0.80 & 59134.21 & $261.7^{+9.9 }_{-8.3 }$& $275.1^{+5.3 }_{-5.2 }$& $255.9^{+8.6}_{-8.5}$& $249.7^{+10.8}_{-10.9}$& $266.6\pm 9.7         $& $210.9^{+26.8}_{-28.5}$\\
		14 Oct 2020 & 1.97 & 0.82 & 59136.17 & $290.8^{+9.2 }_{-10.8}$& $315.5^{+6.0 }_{-5.8 }$& $293.4\pm 9.0       $& $279.1^{+15.2}_{-14.7}$& $192.9^{+13.7}_{-14.2}$& $196.1^{+26.4}_{-30.4}$\\
		15 Oct 2020 & 1.79 & 1.07 & 59137.20 & $299.3^{+12.9}_{-15.5}$& $330.6^{+7.5 }_{-7.1 }$& $302.4\pm 11.1      $& $316.5^{+14.6}_{-13.7}$& $180.4^{+15.1}_{-15.5}$& $133.4^{+21.2}_{-24.1}$\\
		16 Oct 2020 & 1.84 & 0.59 & 59138.19 & $317.3 \pm 6.5        $& $335.7^{+5.9 }_{-5.8 }$& $329.0\pm 9.1       $& $288.9 \pm 13.2       $& $157.6\pm 10.6        $& $166.7^{+23.8}_{-25.5}$\\
		17 Oct 2020 & 1.58 & 0.54 & 59139.24 & $357.7^{+8.6 }_{-8.1 }$& $342.1^{+4.4 }_{-4.2 }$& $346.8\pm 7.8       $& $314.4^{+9.3 }_{-8.9 }$& $152.4\pm 8.4         $& $135.6^{+15.1}_{-16.0}$\\
		19 Oct 2020 & 1.51 & 0.58 & 59141.26 & $359.9^{+8.5 }_{-8.3 }$& $352.0^{+7.6 }_{-7.0 }$& $336.1\pm 8.4       $& $327.2^{+10.2}_{-9.5 }$& $155.3^{+10.9}_{-10.7}$& $111.9^{+16.7}_{-18.0}$\\
		20 Oct 2020 & 1.68 & 1.26 & 59142.21 & $327.6^{+9.7 }_{-8.9 }$& $329.4^{+11.6}_{-12.1}$& $305.2\pm 10.0      $& $316.6^{+15.2}_{-14.1}$& $179.8^{+13.6}_{-13.7}$& $80.8 ^{+33.9}_{-30.0}$\\
		21 Oct 2020 & 1.67 & 0.47 & 59143.21 & $316.1^{+5.6 }_{-5.4 }$& $314.0^{+5.3 }_{-5.4 }$& $297.8^{+8.4}_{-8.3}$& $304.4^{+11.3}_{-10.9}$& $201.6^{+9.6 }_{-9.8 }$& $161.9^{+21.3}_{-22.8}$\\
		22 Oct 2020 & 1.83 & 0.71 & 59144.17 & $263.5^{+7.9 }_{-7.3 }$& $249.4^{+5.7 }_{-5.6 }$& $255.1^{+8.8}_{-8.7}$& $262.4^{+14.4}_{-14.0}$& $272.4^{+10.7}_{-10.6}$& $276.7^{+39.8}_{-40.4}$\\
		24 Oct 2020 & 1.72 & 0.62 & 59146.19 & $129.4^{+7.0 }_{-6.8 }$& $135.9^{+6.2 }_{-6.1 }$& $169.9\pm 8.2       $& $139.4^{+8.3 }_{-8.7 }$& $432.9^{+10.8}_{-11.0}$& $520.2 \pm 14.8       $\\
		25 Oct 2020 & 1.41 & 0.96 & 59147.31 & $126.6^{+7.3 }_{-6.6 }$& $140.3^{+5.2 }_{-5.0 }$& $148.5^{+8.4}_{-8.3}$& $146.0^{+8.6 }_{-9.0 }$& $494.9^{+9.5 }_{-12.8}$& $474.0^{+18.6}_{-19.2}$\\
		26 Oct 2020 & 1.74 & 0.63 & 59148.18 & $146.3^{+5.9 }_{-5.8 }$& $159.1^{+5.5 }_{-5.2 }$& $160.8\pm 8.3       $& $174.4^{+17.2}_{-18.4}$& $416.1\pm 16.2        $& $455.8^{+26.6}_{-28.0}$\\
		\hline
	\end{tabular}
\end{table*}

Visual inspection showed that \ion{N}{iv}\,4058 and \ion{C}{iv}\,5801 hardly show any signatures of the secondary star. With the stellar atmosphere and radiative transfer code {\sc cmfgen} \citep{hillier1998} we computed a synthetic reference spectrum based on estimated stellar parameters from our spectral classification (Sect.~\ref{s:spt}). Using the synthetic spectrum we performed $\chi^2$-minimisation analyses to derive the radial velocities (RV) for \ion{N}{iv}\,4058 and \ion{C}{iv}\,5801.

The observation taken on 25 October (MJD = 59147.3122d) revealed the largest separation of the \ion{He}{ii}\,4542 lines of the primary and secondary (Fig.~\ref{f:largest_sep}). This epoch was used to estimate the optical flux ratio of $\sim 0.25$ for the individual components by scaling the templates of the primary and secondary by a factor of 0.8 and 0.2 to match the \ion{He}{ii} line strength. According to the spectral classification we chose a template for the secondary with an effective temperature of 45\,000\,K. As an initial RV guess for the primary we used the average of the \ion{N}{iv}\,4058 and \ion{C}{iv}\,5801 RV measurements for each epoch. Subsequently, we performed an iterative $\chi^2$-minimisation analysis of \ion{He}{ii}\,4542 and $\mathrm{H}_\eta$ by alternately co-adding and shifting the template spectra of the primary and secondary until convergence. The nebular emission was clipped in the case of $\mathrm{H}_\eta$. Lower excitation levels of the Balmer series showed too strong nebular lines and it was not possible to derive RV for all observations. The results are given in Table~\ref{t:rv}.

\begin{figure}
	\includegraphics[width=\columnwidth]{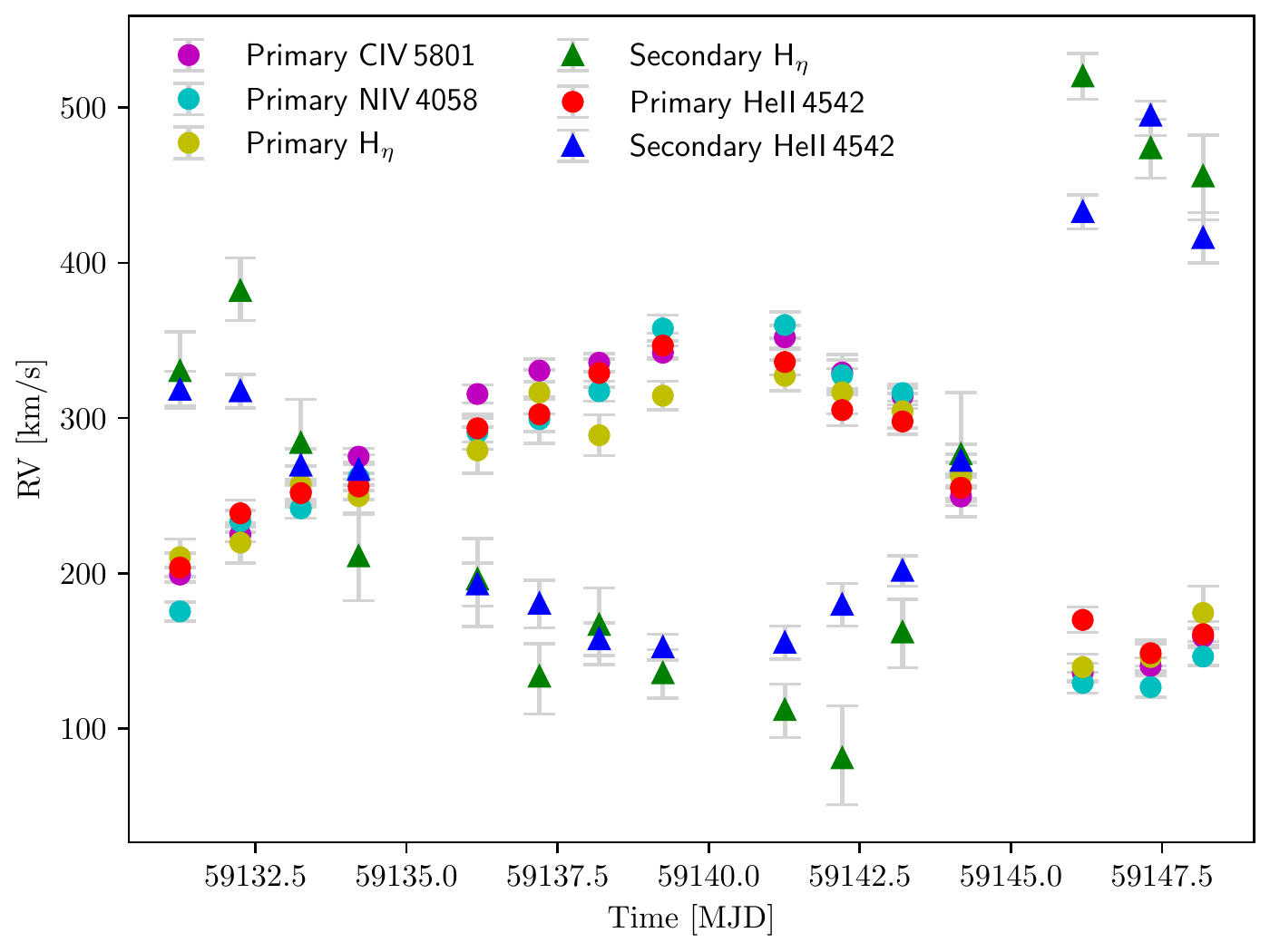}
    \caption{Radial velocity measurements versus MJD for the primary OC2.5\,If* and secondary O4\,V components.}
    \label{f:rv_measurements}
\end{figure}

In Fig.~\ref{f:rv_measurements} we visualise our RV measurements. It clearly shows the anti-phase between the primary and secondary of the Mk\,33Na system. The uncertainties of the RV measurements are larger for the secondary and in particular for $\mathrm{H}_\eta$.

\subsection{Orbit fitting and orbital parameters}\label{s:orbit_par}

\begin{figure*}
	\includegraphics[width=\textwidth]{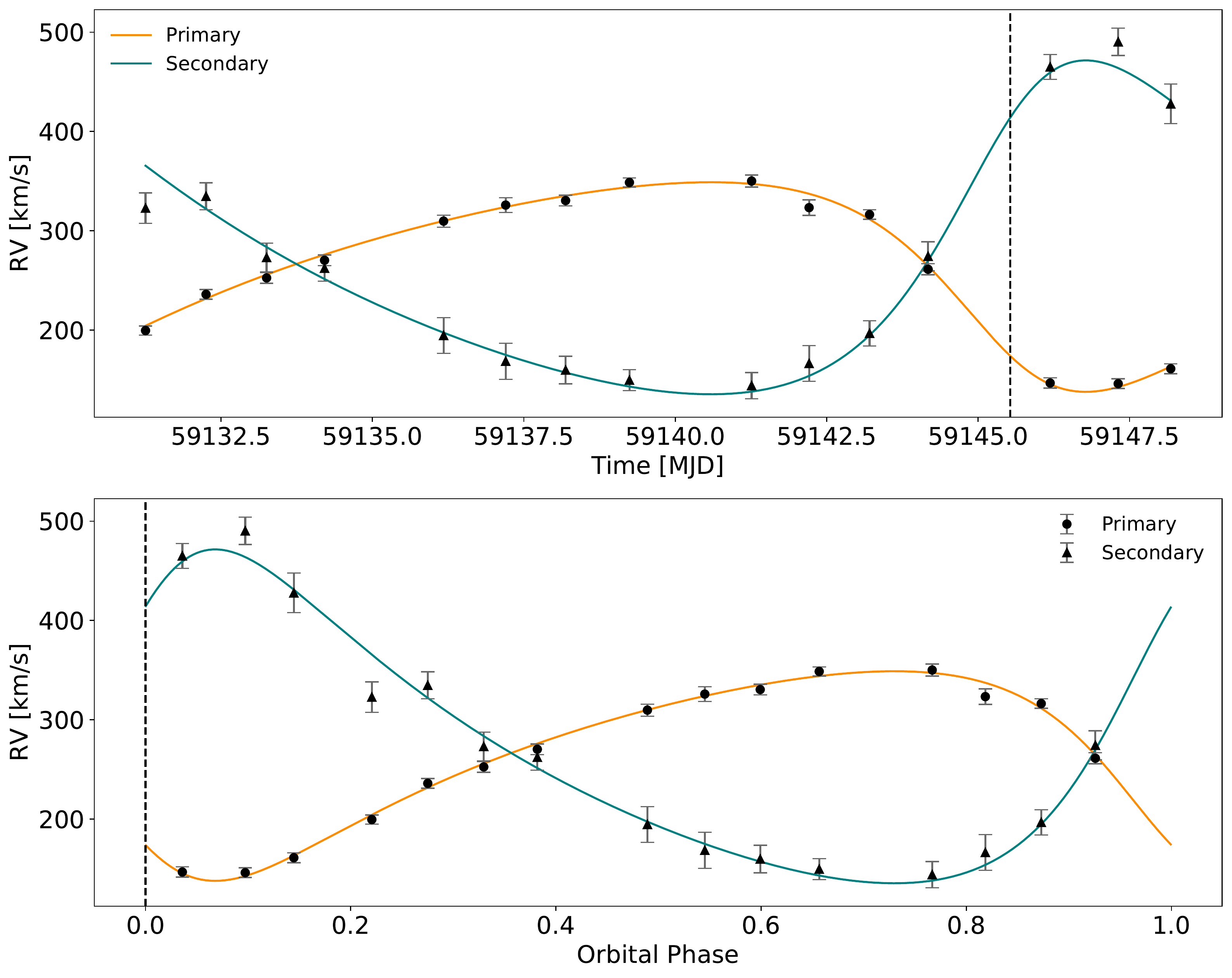}
    \caption{MCMC fit to determine the Keplerian orbital parameters of the Mk\,33Na system. Solution for the primary and secondary are shown by the solid orange and cyan-green lines. Black dots and triangle are the uncertainties weighted averaged RV measurements from Table~\ref{t:rv} for the primary and secondary, respectively.}
    \label{f:orbit_param}
\end{figure*}

\begin{figure*}
	\includegraphics[width=\textwidth]{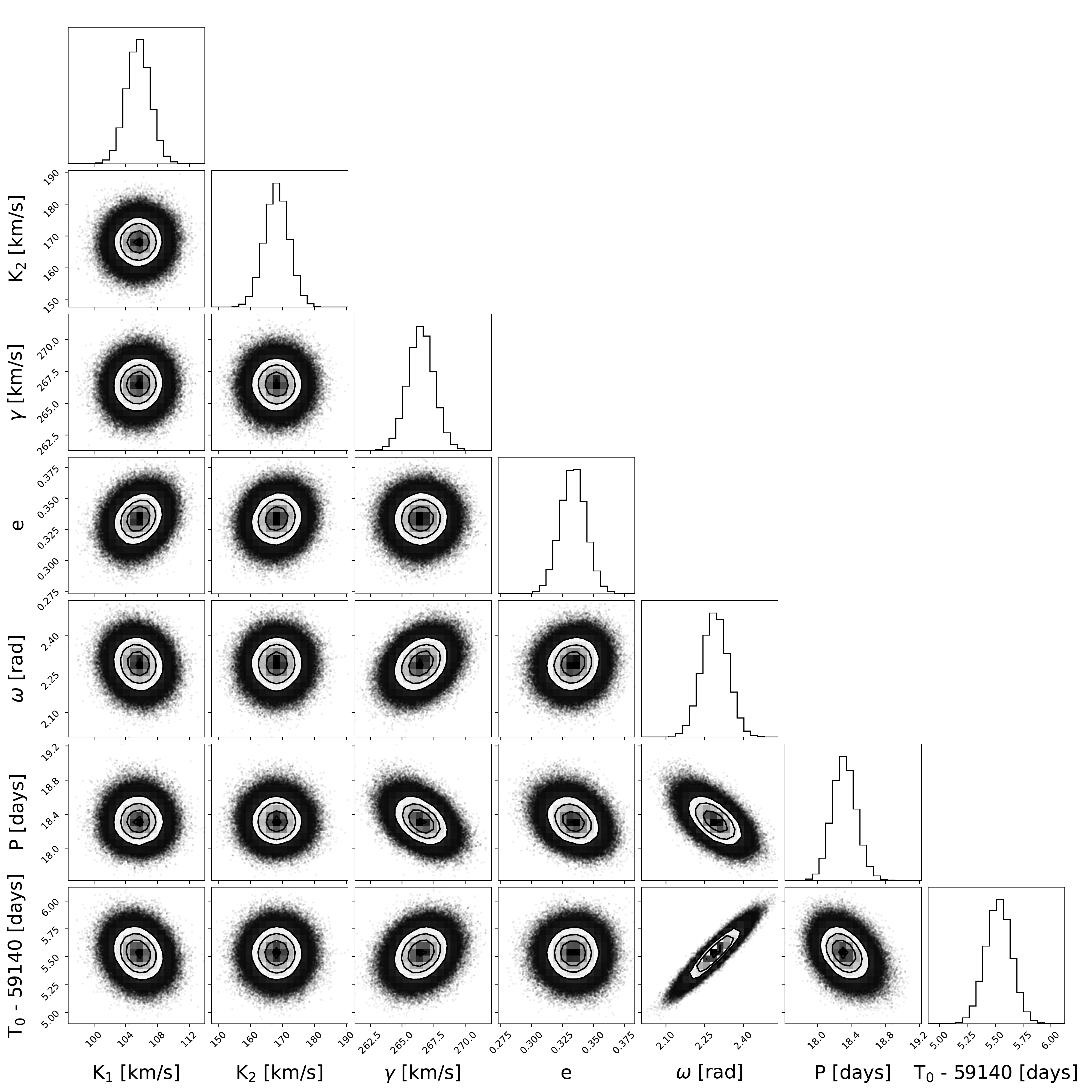}
    \caption{Corner plot showing posterior probabilities for the MCMC fit solution, where $K_1$ and $K_2$ are the semi-amplitudes of the velocities for the primary and secondary respectively, $\gamma$ is the systemic velocity, $e$ is the eccentricity, $\omega$ is the longitude of the periastron, $P$ is the orbital period and $T_0$ is the time of periastron.}
    \label{f:corner_plot}
\end{figure*}

\begin{table}
	\centering
	\caption{Keplerian orbital parameters with 1-$\sigma$ uncertainties derived using the MCMC methods of \citet{tehrani2019}.}
	\label{t:orbit_param}
	\begin{tabular}{lcc} 
		\hline
		Parameter & Unit & MCMC\\
		\hline
		$K_1$ 		& [km/s]	& $105.6\pm1.5$		\\
		$K_2$ 		& [km/s]	& $168.2\pm 3.8$	\\
		$\gamma$ 	& [km/s] 	& $266.5\pm 1.0$	\\
		e 			&			& $0.334\pm0.010$		\\
		$\omega$	& [$^{\circ}$]		& $131.19\pm2.86$	\\
		$P$			& [days]	& $18.319\pm0.139$	\\
		$T_0$		& [MJD]	& $59145.53\pm0.12$	\\
		$M_{\rm orb,1} \sin^3 i$ & [$M_{\odot}$]	& $20.04\pm1.23$	\\
		$M_{\rm orb,2} \sin^3 i$ & [$M_{\odot}$]	& $12.58\pm0.85$	\\
		$q$			&			&	$0.628\pm 0.016$\\
		\hline
	\end{tabular}
\end{table}

To determine the Keplerian orbital parameters we used the Markov Chain Monte Carlo (MCMC) fitting routine developed, utilised and described in \cite{tehrani2019}. As input prior for the period we used a uniform distribution of $18 \pm 2$\,days according to the X-ray period from Fig.~\ref{f:spec_dens_period}. A flat, uninformative prior was adopted for the other parameters. The MCMC fitting was performed using the uncertainties weighted averaged RV measurements from Table~\ref{f:rv_measurements} for the primary and secondary. The routine delivers a full set of Keplerian orbital parameters: semi-amplitudes of the velocities of the primary ($K_1$) and secondary star ($K_2$), system velocity $\gamma$, eccentricity $e$, longitude of periastron $\omega$, orbital period $P$, and time of periastron $T_0$.

In Fig.~\ref{f:orbit_param} we show the best-fit RV curve in time and phase space using the most probable parameters from MCMC fitting. The corresponding posterior probability corner plot is given in Fig.~\ref{f:corner_plot} and shows that each orbital parameter is well constrained. Within the uncertainties the orbital period agrees well with the $P$ derived from the X-ray light-curve (Sect.~\ref{X-ray}). Based on the orbital parameters we are able to infer minimum masses 
through the standard formula
\begin{equation}\label{e:min_mass}
M_{1,2} \sin^3 i = P/(2\pi G) (1- e^2)^{1.5} (K_1+ K_2)^2 K_{2,1}.
\end{equation}
From Eq.~\ref{e:min_mass} we are able to derive the mass ratio of the system, $q = K_1/K_2 = M_2/M_1 = 0.63\pm 0.02$. The eccentricity is low enough that Mk\,33Na can be assumed to be a detached binary system. The system velocity $\gamma$ of $266.5\pm 1.0$\,km/s is consistent with the mean value of 267.7\,km/s within 5\,pc around the stellar cluster R136 \citep{henault2012}.

All orbital parameters and minimum masses including uncertainties are given in Table~\ref{t:orbit_param}.

\subsection{X-ray light curve}\label{X-ray}

\begin{figure}
	\includegraphics[width=\columnwidth]{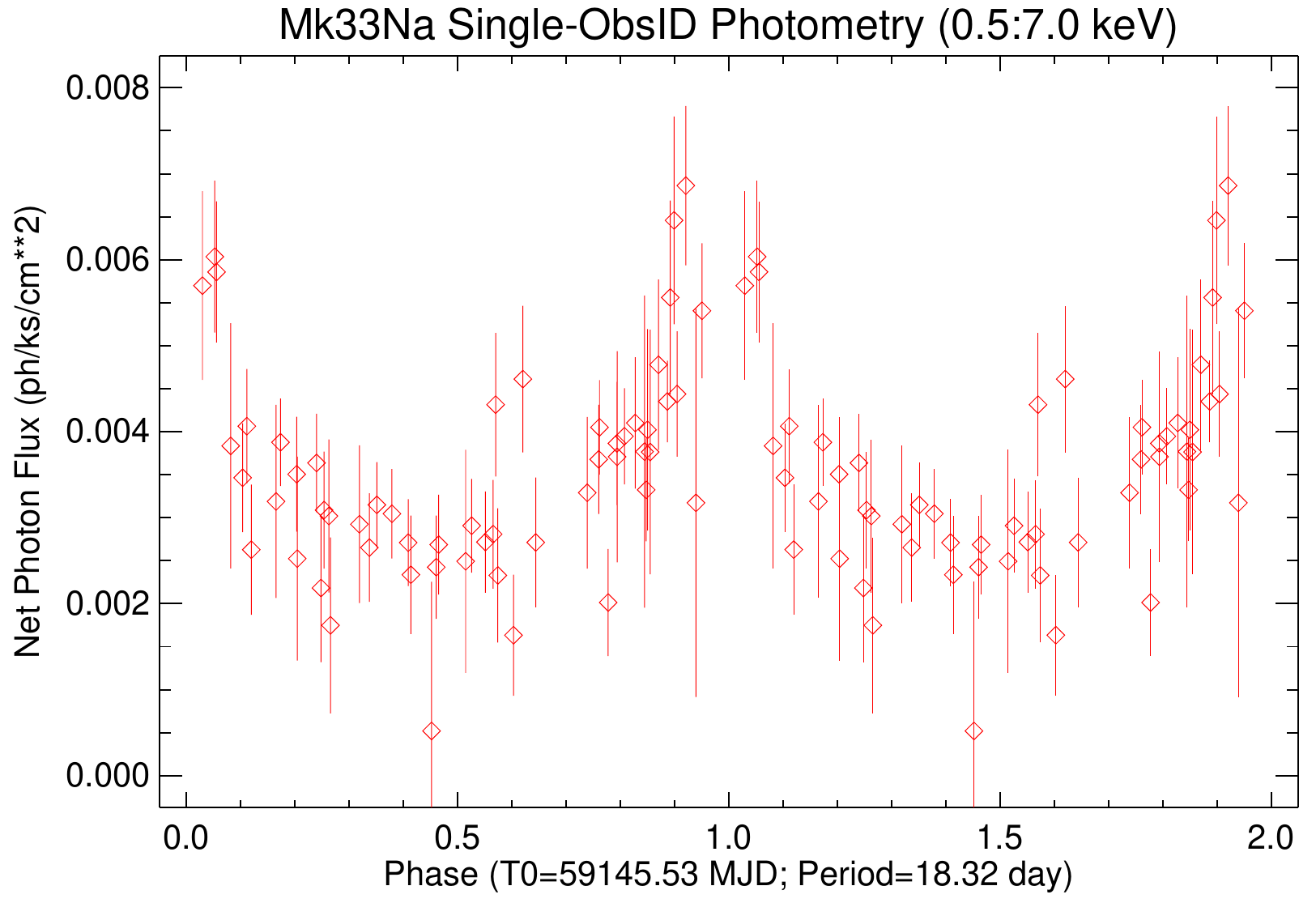}
    \caption{X-ray light curve of Mk\,33Na folded on a period of 18.32 days about the periastron passage.}
    \label{f:lightcurve}
\end{figure}

\begin{figure}
	\includegraphics[width=\columnwidth]{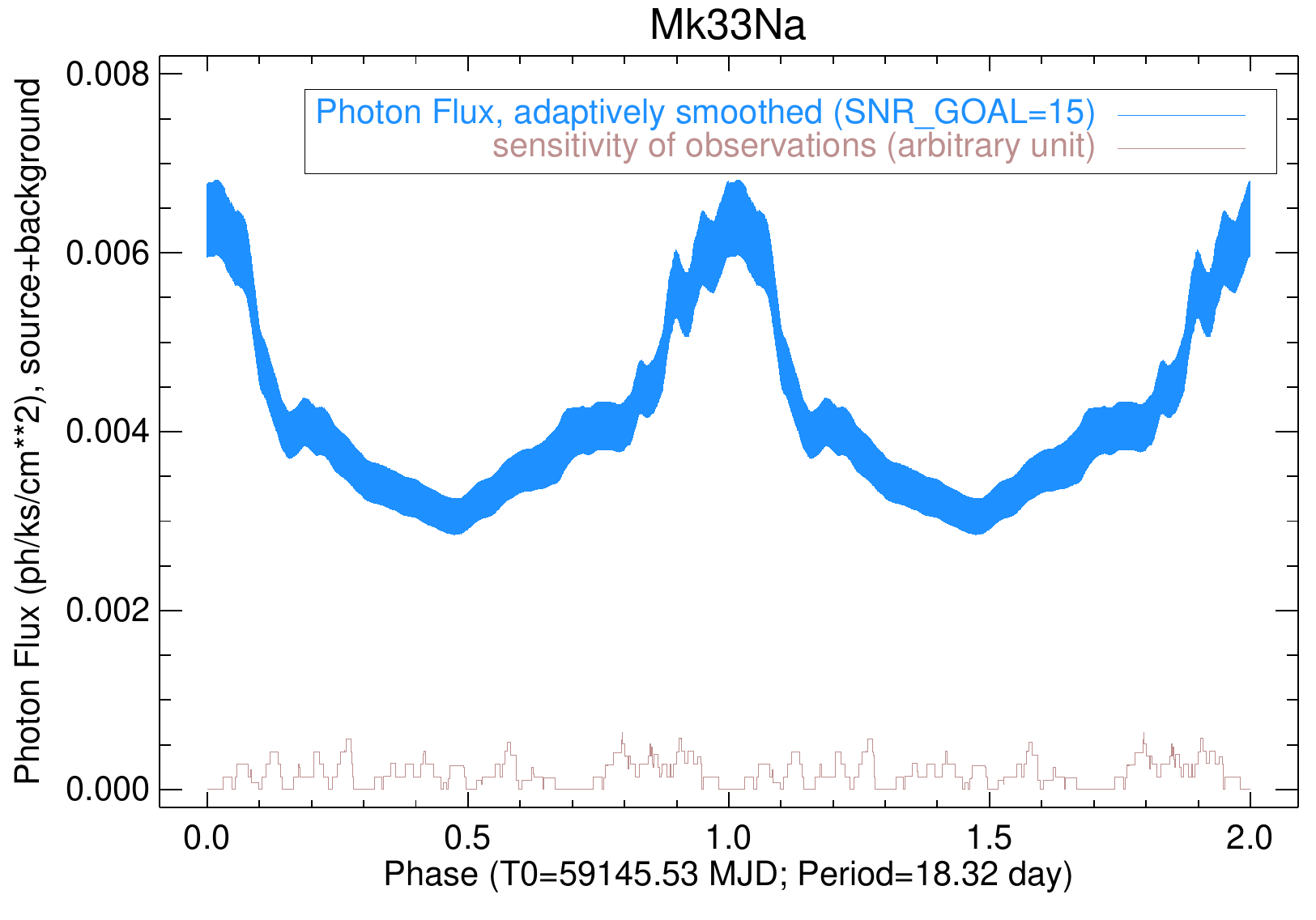}
    \caption{The phases of individual events are adaptively smoothed. The thickness of the blue ribbon represents the uncertainty of the flux estimate at that phase.}
    \label{f:phase_bins}
\end{figure}

\begin{figure}
	\includegraphics[width=\columnwidth]{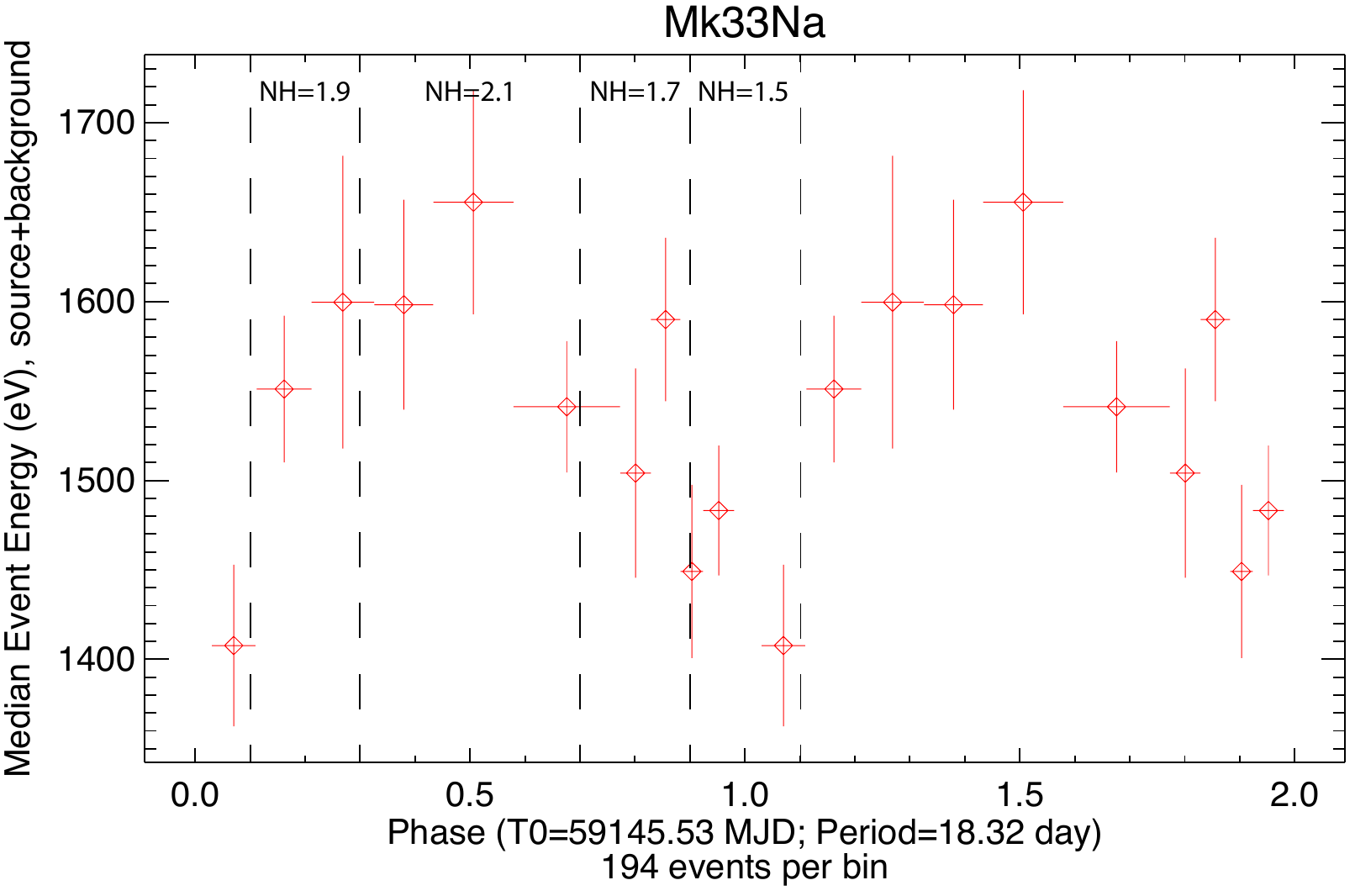}
    \caption{Folded median X-ray energy divided into 10 phase bins. Dashed vertical lines indicate 4 broad phase bins including the absorption column density ($N_{\rm H} / 10^{22}$, see text). The spectrum is softer near periastron as the binary-separation dependent luminosity is modulated by phase-dependent absorption.}
    \label{f:median_energy}
\end{figure}




The 2 Ms {\it Chandra} Visionary Program T-ReX was executed over 630 days between 2014 May 3 and 2016 Jan 22 using the ACIS instrument (16.9$' \times 16.9'$ field of view) centred on R136a, the central star at the heart of the Tarantula Nebula cluster (Townsley et al.; in prep). Here we confirm the 18.3 day orbital period of Mk\,33Na from UVES, supporting the X-ray results from the Kuiper statistic (Fig.~\ref{f:spec_dens_period}) from the T-ReX dataset (Broos \& Townsley, in prep). We present its X-ray light curve folded on the best-fit optical radial-velocity period in Fig.~\ref{f:lightcurve} and \ref{f:phase_bins}. The T-ReX campaign itself covered 34 orbital cycles suggesting a high degree of reproducibility for an X-ray source of mean count rate close to 1 count per kilosecond. This conclusion is reinforced by the 3 relatively bright measurements taken close to an implied periastron about a decade and 165 orbital cycles earlier in 2006 January. The X-ray sourcelist and source properties from the 2006 Chandra data on 30 Doradus were published in \cite{townsley2014}.


At first sight, Fig.~\ref{f:lightcurve} and \ref{f:phase_bins} appear consistent with the luminosity variations expected from a colliding-wind binary system in which the X-ray cooling time exceeds the flow time out of the system leading to a luminosity inversely proportional to the binary separation \citep[e.g.][]{stevens1992}. However, the shape of the observed X-ray spectrum varies with phase as shown in Fig.~\ref{f:median_energy}. The rise in flux in the approach to periastron is mostly due to soft X-rays near 1 keV. Spectral analysis performed in four broad phase bins reveals that the data are consistent with a model in which a separation-dependent luminosity is further modulated by a phase-dependent absorbing column density, where $N_{\rm H} / 10^{22}$ = 1.5, 1.9, 2.1, 1.7 cm$^{-2}$ in phase bins -0.1:0.1, 0.1:0.3, 0.3:0.7, 0.7:0.9 (Fig.~\ref{f:median_energy}).

Although other models can fit the data, this variable-absorber model offers a simple interpretation. The lowest absorbing column (maximum count rate) occurs when the shocked X-ray emission occurring in the space between the stars is viewed through the weaker wind of the secondary O4V star which passes in front across the line-of-sight 1.4 days before periastron, according to the orbital elements in Table~\ref{t:orbit_param}. High reddening shows that the interstellar medium accounts for much of the observed X-ray absorption. The time-averaged X-ray luminosity of Mk\,33Na is $\log L_{\rm X} = 33.95 \pm 0.05$ erg\,s$^{-1}$ (Crowther et al. in prep.), a value more typical of binary systems than the lower intrinsic luminosities of single stars.



\section{Spectral type and spectroscopic analysis}
\label{s:spt_sa} 
Before we determine the spectral types (Sect.~\ref{s:spt}) and perform the spectroscopic analysis (Sect.~\ref{s:sa}) we disentangled the spectra into primary and secondary components (Sect.~\ref{s:dis}). Line broadening and line-profile variations are described in Sect.~\ref{s:lb} and \ref{s:lpv}). The section concludes with Sect.~\ref{s:lum} where we obtain reddening parameters and luminosities for primary and secondary star. 

\subsection{Disentangling}\label{s:dis}

For an accurate spectral classification and reliable spectroscopic analysis, we disentangled the spectra into primary and  secondary components. Individual components exhibit large RV variations which are enhanced by their anti-phase (Fig.~\ref{f:rv_measurements}) making disentanglement relatively straightforward. 

Based on the orbital solution from Sect.~\ref{s:orbit_par}, we calculate more precise radial velocities for each epoch of the primary and secondary than measured in Sect.~\ref{s:rv}. In the first step we shifted the epochs to the rest-frame of the primary according to its RV. Co-adding and averaging the spectra would have included signatures of the secondary and for example cosmics which are not removed by 2D slit extraction (Sect.~\ref{s:opt_obs}). Therefore, we calculated a median spectrum which should largely remove the secondary component. The difference between the both methods is shown in Fig.~\ref{f:dis_comp}. To obtained the spectrum for the secondary we applied the RV-shift for each epoch to the primary median spectrum from the first step and divided the epoch by the primary spectrum (Fig.~\ref{f:dis_med_sec}). Then we shifted the epochs to the rest-frame of the secondary and calculated the median spectrum of the secondary.

For an improved disentangling we applied the same methodology for the primary by dividing all epochs by the secondary median spectrum before calculating the median spectrum of the primary. Those steps were repeated until no changes to the previous iteration were observed. The spectra of Mk\,33Na$_1$ and Mk\,33Na$_2$ are rescaled according to their flux ratio (Sect.~\ref{s:lum}).

A similar methodology was applied to remove the broad \ion{C}{iv}\,5801-12 emission caused by the WC4 star Mk33Sb (Fig.~\ref{f:Mk33} and \ref{f:dis_wc}). The disentangled spectra are shown in Fig.~\ref{f:pri_dis} and \ref{f:sec_dis}. A detailed description of the spectra disentangling methodology is given in Appendix~\ref{s:dsd}.

\begin{table*}
	\centering
	\caption{Stellar parameters obtained from the spectroscopic analysis. Masses under the assumption of chemically homogeneous evolution ($M_{\rm hom}$) are derived with the mass-luminosity relation by \citet{graefener2011}. Most probable stellar parameters derived with BONNSAI \citep{schneider2014} based on stellar evolutionary models by \citet{brott2011, koehler2015}.}
	\label{t:stel_param}
	\begin{tabular}{@{}l@{~~~}l@{~~~}c@{~~~}c@{~~~}c@{~~~}c@{~~~}c@{~~~}c@{~~~}c@{~~~}c@{~~~}c@{~~~}c@{~~~}c@{~~~}c@{~~~}c@{~~~}c@{~~~}c@{~~~}c@{~~~}c@{~~~}c@{~~~}c@{}} 
	\hline
			\multicolumn{11}{c}{Spectroscopic analysis} \\
		\hline
		 & SpT &$\log L/L_{\odot}$ & $T_{\rm eff}$ & $R_{\rm eff}$ & $\log g$ & $\log\dot{M}/\sqrt{f_{\rm V}}$ & $\epsilon_{\mathrm{C}}$ & $\epsilon_{\mathrm{N}}$ &$M_{\rm hom}$ & $M_{\rm sp}$ \\
		 &&                    & [K]               &$[R_{\odot}]$  & [cm\,s$^{-2}$] & [$M_{\odot}$yr$^{-1}$] & $\log(\mathrm{C/H})+12$ & $\log(\mathrm{N/H})+12$ & [$M_{\odot}$] & [$M_{\odot}$]\\
		\hline
		Primary  & OC2.5\,If* &$6.15\pm 0.18$ & $50\,000\pm 2500$ & $15.8\pm2.1$ & $4.0\pm 0.1$& --5.6 to --5.2 $\pm$ 0.2 & $7.7\pm0.3$ & $7.2\pm0.3$ & 106 & $90^{+25}_{-18}$\\
		Secondary& O4\,V  &$5.78\pm 0.18$ & $45\,000\pm 2500$ & $12.8\pm 1.7$ & --$^{\rm a}$        & --6.5$^{\rm b}$ & -- & -- & 66 & --\\
		\hline
		\hline
				\multicolumn{11}{c}{Recovered stellar parameters by BONNSAI} \\
		\hline
		 & SpT & $\log L/L_{\odot}$ & $T_{\rm eff}$ & $R_{\rm eff}$ & $\log g$ & Age & $\epsilon_{\mathrm{C}}$ & $\epsilon_{\mathrm{N}}$ & $ M_{\rm evo}$ & $ M_{\rm evo,ini}$\\
		 &                    & & [K]   &$[R_{\odot}]$        & [cm\,s$^{-2}$] & [$\mathrm{Myr}$] &  &  & [$M_{\odot}$] & [$M_{\odot}$]\\
		\hline
		Primary  & OC2.5\,If* & $6.08\pm 0.14$ & $50\,700^{+2500}_{-2200}$ & $13.9^{+2.5}_{-2.0}$ & $4.15^{+0.01}_{-0.17}$ & $0.9 \pm 0.6$ & 7.75$^{\rm c}$ & 6.9$^{\rm c}$ & $83\pm19$ & $84^{+21}_{-19}$\\
		Secondary& O4\,V & $5.66^{+0.17}_{-0.19}$ & $45\,800^{+2800}_{-2700}$ & $10.1^{+2.5}_{-2.0}$  & $4.17^{+0.05}_{-0.21}$ & $1.6^{+0.7}_{-1.3}$ & 7.75$^{\rm c}$ & 6.9$^{\rm c}$ & $48\pm11$ & $47^{+13}_{-9}$\\
		\hline
	\end{tabular}\\
	\begin{tablenotes}\footnotesize
	\item { $^{\rm a}$ $\log g$ could not be derived, because the wings of the Balmer lines are in emission. $^{\rm b}$ Approximate upper limit for the mass-loss rate. 
	$^{\rm c}$ The probability distribution function shows a wide spread of possible composition, but the most probable value is the initial chemical composition for the LMC \citep{brott2011}.} 
	\end{tablenotes}
\end{table*}

\subsection{Spectral type\label{s:spt}}

\begin{figure}
	\includegraphics[width=\columnwidth]{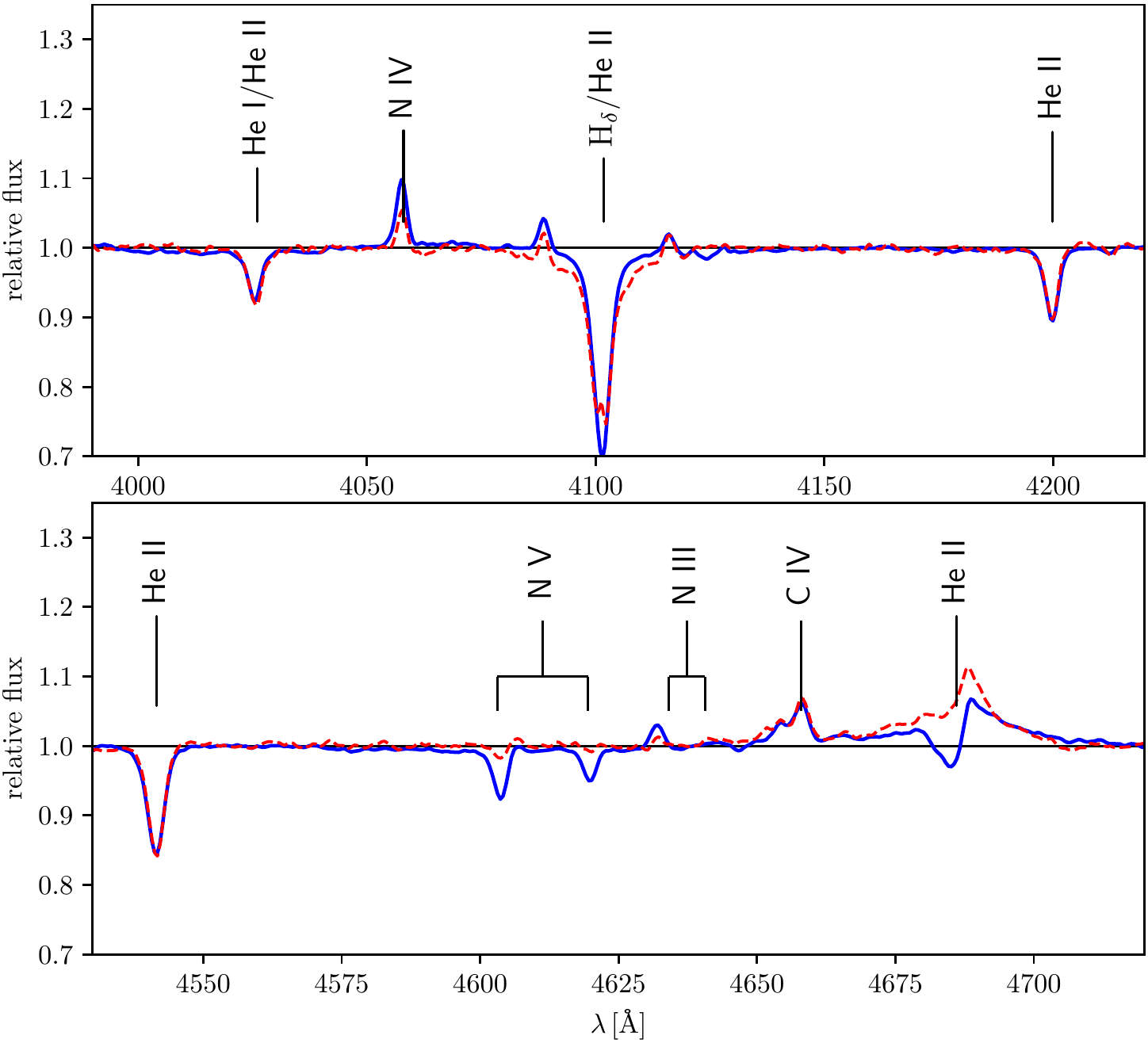}
    \caption{Spectral comparison of Mk\,33Na$_1$ (red dashed line) with VFTS\,016 \citep[blue solid line][]{evans2010}. \ion{C}{iv}\,4658 has a similar line strength in both stars, but the nitrogen lines are much weaker for Mk\,33Na$_1 \Rightarrow$ OC2.5\,If*.}  
    \label{f:vfts016}
\end{figure}

Spectral types for the unresolved source Mk\,33N from the literature include O4 from \cite{melnick1985}, with spatial crowding hindering further progress until \cite{massey1998} utilised HST/FOS to obtain spectra of Mk\,33Na and b, from which classifications of O3\,If* and O6.5\,V, respectively, were obtained. From classification of our UVES spectroscopy, the criteria of \cite{sota2011} were followed and verified with the spectral templates from \cite{walborn2014}. The most unusual feature of Mk\,33Na is the clear presence of \ion{C}{iv}\,4658 and weakness of \ion{N}{v} 4603-20 plus the strength of the \ion{C}{iv}\,5801-12 doublet in emission suggestive of an OC star for the primary component. The line ratio of the nitrogen lines is similar to VFTS\,169 (O2.5\,V(n)((f*)), but with a much stronger stellar wind as seen in the emission line strength of \ion{He}{ii}\,4686 and $\mathrm{H}_{\alpha}$ which exceed the line strength of VFTS\,016 (O2\,III-If*, Fig.~\ref{f:vfts016}). VFTS\,016 and Mk33Na$_1$ have similar \ion{C}{iv}\,4658 line strength, but the nitrogen lines are stronger in VFTS\,016 while weaker in the Mk\,33Na$_1$ spectrum (\ion{C}{iv} 4658 $>$ \ion{N}{v} 4603--20). The carbon abundances ($\epsilon_{\rm C} = \log(\mathrm{C/H})+12 = 7.82$) of VFTS\,016 and Mk33Na$_1$ are comparable, but the nitrogen abundance ($\epsilon_{\rm N} = \log(\mathrm{C/H})+12 = 7.76$) of VFTS\,016 is higher by $\sim$0.5\,dex \citep[Table~\ref{t:stel_param}]{evans2010}. Therefore, we classified Mk\,33Na$_1$ as OC2.5\,If*.

The spectral type of the secondary (Mk\,33Na$_2$) is more difficult to determine. The absence of strong signatures of carbon and nitrogen lines is indicative for dwarfs with a relatively weak stellar wind. Based on the spectral types by \cite{walborn2014} we noticed that VFTS\,797 (O3.5\,V((n))((fc))) shows similar spectral properties such as weak C, N  and \ion{He}{i}\,4471 lines and relatively strong \ion{He}{ii}\,4542 absorption line. \ion{N}{iv}\,4058 is not or hardly visible in the spectrum and there might be a hint of \ion{N}{iii} 4634-41 and \ion{C}{iii} 4650. We concluded that the spectral type of Mk\,33Na$_2$ is approximately O4\,V.

\subsection{Line broadening}
\label{s:lb}
\begin{figure}
	\includegraphics[width=\columnwidth]{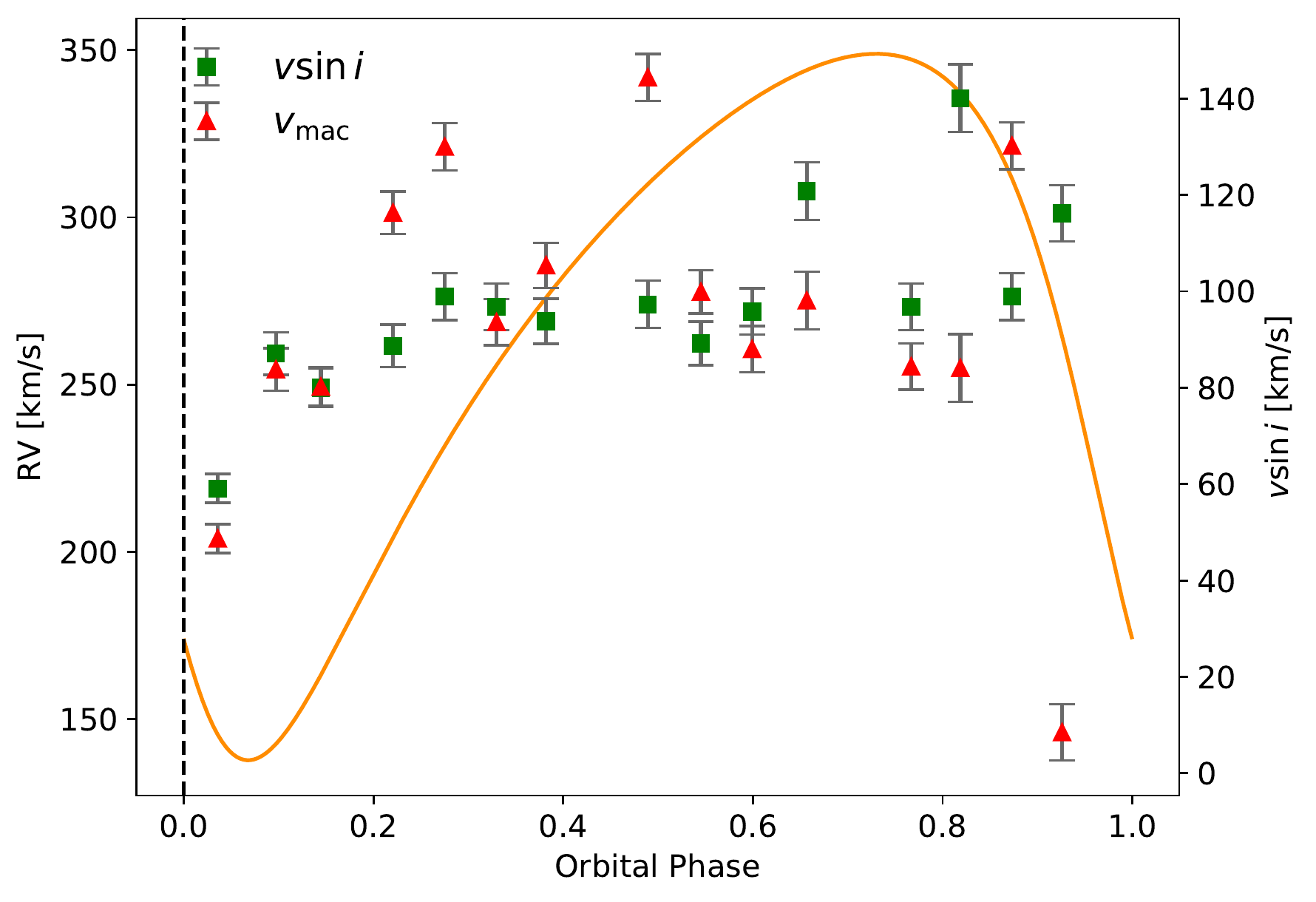}
    \caption{Projected rotational and macro-turbulent velocity of Mk\,33Na$_1$ as a function of orbital phase. The fit to the radial velocity measurements (solid orange line) is shown for reference.}  
    \label{f:vsini}
\end{figure}

If the binary system is tidally locked, the mean rotational velocity is $2\upi R_{\rm eff}/P$ based on stellar radius and period of the system (Table~\ref{t:orbit_param} and \ref{t:stel_param}). In an eccentric orbit, the projected rotational velocity ($\varv \sin i$) should vary throughout the orbit and lower or equal to the mean rotational velocity, e.g. $\lesssim 50$\,km/s for the Mk33Na$_1$. We used IACOB-BROAD \citep{simon-diaz2014} to derive $\varv \sin i$ and macro-turbulent velocity ($\varv_{\rm mac}$). IACOB-BROAD is an interactive analysis tool that combines Fourier transformation (FT) and goodness-of-fit methods. The code is publicly available and is written in the interactive data language (IDL).

We were only able to derive the line broadening for the primary based on the \ion{C}{iv}\,5801 line. Unfortunately, we were unable to employ metal lines in absorption to determine the line broadening which would provide more accurate measurements. We inferred $\varv \sin i$ using the first minimum of the FT. The macro-turbulent velocity was determined based on the goodness-of-fit while $\varv \sin i$ was set to the FT value. The measurements are listed in the Appendix (Table~\ref{t:line-broad}).

Figure~\ref{f:vsini} shows $\varv \sin i$ and $\varv_{\rm mac}$ as a function of phase. For reference the RV variations for the primary are included as well. Shortly after periastron both broadening velocities drop below 60\,km/s. With the exception of a couple of outliers the average for $\varv \sin i$ and $\varv_{\rm mac}$ lies around 95\,km/s. Averaged $\varv \sin i = 75.5\pm 0.6$ and $\varv_{\rm mac} = 110.3\pm4.1$ based on the measurements of \ion{N}{iv} and \ion{C}{iv}\,5801-12 (Table~\ref{t:line-broad}). The stellar spectra of the primary show large line profile variations (LPV) including the \ion{C}{iv}\,5801-12 doublet (Sect.~\ref{s:lpv}). The outliers could be due to the dynamics within the stellar atmosphere and wind. However, we can conclude that the primary is not tidally locked.

For the secondary we were only able to measure $\varv \sin i \approx 125$\,km/s and $\varv_{\rm mac}\approx10$\,km/s at the orbital phase $\sim 0.1$ using \ion{He}{ii}\,4542. The \ion{He}{ii} line is also broadened by the Stark effect and could lead to an overestimation of $\varv \sin i$ and $\varv_{\rm mac}$. The $\varv \sin i \approx 125$\,km/s can be considered as an upper limit for the secondary. $\varv \sin i$ and $\varv_{\rm mac}$ based on the median spectrum is around 78 and 56\,km/s for the \ion{C}{iv}\,5812 line (Table~\ref{t:line-broad}).

\subsection{Line profile variations\label{s:lpv}}

Line profiles of some lines varied significantly throughout the orbit, e.g. \ion{He}{ii}\,4686, \ion{N}{iv}\,4058 and wings of Balmer lines. 
Shortly after periastron \ion{He}{ii}\,4686 is weakest and \ion{N}{iv}\,4058 has nearly disappeared as well. At phase $\sim 0.1$ the line strength suddenly increases and settles down again after 1 day. This could be an outburst of material, but \ion{N}{iv}\,4058 is hardly affected. The emission-line strength increases until apastron at orbital phase 0.5 and decreases again with some variation towards periastron (Fig.~\ref{f:heii4684_ew}).

\begin{figure*}
	\includegraphics[width=0.9\textwidth]{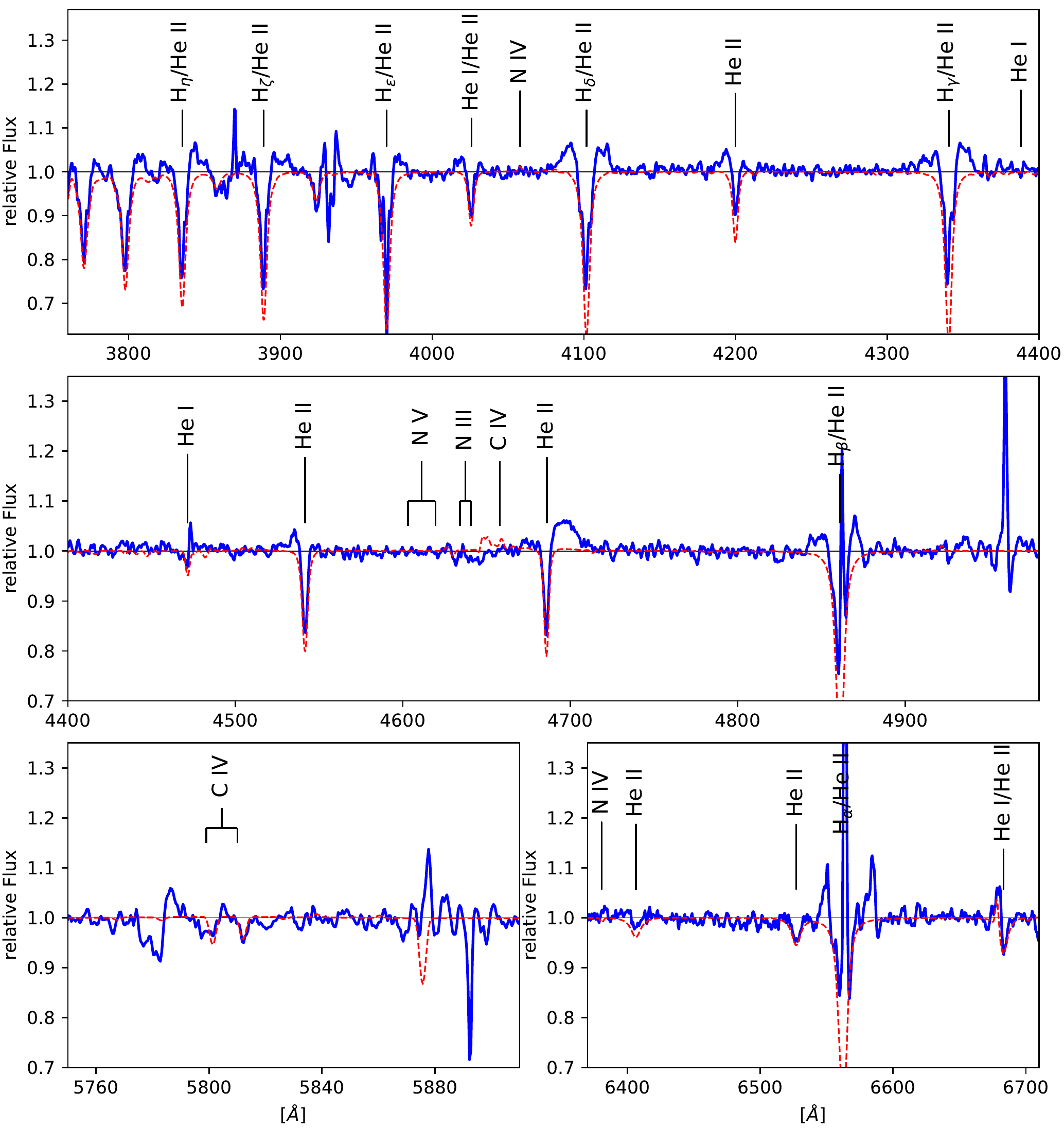}
    \caption{Spectroscopic analysis of Mk\,33Na$_2$ (O4\,V). The blue line is the disentangled median spectrum (Sect.~\ref{s:dis}) while red the theoretical spectrum.	\label{f:sec_dis}}
\end{figure*}

There are noticeable variations in the wings of the absorption lines $\mathrm{H}_{\beta}$, $\mathrm{H}_{\gamma}$, \ion{He}{ii}\,4200 and 4542 with filling in of an emission line component. As seen in Fig.~\ref{f:sec_dis} those features seem to belong to the secondary. Either the mass-loss rate varies significantly or those LPVs are caused by the stellar wind and ionising flux of the primary star. However, the \ion{C}{iv}\,5801-12 doublet is very sensitive to the mass-loss rate and a small increase would turn the lines from absorption into emission.

A future study will try to understand the nature and physical processes of those variations including the X-ray variability (Sect.~\ref{X-ray}).

\subsection{Spectroscopic analysis\label{s:sa}}

\begin{figure*}
	\includegraphics[width=0.9\textwidth]{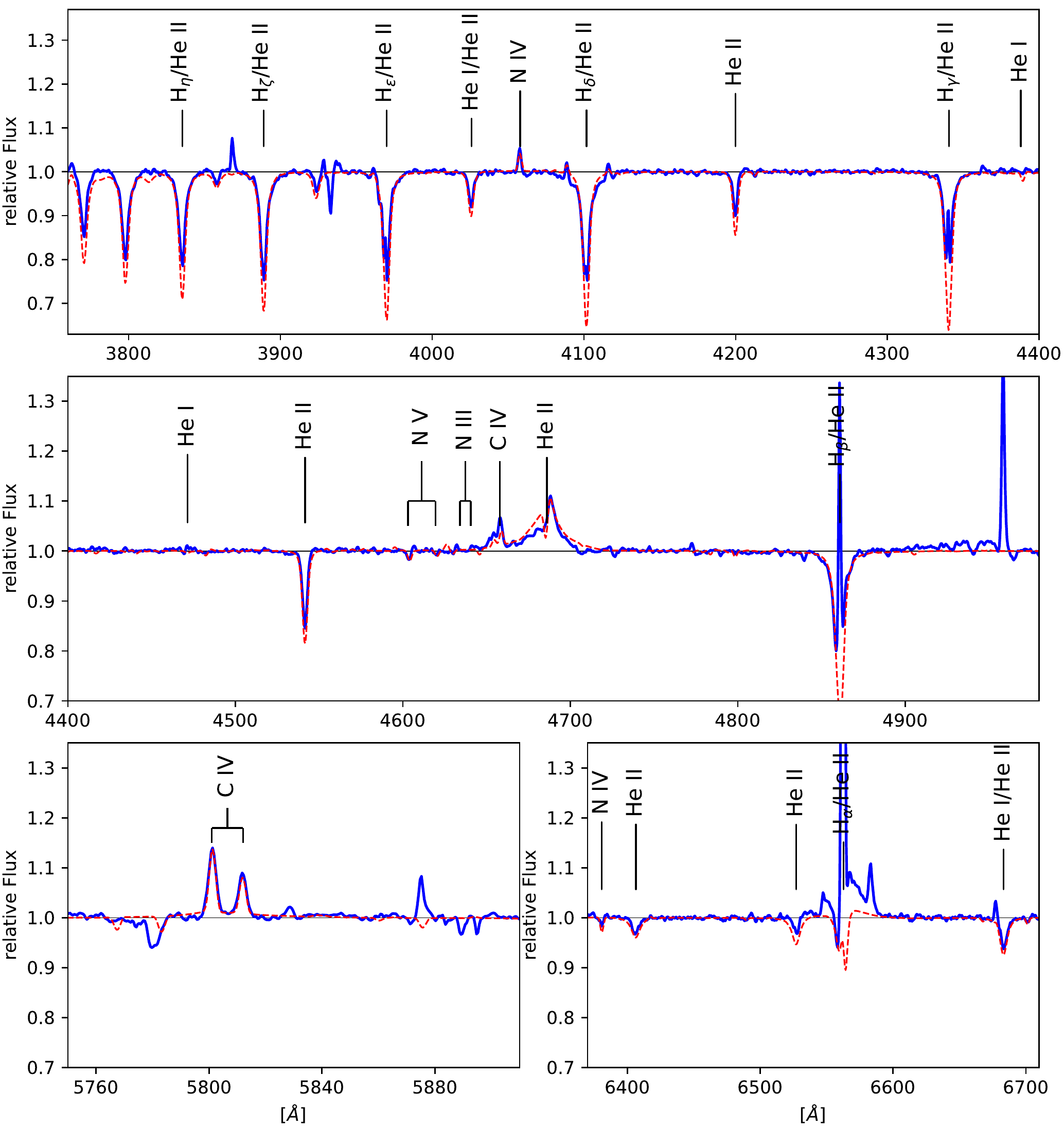}
    \caption{Spectroscopic analysis of Mk\,33Na$_1$ (OC2.5\,If*). The blue line is the disentangled median spectrum (Sect.~\ref{s:dis}) while red the theoretical spectrum.	\label{f:pri_dis}}
\end{figure*}

The spectra were analysed with the spherical, non-LTE stellar atmosphere and radiative transfer code CMFGEN \citep{hillier1998}. We used the grid of stellar atmospheres from \cite{bestenlehner2014} to roughly estimate the stellar parameters. Based on the initial guesses we computed a grid with an extended atomic model by varying the effective temperature $T_{\rm eff}$ defined at an optical depth $\tau = 2/3$, surface gravity $\log g$, mass-loss rate $\dot{M}$ and chemical abundances of carbon and nitrogen. The terminal velocity ($\varv_{\infty}$) could only be derived for Mk\,33Na$_1$ based on the line width of \ion{He}{ii}\,4686 and $\mathrm{H}_{\alpha}$ ($\varv_{\infty} \sim 2200$\,km/s). The wind volume filling factor ($f_{\mathrm{v}}$) was set to 0.1 and $\varv_{\rm mic}$ was set to 10\,km/s. The grid of stellar atmospheres contains the following element ions: \ion{H}{i}, \ion{He}{i-ii}, \ion{C}{ii-v}, \ion{N}{ii-v}, \ion{O}{ii-vi}, \ion{Ne}{ii-v}, \ion{Mg}{ii-iv}, \ion{Ca}{iii-v}, \ion{Si}{ii-vi}, \ion{P}{iii-v}, \ion{S}{iii-vi}, \ion{Fe}{iii-vii} and \ion{Ni}{iii-vi}.

We were only able to confidently derive stellar parameters for the primary. The effective temperature is based on the presence of \ion{C}{iv}\,4658 and the weakness of \ion{N}{v}\,4604-20 relative to \ion{N}{iv}\,4058. The surface gravity is derived from wings of $\mathrm{H} {\gamma-\epsilon}$. 
The mass-loss rate is determined from the emission line strength of \ion{He}{ii}\,4686 which varies throughout the orbit. $\mathrm{H}_{\alpha}$ was strongly contaminated by nebular lines and was not suitable. Due to CN cycle, the C abundance decreases and N increases during the main-sequence lifetime. However, the primary is still carbon-rich  close to the LMC baseline of ($\epsilon_{C} - \epsilon_{N})_{\rm LMC} = 0.85$ dex \citep{brott2011}, with $\epsilon_{C} - \epsilon_{N} = (\epsilon_{C} - \epsilon_{N})_{\rm LMC} - 0.35 \pm 0.4$ dex, suggesting a young age for the binary system of less than $\sim 1$\,Myr.

The determination of stellar parameters for the secondary is less straightforward because of the emission line components in the wings of the Balmer and helium lines. As the $\log g$ is derived from the broadening of the wings of the Balmer lines, we were unable to determine the surface gravity for the secondary.
The absence or the difficult identification of \ion{N}{iii}\,4634-41, \ion{N}{iv}\,4058, \ion{N}{v}\,4604-20, \ion{C}{iii}\,4650, the absorption of  \ion{C}{iv}\,5801-12 and the presence of \ion{He}{i}\,4471 suggest a $T_{\rm eff} = 45\,000\pm 2500$\,K for the secondary. \ion{C}{iv}\,5801-12 turns from absorption into emission if the mass-loss rate is increased. Other lines are hardly effected and we are only able to determine an upper mass-loss rate limit of $\log \dot{M} [M_{\odot}/yr] < -6.5$.

Physical, wind and chemical properties of Mk\,33Na are provided in Table~\ref{t:stel_param}. The only previous spectroscopic analysis of the system was undertaken by \cite{castro2021} who analysed \ion{He}{i} 4921 and \ion{He}{ii} 5411 lines of a large sample of 30 Doradus stars using VLT/MUSE spectroscopy \citep[CCE]{castro2018} using a grid of FASTWIND models to obtain $T_{\rm eff}$ = 36kK and $\log L/L_{\odot}$ = 5.93, $\varv \sin i$ = 170 km/s for Mk\,33Na (= CCE 1943) and an approximate spectral type of "O5".

\subsection{Stellar luminosities}\label{s:lum}

\begin{figure*}	
	\includegraphics[width=0.9\textwidth]{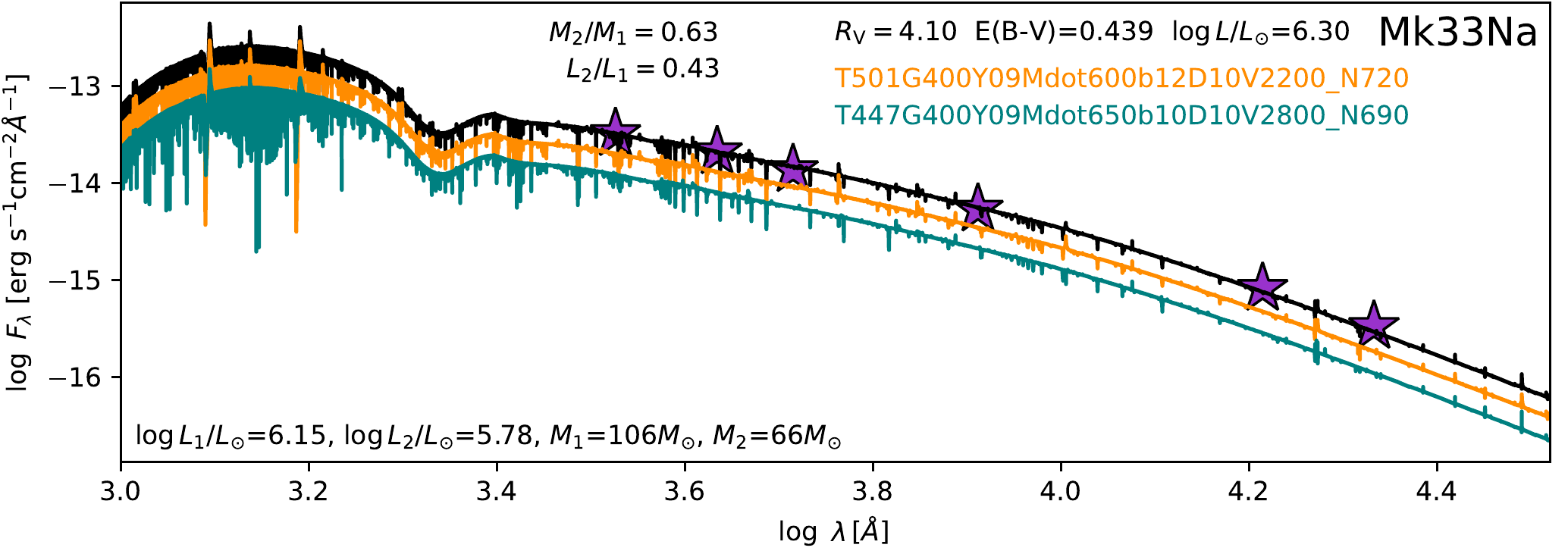}
    \caption{Combined and individual SEDs of Mk\,33Na$_{1,2}$ (solid black, orange and cyan-green line). Derived luminosities based on the mass ratio derived using the orbital parameters (Sect.~\ref{s:orbit_par}). Purple stars represent HST/WFC3 F336W, F438W and F555W and VLT/MAD H and K-band photometry.\label{f:sed}}
\end{figure*}

\begin{table}
	\centering
	\caption{Total bolometric luminosity of Mk\,33Na and the reddening parameters. Note: the error given for the combined $L$ only represents the uncertainties of the reddening parameters.
}
	\label{t:lum_red}
	\begin{tabular}{lcccc}
		\hline
		$L_{\rm Bol,2}$	&$\log L_{\rm Bol,1+2}$& $R_{5495}$ &  $E(4405-5495)$ & $M_{5495}$ \\
		$/L_{\rm Bol,1}$          & $/L_{\odot}$                  & mag             & mag                     & mag \\
		\hline
		0.43		&$6.30\pm0.06$		& $4.10\pm0.34$	& $0.439\pm0.007$	& --6.63	\\
		\hline
	\end{tabular}
\end{table}

Bolometric luminosities ($L$) of both components are derived by matching optical and near-infrared (near-IR) photometric data with the combined theoretical spectral energy distribution (SED). We employed optical HST/WFC3 photometry $F336W = 12.506$\,mag, $F438W = 13.765$\,mag, $F555W = 13.647$\,mag and $F814W = 13.221$\,mag from \cite{deMarchi2011} and near-IR VLT/MAD $H = 12.85$\,mag and $K_{\rm s} = 12.77$\,mag \citep{campbell2010}. We corrected the observed magnitudes for the distance to the LMC (49.6 kpc) by adopting the distance modulus of 18.48\,mag \citep{pietrzynski2019} before converting them into fluxes.

The combined bolometric luminosity as well as the reddening parameters $R_{5495}$ and $E(4405-5495)$ were obtained by fitting the observed and distance corrected fluxes with the combined model SED and reddening law by \cite{ma-ap2014} with the {\sc lmfit} routine: non-linear least-squares minimization and curve-fitting for Python\footnote{\url{https://lmfit.github.io/lmfit-py/index.html}}. We used the nomenclature according to \cite{ma-ap2014} instead of the more common $R_{V}$ and $E(B-V)$.

The combined model SED is inferred using the mass ratio of the Mk\,33Na system (Sect.~\ref{s:orbit_par}) and the luminosity\,--\,mass relation (LMR) by \citet[Eq. 9 with coefficients from row 1 of Table A.1]{graefener2011} for chemical-homogeneously core-hydrogen burning stars to calculate the flux ratio of the binary system for a given primary stellar mass (Fig.~\ref{f:sed}). The LMR requires a hydrogen surface abundances as input. We assumed the baseline abundances for hydrogen in mass-fraction according to \citet[$X=0.7391$]{brott2011}, which is supported by the unprocessed C and N abundances (Section.~\ref{s:sa} and \ref{s:mass_age_i}).

Under these assumptions the luminosity ratio is $L_2/L_1 = 0.43$ with a flux ratio in the optical of 0.62. The flux ratio is larger than the value of 0.25 assumed for the RV measurements in Sect.~\ref{s:rv}. Looking at Fig.~\ref{f:sec_dis} and the spectroscopic fit we can see that the Balmer and helium absorption lines of the secondary are filled in by an emission line component.


Reddening parameters and total bolometric luminosity of the Mk\,33Na system are given in Table~\ref{t:lum_red} while individual $L$ for the primary and secondary are listed in Table~\ref{t:stel_param}. 
The error provided for the combined $L$ only represents the uncertainties of the reddening parameters. $R_{5495}$ and $E(4405-5495)$ are typical values for the stellar cluster R136 and its immediate surroundings \citep[e.g.][]{doran2013, bestenlehner2020b}. Armed with $\log L_{\rm X} = 34.01 \pm 0.05$ erg\,s$^{-1}$ from Sect.~\ref{X-ray}, we confirm that Mk\,33Na has an usually high $L_{\rm X}/L_{\rm Bol}$ = 10$^{-5.9}$ consistent with excess X-ray emission from wind-wind collisions. 

\section{Stellar masses, ages and orbital inclination}\label{s:mass_age_i}

Our spectroscopic  results for Mk\,33Na favour masses of 106 $M_{\odot}$ and 66 $M_{\odot}$ for the primary and secondary based on the LMR for chemical homogeneously evolving hydrogen burning stars of \cite{graefener2011}. Therefore, those masses can be seen as an upper mass-limit for the Mk\,33Na system. This can be reconciled with results from our orbital solution, $M_{1,2} \sin^3 i = 20.0$ and $12.6 M_{\odot}$ if the orbital inclination of Mk\,33Na is $\sim 35^{\circ}$.

For LMC metallicity and He baseline chemical abundances the star approaches the Eddington limit at an Eddington parameter $\Gamma_{\rm e} \approx 0.7$ considering only the electron opacity $\chi_{\rm e}$. With the assumption that a star cannot exist above the Eddington limit we derive lower mass limits of $\sim 39$ and $\sim 17\,M_{\odot}$ for Mk\,33Na$_{1,2}$. Taking the dynamical mass ratio into account the minimum for Mk\,33Na$_{2}$ would correspond to $\sim 25\,M_{\odot}$. This results in a maximum orbital inclination of $< 53^{\circ}$ which is a true upper limit as stars in proximity to Eddington limit experience enhanced mass loss with strong emission lines in their spectra and would be of spectral type WN rather than Of* \citep[e.g.][]{bestenlehner2014, bestenlehner2020}. We conclude that the orbital inclination is in the range between 35 and 53$^{\circ}$.

We have also estimated the physical properties of the Mk\,33Na components by utilising BONNSAI\footnote{\url{https://www.astro.uni-bonn.de/stars/bonnsai/}} \cite{schneider2014} which applies a Bayesian approach to the determination of stellar masses and ages based on evolutionary models, for which we employ LMC metallicity models from \citet{brott2011} and \citet{koehler2015}. BONNSAI results are presented in Table~\ref{t:stel_param} and reveal evolutionary masses of 83 and 48 $M_{\odot}$ for the components of Mk\,33Na plus individual ages of 0.9$\pm$0.6 and $1.6^{+0.7}_{-1.3}$ Myr, such that the favoured age of the system is $\sim$1.1 Myr. Evolutionary masses favour an orbital inclination of $\sim 38^{\circ}$. Alternate evolutionary masses and ages based on the optical flux ratio solution are also presented in Table~\ref{t:stel_param}, from which a similar age is obtained, albeit with 91 and 42 $M_{\odot}$ for systemic components, contrary to the dynamical mass ratio.

Near Mk\,33Na is the star cluster R136. By analysing the stellar population of R136 \cite{bestenlehner2020b} discovered a positive mass-discrepancy, where the spectroscopic masses are systematically larger than evolutionary masses for stars more massive than $\sim 40\,M_{\odot}$. Based on a sample of eclipsing binaries with known inclination \cite{mahy2020} showed that spectroscopic masses are in better agreement with dynamical than evolutionary masses. Therefore, the actual masses of both components might be larger than the values obtained above with BONNSAI. 

Unusually for an O supergiant, the primary in Mk\,33Na has an OC spectral type since \ion{C}{iv} 4658 $>$ \ion{N}{v} 4603--20. The LMC baseline abundance difference between carbon and nitrogen is ($\epsilon_{C} - \epsilon_{N})_{\rm LMC} = 0.85$ dex, with $\epsilon_{C} - \epsilon_{N} = (\epsilon_{C} - \epsilon_{N})_{\rm LMC} - 0.35 \pm 0.4$ dex obtained from our spectroscopic analysis. The favoured BONNSAI solutions involve no chemical enrichment for Mk\,33Na, as summarised in Table~\ref{t:stel_param}.

\begin{table*}
	\centering
	\caption{Physical properties of  double-lined spectroscopy binaries in the LMC whose minimum primary mass exceeds $\sim 20 M_{\odot}$. These have been identified from 
	optical photometric eclipses (Phot: e.g. MACHO, OGLE), optical spectroscopic radial velocity variability (Spec: e.g. VFTS, \citealp{evans2011}; TMBM: \citealp{almeida2017}) or X-ray variability (X-ray: Broos \& Townsley, in prep). Mk\,33Na is amongst the highest mass systems, according to spectroscopic or evolutionary models, ranking joint third with R139 after Mk\,34 and R144.}
	\label{t:massive_binaries}
	\begin{tabular}{l@{\hspace{0.5mm}}l@{\hspace{0.5mm}}l@{\hspace{0.5mm}}c@{\hspace{0.5mm}}c@{\hspace{0.5mm}}c@{\hspace{0.5mm}}c@{\hspace{0.5mm}}c@{\hspace{0.5mm}}c@{\hspace{0.5mm}}c@{\hspace{0.5mm}}c@{\hspace{0.5mm}}c@{\hspace{0.5mm}}c@{\hspace{0.5mm}}l@{\hspace{0.5mm}}l} 
		\hline
System & Alias & Spect.Type& $P_{\rm orb}$      & $e$ & $M_{\rm orb,1} \sin^{3} i$ & $q$  & $i_{\rm orb}$        & $M_{\rm orb,1}$ & $M_{\rm sp,1}$ & $M_{\rm evol, 1}$ & $i_{\rm evol}$ & Survey & Ref\\
             &          &                   & days &    & $M_{\odot}$         &                                & & $M_{\odot}$ & $M_{\odot}$ &  $M_{\odot}$ &    & & \\
		\hline
R139      &  VFTS 527  & O6.5\,Iafc+O6\,Iaf & 153.94$\pm$0.01 & 0.38$\pm$0.02 & 69.4$\pm$4.1 & 0.78$\pm$0.02 & --                   &        --    & 89.7$\pm$19.1         & 81.6$^{+7.5}_{-7.2}$ & 71$^{\circ}$ & Spec  & a\\
Mk\,34    &  HSH95 8  & WN5h+WN5h          & 154.55$\pm$0.05 & 0.68$\pm$0.02 & 65$\pm$7     & 0.92$\pm$0.07 & --                   &        --    & 147$\pm$22            & 139$^{+21}_{-18}$    & 50$^{\circ}$ & X-ray & b\\
		HSH95 38 &  VFTS 1019 & O3\,V+O6\,V        & 3.39$^{\dagger}$& 0.00$^{\ddagger}$& 53.8         & 0.41          & 79$\pm1^{\circ}$     & 56.9$\pm$0.6 & --                    & 53$\pm$5             & 90$^{\circ}$ & Phot  & c\\
R144      &  HD 38282  & WN5-6h+WN6-7h     & 74.207$\pm$0.004&0.506$\pm$0.004& 48.3$\pm$1.8 & 0.94$\pm$0.02 & 60.4$\pm1.5^{\circ}$ & 74$\pm$4     & --                    & 111$\pm$12           & 50$^{\circ}$ & Spec  & d\\
VFTS 508  &            & O9.5\,V            &128.586$\pm$0.025& 0.44$\pm$0.03 & 44.9$\pm$5.5 & 0.74$\pm$0.05 & --                   & --           &  30.2$\pm$6.8         & 19.8$^{+1.0}_{-0.7}$ &          --  & Spec  & a\\
		HSH95 42 &            & O3\,V+O3--4\,V     & 2.89$^{\dagger}$& 0.00$^{\ddagger}$& 39.9$\pm$0.1 & 0.81          & 85.4$\pm0.5^{\circ}$ & 40.3$\pm$0.1 & --                    & 42$\pm$2             & 79$^{\circ}$ & Phot  & c\\
		Sk --67$^{\circ}$ 105 &&  O4f + O6\,V       & 3.30$^{\dagger}$& 0.00$^{\ddagger}$& 38.5$\pm$0.6 & 0.65$\pm$0.04 & 68$\pm 1^{\circ}$    & 48.3$\pm$0.7 & --                    & --                   &           -- & Phot  & g\\ 
VFTS 63   &            & O5\,III(n)(fc)+    & 85.77$\pm$0.07  & 0.65$\pm$0.04 & 38.0$\pm$13.5& 0.54$\pm$0.09 & --                   & --           &$68.2^{+21.4}_{-17.1}$ & 47.8$^{+3.8}_{-3.9}$ & 68$^{\circ}$ & Spec  & a\\
		W61 28-22 & LH 81-72   & O7\,V+O8\,V        & 4.25$^{\dagger}$& 0.00$^{\ddagger}$& 30.9$\pm$1.0 & 0.42$\pm$0.02 & 89.9$\pm 0.9^{\circ}$& 30.9$\pm$1.0 & --                    & --                   &           -- & Phot  & f\\
		VFTS 176  &          & O6\,V:((f))+O9.5:\,V:& 1.78$^{\dagger}$& 0.00$^{\ddagger}$& 28.3$\pm$1.5 & 0.62$\pm$0.02 & --                   & --           & 24.6$\pm$1.3          & 32.2$^{+1.2}_{-1.4}$ & 73$^{\circ}$ & Spec  & a\\ 
		HSH95 77 &            & O5.5\,V+O5.5\,V    & 1.88$^{\dagger}$& 0.00$^{\ddagger}$& 28.3$\pm$0.1 & 0.90          & 83$\pm 1^{\circ}$    & 28.9$\pm$0.3 & --                    & 28$\pm$1             & 90$^{\circ}$ & Phot  & c\\
		Mk\,30    & VFTS 542   & O2\,If*/WN5+B0\,V  & 4.6965$^{\dagger}$&0.00$^{\ddagger}$& 25.3$\pm$2.8 & 0.32$\pm$0.04 & --               & --           & 101         & 53$^{+20}_{-15}$      & 52$^{+6\circ}_{-5}$   & Spec  & h\\ 
		HSH95 39 & VFTS 1005  & O3\,V+O5.5\,V      & 4.06$^{\dagger}$& 0.00$^{\ddagger}$& 24.5$\pm$0.1 & 0.75          &$<75^{\circ}$         & $>27.2$      & --                    & 46$\pm$2             & 54$^{\circ}$ & Phot  & c\\
BAT99-019 & HD 34169   & WN3+O6\,V          & 17.998(2)       & 0.00$^{\ddagger}$& 22.1$\pm$2.6 & 1.79$\pm$0.05 & 86$^{+4\circ}_{-3}$&22$\pm$3&  21                   &  --                  & --           & Spec & h\\  
		VFTS 450  &            & O9.7\,III: + O7::  & 6.89$^{\dagger}$& 0.00$^{\ddagger}$& 20.8$\pm$2.8 & 0.96$\pm$0.08 & --                   & --           &36.1$^{+10.4}_{-11.7}$ & 25.0$^{+4.0}_{-1.3}$ & 70$^{\circ}$ & Spec  & a\\
		VFTS 661  &          & O6.5\,V(n)+O9.7:\,V: & 1.266$^{\dagger}$& 0.00$^{\ddagger}$& 20.1$\pm$0.6 & 0.71$\pm$0.01 & --                   & --           &  25.8$^{+4.2}_{-2.4}$ & 26.0$^{+1.3}_{-1.0}$ & 67$^{\circ}$ & Spec  & a\\ 
Mk\,33Na  & HSH95 16  & OC2.5\,If*+O4\,V   & 18.32$\pm$0.14  & 0.33$\pm$0.01 & 20.0$\pm$1.2 & 0.63$\pm$0.02 &  --                  &          --  & 90$^{+25}_{-18}$      & 83$\pm$19            & 38$^{\circ}$ &  X-ray & i\\
		\hline
	\end{tabular}
	$^{\dagger}$: undefined uncertainties; $^{\ddagger}$: eccentricities fixed at zero; a: \citealp{mahy2020}; b: \citealp{tehrani2019}; c: \citealp{massey2002}; d: \citealp{shenar2021}; e: \citealp{massey2012}; f: \citealp{bonanos2009}; g: \citealp{ostrov2003}; h:~\citealp{shenar2019}; i: This work
\end{table*}

\section{Massive binaries in the LMC}
\label{s:disc}

\begin{figure*}
	\includegraphics[width=0.9\textwidth]{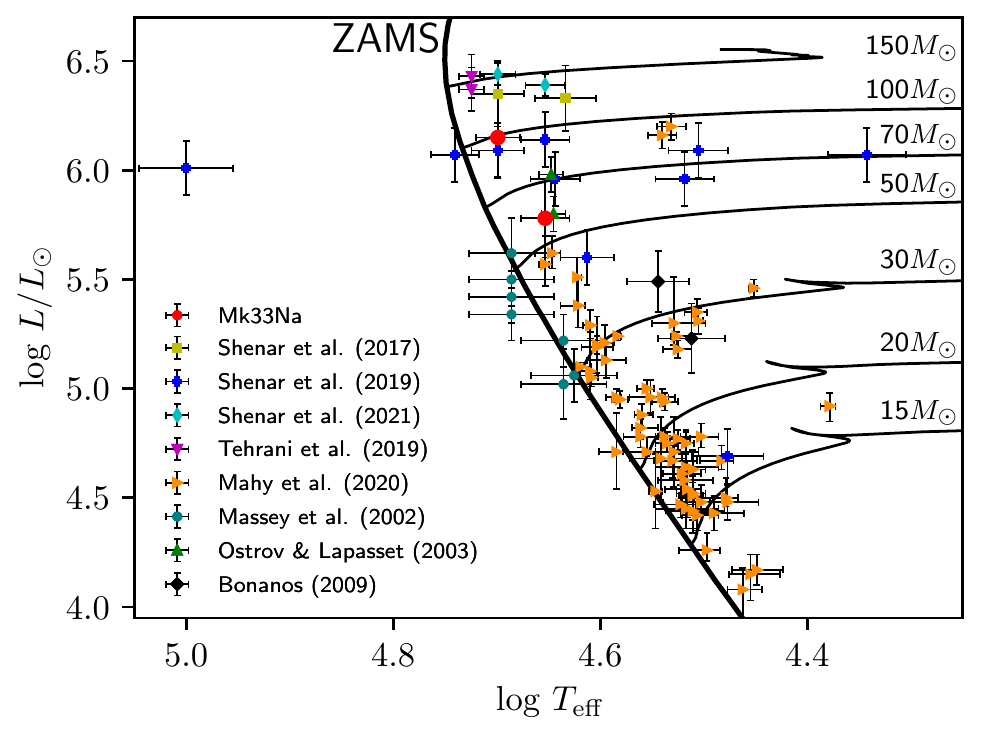}
    \caption{HRD of massive binary systems in the LMC. Binaries are included from this work (Mk\,33Na), \citet{shenar2017, shenar2019, shenar2021, tehrani2019, mahy2020, massey2002, ostrov2003, bonanos2009}. Non-rotating evolutionary tracks and zero-age main-sequence isochrone are from \citet{brott2011,koehler2015}.	\label{f:hrd}}
\end{figure*}

Historically HSH95 38 (O3\,V+O6\,V) was the only massive eclipsing binary system in the LMC with a primary whose mass exceeded $\sim 50 M_{\odot}$ \citep{massey2002}. A number of other eclipsing binary systems have been discovered via OGLE, MACHO or other photometric surveys, including Sk --67$^{\circ}$ 105 \citep{ostrov2003} and W61 28-22 \citep{bonanos2009}. The advent of multi-epoch optical spectroscopic surveys involving hundreds of OB stars, such as VFTS \citep{evans2011}, and subsequent follow up studies \citep[TMBM,][]{almeida2017}, has led to the discovery of several non-eclipsing systems involving O supergiants or H-rich WN stars such as R139 \citep{taylor2011, mahy2020} and R144 \citep{shenar2021}. A compilation of massive binary systems is shown in Fig.~\ref{f:hrd} and listed in Table~\ref{t:massive_binaries}.

Optical spectroscopic studies are ideal methods of identifying relatively short period binaries with high inclinations. X-ray monitoring, such as T-ReX (Broos \& Townsley, in prep.), provides an alternative approach to identifying binaries via the detection of excess X-ray emission, arising from wind-wind collisions \citep{pollock1987}. Crowther et al. (2021) have shown that SB2 systems have higher $L_{X}/L_{\rm Bol}$ ratios than single or SB1 systems detected in T-ReX.

T-ReX has previously identified Mk34 as an exceptional colliding wind binary involving two luminous H-rich WN stars in a 155 day orbit, with evolutionary masses of 139 and 127 $M_{\odot}$. Here, we have confirmed the suspicion from T-ReX that Melnick 33Na is another massive short-period (18.3 day) binary comprising an OC2.5\,If* primary with an evolutionary mass of  $83 \pm 19 M_{\odot}$ and an O4\,V secondary with a mass of $48 \pm 11 M_{\odot}$. Table~\ref{t:massive_binaries} provides a summary of the most massive double-lined spectroscopic binaries in the LMC, for which the primary of Mk\,33Na has the (joint) third highest mass according to spectroscopic/evolutionary models, after Mk34 and R144, and comparable to R139.

Figure~\ref{f:hrd} shows a Hertzsprung-Russell diagram (HRD) of binary systems in the LMC, where stellar parameters are available for both components. Mk33\,Na$_1$ has the earliest O star spectral type known in a binary system. In the HRD the star is located in a region which is populated by Of/WN and WNh stars \citep{shenar2017, tehrani2019, shenar2019, shenar2021}. The only O stars with a similar luminosity is the O supergiant binary system R139 \citep{taylor2011, mahy2020}, but of later spectral type. Mk33\,Na$_1$ still has a high carbon abundance at the surface while stars of similar luminosity show chemical enrichment due to the CNO-cycle, N enriched and C deficient. Some of those luminous stars are even He enriched or potentially He core-burning. This supports the young age of $\sim$1\,Myr for the Mk\,33Na system.

Looking at the top and high luminosity end of the HRD (Fig.~\ref{f:hrd}) and the stellar masses and mass ratio of those objects (Table~\ref{t:massive_binaries}) it is apparent that most of those binary systems have $q$ close to unity and consist of stars with similar luminosities which is likely an observational bias. Binaries can be easier identified if either both components have similar luminosities or the spectrum has high SNR and high spectral resolution (SB2).  Photometrically eclipsing binaries with high inclinations or a large monitoring program such as TMBM are able to identify and characterise binary systems over a wide range of mass ratios with $q$ between 0.35 and 1 \citep[e.g.][Table~\ref{t:massive_binaries}]{massey2002, bonanos2009, mahy2020}. Those circumstances have left Mk\,33Na undetected as a binary system for several decades. Only now has the X-ray variability due to colliding winds finally revealed the multiple nature of Mk\,33Na.


\section{Conclusions}\label{conclusions}

We establish Melnick 33Na (Mk\,33Na) in the Tarantula Nebula as a double-lined spectroscopic binary, from an analysis of VLT/UVES time-series spectroscopy,
with the following properties
\begin{itemize}
\item An orbital period of 18.3$\pm$0.1 days, supporting the period estimate from analysis of X-ray observations obtained with the T-ReX survey, eccentricity of $\sim$0.33 with a mass ratio of $q = M_{2}/M_{1} = 0.63$. 
\item A primary component with spectral type OC2.5\,If* with a minimum dynamical mass of 20 $M_{\odot}$, although a much higher mass obtained from spectroscopic or evolutionary models (83$\pm19 M_{\odot}$ and an age of 0.9$\pm$0.6 Myr via BONNSAI) since our favoured spectroscopic luminosity is $\log L/L_{\odot} = 6.15 \pm 0.18$. Unusually, $\epsilon_{C} - \epsilon_{N} = (\epsilon_{C} - \epsilon_{N})_{\rm LMC} - 0.35 \pm 0.4$ dex, also supporting a young age for the system of $\leq$1 Myr.
\item A secondary component with spectral type O4\,V with a minimum dynamical mass of 12.6 $M_{\odot}$, and again a substantially higher mass from spectroscopic or evolutionary models 
(48$\pm11 M_{\odot}$ and an age of 1.6$^{+0.7}_{-1.3}$ Myr via BONNSAI) since our favoured spectroscopic luminosity is $\log L/L_{\odot} = 5.78 \pm 0.18$. 
\item Evolutionary mass estimates of the primary of Mk\,33Na rank it (joint) third in terms of the most massive SB2 systems in the LMC after Mk\,34 (WN5h+WN5h) and R144 (WN5--6+WN6--7h) and comparable to R139 (O6.5\,Iafc+O6\,Iaf). Evolutionary and dynamical solutions can be reconciled if $i \sim 38^{\circ}$. Mk\,33Na is confirmed as a colliding wind binary with excess X-ray emission from T-ReX and $L_{\rm X}/L_{\rm Bol}$ = 10$^{-5.9}$.
\end{itemize}

\section*{Acknowledgements}
We thank the referee, Paco Najarro, for detailed and helpful comments which improved the clarity and content of the manuscript. JMB would like to thank STFC for financial support through STFC consolidated grant ST/V000853/1. PSB and LKT are supported by the Penn State ACIS Instrument Team Contract SV4-74018, issued by the Chandra X-ray Center, which is operated by the Smithsonian Astrophysical Observatory for and on behalf of NASA under contract NAS8-03060.

\section*{Data Availability}
The data underlying this article are available in the article and in its supplementary material. Spectroscopic data will be available via the ESO archive facility while synthetic spectra can be requested from the lead author. X-ray data are available in the Chandra Data Archive and the Chandra Source Catalogue and will be published in Townsley et al., Broos et al., in preparation.
 



\bibliographystyle{mnras}
\bibliography{reference} 

\begin{thebibliography}{}
\makeatletter
\relax
\def\mn@urlcharsother{\let\do\@makeother \do\$\do\&\do\#\do\^\do\_\do\%\do\~}
\def\mn@doi{\begingroup\mn@urlcharsother \@ifnextchar [ {\mn@doi@}
  {\mn@doi@[]}}
\def\mn@doi@[#1]#2{\def\@tempa{#1}\ifx\@tempa\@empty \href
  {http://dx.doi.org/#2} {doi:#2}\else \href {http://dx.doi.org/#2} {#1}\fi
  \endgroup}
\def\mn@eprint#1#2{\mn@eprint@#1:#2::\@nil}
\def\mn@eprint@arXiv#1{\href {http://arxiv.org/abs/#1} {{\tt arXiv:#1}}}
\def\mn@eprint@dblp#1{\href {http://dblp.uni-trier.de/rec/bibtex/#1.xml}
  {dblp:#1}}
\def\mn@eprint@#1:#2:#3:#4\@nil{\def\@tempa {#1}\def\@tempb {#2}\def\@tempc
  {#3}\ifx \@tempc \@empty \let \@tempc \@tempb \let \@tempb \@tempa \fi \ifx
  \@tempb \@empty \def\@tempb {arXiv}\fi \@ifundefined
  {mn@eprint@\@tempb}{\@tempb:\@tempc}{\expandafter \expandafter \csname
  mn@eprint@\@tempb\endcsname \expandafter{\@tempc}}}

\bibitem[\protect\citeauthoryear{{Almeida} et~al.,}{{Almeida}
  et~al.}{2017}]{almeida2017}
{Almeida} L.~A.,  et~al., 2017, \mn@doi [\aap] {10.1051/0004-6361/201629844},
  \href {https://ui.adsabs.harvard.edu/abs/2017A&A...598A..84A} {598, A84}

\bibitem[\protect\citeauthoryear{{Andersen}}{{Andersen}}{1991}]{andersen1991}
{Andersen} J.,  1991, \mn@doi [\aapr] {10.1007/BF00873538}, \href
  {https://ui.adsabs.harvard.edu/abs/1991A&ARv...3...91A} {3, 91}

\bibitem[\protect\citeauthoryear{{Bestenlehner}}{{Bestenlehner}}{2020}]{bestenlehner2020}
{Bestenlehner} J.~M.,  2020, \mn@doi [\mnras] {10.1093/mnras/staa474}, \href
  {https://ui.adsabs.harvard.edu/abs/2020MNRAS.493.3938B} {493, 3938}

\bibitem[\protect\citeauthoryear{{Bestenlehner} et~al.,}{{Bestenlehner}
  et~al.}{2014}]{bestenlehner2014}
{Bestenlehner} J.~M.,  et~al., 2014, \mn@doi [\aap]
  {10.1051/0004-6361/201423643}, \href
  {http://adsabs.harvard.edu/abs/2014A%26A...570A..38B} {570, A38}

\bibitem[\protect\citeauthoryear{{Bestenlehner} et~al.,}{{Bestenlehner}
  et~al.}{2020}]{bestenlehner2020b}
{Bestenlehner} J.~M.,  et~al., 2020, \mn@doi [\mnras] {10.1093/mnras/staa2801},
  \href {https://ui.adsabs.harvard.edu/abs/2020MNRAS.499.1918B} {499, 1918}

\bibitem[\protect\citeauthoryear{{Bonanos}}{{Bonanos}}{2009}]{bonanos2009}
{Bonanos} A.~Z.,  2009, \mn@doi [\apj] {10.1088/0004-637X/691/1/407}, \href
  {https://ui.adsabs.harvard.edu/abs/2009ApJ...691..407B} {691, 407}

\bibitem[\protect\citeauthoryear{{Bonanos} et~al.,}{{Bonanos}
  et~al.}{2004}]{bonanos2004}
{Bonanos} A.~Z.,  et~al., 2004, \mn@doi [\apjl] {10.1086/423671}, \href
  {https://ui.adsabs.harvard.edu/abs/2004ApJ...611L..33B} {611, L33}

\bibitem[\protect\citeauthoryear{{Brott} et~al.,}{{Brott}
  et~al.}{2011}]{brott2011}
{Brott} I.,  et~al., 2011, \mn@doi [\aap] {10.1051/0004-6361/201016113}, \href
  {http://adsabs.harvard.edu/abs/2011A%26A...530A.115B} {530, A115}

\bibitem[\protect\citeauthoryear{{Campbell}, {Evans}, {Mackey}, {Gieles},
  {Alves}, {Ascenso}, {Bastian}  \& {Longmore}}{{Campbell}
  et~al.}{2010}]{campbell2010}
{Campbell} M.~A.,  {Evans} C.~J.,  {Mackey} A.~D.,  {Gieles} M.,  {Alves} J.,
  {Ascenso} J.,  {Bastian} N.,   {Longmore} A.~J.,  2010, \mn@doi [\mnras]
  {10.1111/j.1365-2966.2010.16447.x}, \href
  {https://ui.adsabs.harvard.edu/abs/2010MNRAS.405..421C} {405, 421}

\bibitem[\protect\citeauthoryear{{Castro}, {Crowther}, {Evans}, {Mackey},
  {Castro-Rodriguez}, {Vink}, {Melnick}  \& {Selman}}{{Castro}
  et~al.}{2018}]{castro2018}
{Castro} N.,  {Crowther} P.~A.,  {Evans} C.~J.,  {Mackey} J.,
  {Castro-Rodriguez} N.,  {Vink} J.~S.,  {Melnick} J.,   {Selman} F.,  2018,
  \mn@doi [\aap] {10.1051/0004-6361/201732084}, \href
  {https://ui.adsabs.harvard.edu/abs/2018A&A...614A.147C} {614, A147}

\bibitem[\protect\citeauthoryear{{Castro} et~al.,}{{Castro}
  et~al.}{2021}]{castro2021}
{Castro} N.,  et~al., 2021, \mn@doi [\aap] {10.1051/0004-6361/202040008}, \href
  {https://ui.adsabs.harvard.edu/abs/2021A&A...648A..65C} {648, A65}

\bibitem[\protect\citeauthoryear{{Crowther}}{{Crowther}}{2019}]{crowther2019}
{Crowther} P.~A.,  2019, \mn@doi [Galaxies] {10.3390/galaxies7040088}, \href
  {https://ui.adsabs.harvard.edu/abs/2019Galax...7...88C} {7, 88}

\bibitem[\protect\citeauthoryear{{Crowther} \& {Dessart}}{{Crowther} \&
  {Dessart}}{1998}]{crowther1998}
{Crowther} P.~A.,  {Dessart} L.,  1998, \mn@doi [\mnras]
  {10.1046/j.1365-8711.1998.01400.x}, \href
  {http://adsabs.harvard.edu/abs/1998MNRAS.296..622C} {296, 622}

\bibitem[\protect\citeauthoryear{{De Marchi} et~al.,}{{De Marchi}
  et~al.}{2011}]{deMarchi2011}
{De Marchi} G.,  et~al., 2011, \mn@doi [\apj] {10.1088/0004-637X/739/1/27},
  \href {http://adsabs.harvard.edu/abs/2011ApJ...739...27D} {739, 27}

\bibitem[\protect\citeauthoryear{{Dekker}, {D'Odorico}, {Kaufer}, {Delabre}  \&
  {Kotzlowski}}{{Dekker} et~al.}{2000}]{dekker2000}
{Dekker} H.,  {D'Odorico} S.,  {Kaufer} A.,  {Delabre} B.,   {Kotzlowski} H.,
  2000, in {Iye} M.,  {Moorwood} A.~F.,  eds,  Society of Photo-Optical
  Instrumentation Engineers (SPIE) Conference Series Vol. 4008, Optical and IR
  Telescope Instrumentation and Detectors. pp 534--545,
  \mn@doi{10.1117/12.395512}

\bibitem[\protect\citeauthoryear{{Doran} et~al.,}{{Doran}
  et~al.}{2013}]{doran2013}
{Doran} E.~I.,  et~al., 2013, \mn@doi [\aap] {10.1051/0004-6361/201321824},
  \href {http://adsabs.harvard.edu/abs/2013A%26A...558A.134D} {558, A134}

\bibitem[\protect\citeauthoryear{{Evans} et~al.,}{{Evans}
  et~al.}{2010}]{evans2010}
{Evans} C.~J.,  et~al., 2010, \mn@doi [\apjl] {10.1088/2041-8205/715/2/L74},
  \href {http://adsabs.harvard.edu/abs/2010ApJ...715L..74E} {715, L74}

\bibitem[\protect\citeauthoryear{{Evans} et~al.,}{{Evans}
  et~al.}{2011}]{evans2011}
{Evans} C.~J.,  et~al., 2011, \mn@doi [\aap] {10.1051/0004-6361/201116782},
  \href {http://adsabs.harvard.edu/abs/2011A%26A...530A.108E} {530, A108}

\bibitem[\protect\citeauthoryear{{Freudling}, {Romaniello}, {Bramich},
  {Ballester}, {Forchi}, {Garc{\'\i}a-Dabl{\'o}}, {Moehler}  \&
  {Neeser}}{{Freudling} et~al.}{2013}]{freudling2013}
{Freudling} W.,  {Romaniello} M.,  {Bramich} D.~M.,  {Ballester} P.,  {Forchi}
  V.,  {Garc{\'\i}a-Dabl{\'o}} C.~E.,  {Moehler} S.,   {Neeser} M.~J.,  2013,
  \mn@doi [\aap] {10.1051/0004-6361/201322494}, \href
  {https://ui.adsabs.harvard.edu/abs/2013A&A...559A..96F} {559, A96}

\bibitem[\protect\citeauthoryear{{Gaia Collaboration} et~al.,}{{Gaia
  Collaboration} et~al.}{2018}]{gaia2018}
{Gaia Collaboration} et~al., 2018, \mn@doi [\aap]
  {10.1051/0004-6361/201833051}, \href
  {https://ui.adsabs.harvard.edu/abs/2018A&A...616A...1G} {616, A1}

\bibitem[\protect\citeauthoryear{{Gonz{\'a}lez} \& {Levato}}{{Gonz{\'a}lez} \&
  {Levato}}{2006}]{gonzalez2006}
{Gonz{\'a}lez} J.~F.,  {Levato} H.,  2006, \mn@doi [\aap]
  {10.1051/0004-6361:20053177}, \href
  {https://ui.adsabs.harvard.edu/abs/2006A&A...448..283G} {448, 283}

\bibitem[\protect\citeauthoryear{{Gr{\"a}fener}, {Vink}, {de Koter}  \&
  {Langer}}{{Gr{\"a}fener} et~al.}{2011}]{graefener2011}
{Gr{\"a}fener} G.,  {Vink} J.~S.,  {de Koter} A.,   {Langer} N.,  2011, \mn@doi
  [\aap] {10.1051/0004-6361/201116701}, \href
  {http://adsabs.harvard.edu/abs/2011A%26A...535A..56G} {535, A56}

\bibitem[\protect\citeauthoryear{{H{\'e}nault-Brunet}
  et~al.,}{{H{\'e}nault-Brunet} et~al.}{2012}]{henault2012}
{H{\'e}nault-Brunet} V.,  et~al., 2012, \mn@doi [\aap]
  {10.1051/0004-6361/201219471}, \href
  {http://adsabs.harvard.edu/abs/2012A%26A...546A..73H} {546, A73}

\bibitem[\protect\citeauthoryear{{Higgins} \& {Vink}}{{Higgins} \&
  {Vink}}{2019}]{higgins2019}
{Higgins} E.~R.,  {Vink} J.~S.,  2019, \mn@doi [\aap]
  {10.1051/0004-6361/201834123}, \href
  {https://ui.adsabs.harvard.edu/abs/2019A&A...622A..50H} {622, A50}

\bibitem[\protect\citeauthoryear{{Hillier} \& {Miller}}{{Hillier} \&
  {Miller}}{1998}]{hillier1998}
{Hillier} D.~J.,  {Miller} D.~L.,  1998, \mn@doi [\apj] {10.1086/305350}, \href
  {http://adsabs.harvard.edu/abs/1998ApJ...496..407H} {496, 407}

\bibitem[\protect\citeauthoryear{{Hunter}, {Shaya}, {Holtzman}, {Light},
  {O'Neil}  \& {Lynds}}{{Hunter} et~al.}{1995}]{hunter1995}
{Hunter} D.~A.,  {Shaya} E.~J.,  {Holtzman} J.~A.,  {Light} R.~M.,  {O'Neil}
  Jr. E.~J.,   {Lynds} R.,  1995, \mn@doi [\apj] {10.1086/175950}, \href
  {http://adsabs.harvard.edu/abs/1995ApJ...448..179H} {448, 179}

\bibitem[\protect\citeauthoryear{{K{\"o}hler} et~al.,}{{K{\"o}hler}
  et~al.}{2015}]{koehler2015}
{K{\"o}hler} K.,  et~al., 2015, \mn@doi [\aap] {10.1051/0004-6361/201424356},
  \href {http://adsabs.harvard.edu/abs/2015A%26A...573A..71K} {573, A71}

\bibitem[\protect\citeauthoryear{{Langer}}{{Langer}}{2012}]{langer2012}
{Langer} N.,  2012, \mn@doi [\araa] {10.1146/annurev-astro-081811-125534},
  \href {http://adsabs.harvard.edu/abs/2012ARA%26A..50..107L} {50, 107}

\bibitem[\protect\citeauthoryear{{Lohr}, {Clark}, {Najarro}, {Patrick},
  {Crowther}  \& {Evans}}{{Lohr} et~al.}{2018}]{lohr2018}
{Lohr} M.~E.,  {Clark} J.~S.,  {Najarro} F.,  {Patrick} L.~R.,  {Crowther}
  P.~A.,   {Evans} C.~J.,  2018, \mn@doi [\aap] {10.1051/0004-6361/201832670},
  \href {https://ui.adsabs.harvard.edu/abs/2018A&A...617A..66L} {617, A66}

\bibitem[\protect\citeauthoryear{{Mahy}, {Damerdji}, {Gosset}, {Nitschelm},
  {Eenens}, {Sana}  \& {Klotz}}{{Mahy} et~al.}{2017}]{mahy2017}
{Mahy} L.,  {Damerdji} Y.,  {Gosset} E.,  {Nitschelm} C.,  {Eenens} P.,  {Sana}
  H.,   {Klotz} A.,  2017, \mn@doi [\aap] {10.1051/0004-6361/201730674}, \href
  {https://ui.adsabs.harvard.edu/abs/2017A&A...607A..96M} {607, A96}

\bibitem[\protect\citeauthoryear{{Mahy} et~al.,}{{Mahy}
  et~al.}{2020}]{mahy2020}
{Mahy} L.,  et~al., 2020, \mn@doi [\aap] {10.1051/0004-6361/201936151}, \href
  {https://ui.adsabs.harvard.edu/abs/2020A&A...634A.118M} {634, A118}

\bibitem[\protect\citeauthoryear{{Ma{\'{\i}}z Apell{\'a}niz}
  et~al.,}{{Ma{\'{\i}}z Apell{\'a}niz} et~al.}{2014}]{ma-ap2014}
{Ma{\'{\i}}z Apell{\'a}niz} J.,  et~al., 2014, \mn@doi [\aap]
  {10.1051/0004-6361/201423439}, \href
  {http://adsabs.harvard.edu/abs/2014A%26A...564A..63M} {564, A63}

\bibitem[\protect\citeauthoryear{{Martins} \& {Palacios}}{{Martins} \&
  {Palacios}}{2013}]{martins2013}
{Martins} F.,  {Palacios} A.,  2013, \mn@doi [\aap]
  {10.1051/0004-6361/201322480}, \href
  {http://adsabs.harvard.edu/abs/2013A%26A...560A..16M} {560, A16}

\bibitem[\protect\citeauthoryear{{Massey} \& {Hunter}}{{Massey} \&
  {Hunter}}{1998}]{massey1998}
{Massey} P.,  {Hunter} D.~A.,  1998, \mn@doi [\apj] {10.1086/305126}, \href
  {http://adsabs.harvard.edu/abs/1998ApJ...493..180M} {493, 180}

\bibitem[\protect\citeauthoryear{{Massey}, {Penny}  \& {Vukovich}}{{Massey}
  et~al.}{2002}]{massey2002}
{Massey} P.,  {Penny} L.~R.,   {Vukovich} J.,  2002, \mn@doi [\apj]
  {10.1086/324783}, \href {http://adsabs.harvard.edu/abs/2002ApJ...565..982M}
  {565, 982}

\bibitem[\protect\citeauthoryear{{Massey}, {Puls}, {Pauldrach}, {Bresolin},
  {Kudritzki}  \& {Simon}}{{Massey} et~al.}{2005}]{massey2005}
{Massey} P.,  {Puls} J.,  {Pauldrach} A.~W.~A.,  {Bresolin} F.,  {Kudritzki}
  R.~P.,   {Simon} T.,  2005, \mn@doi [\apj] {10.1086/430417}, \href
  {http://adsabs.harvard.edu/abs/2005ApJ...627..477M} {627, 477}

\bibitem[\protect\citeauthoryear{{Massey}, {Morrell}, {Neugent}, {Penny},
  {DeGioia-Eastwood}  \& {Gies}}{{Massey} et~al.}{2012}]{massey2012}
{Massey} P.,  {Morrell} N.~I.,  {Neugent} K.~F.,  {Penny} L.~R.,
  {DeGioia-Eastwood} K.,   {Gies} D.~R.,  2012, \mn@doi [\apj]
  {10.1088/0004-637X/748/2/96}, \href
  {https://ui.adsabs.harvard.edu/abs/2012ApJ...748...96M} {748, 96}

\bibitem[\protect\citeauthoryear{{Melnick}}{{Melnick}}{1985}]{melnick1985}
{Melnick} J.,  1985, \aap, \href
  {http://adsabs.harvard.edu/abs/1985A%26A...153..235M} {153, 235}

\bibitem[\protect\citeauthoryear{{Ostrov} \& {Lapasset}}{{Ostrov} \&
  {Lapasset}}{2003}]{ostrov2003}
{Ostrov} P.~G.,  {Lapasset} E.,  2003, \mn@doi [\mnras]
  {10.1046/j.1365-8711.2003.06025.x}, \href
  {https://ui.adsabs.harvard.edu/abs/2003MNRAS.338..141O} {338, 141}

\bibitem[\protect\citeauthoryear{{Paltani}}{{Paltani}}{2004}]{paltani2004}
{Paltani} S.,  2004, \mn@doi [\aap] {10.1051/0004-6361:20034220}, \href
  {https://ui.adsabs.harvard.edu/abs/2004A&A...420..789P} {420, 789}

\bibitem[\protect\citeauthoryear{{Parker}}{{Parker}}{1993}]{parker1993}
{Parker} J.~W.,  1993, \mn@doi [\aj] {10.1086/116661}, \href
  {http://adsabs.harvard.edu/abs/1993AJ....106..560P} {106, 560}

\bibitem[\protect\citeauthoryear{{Pavlovski}, {Southworth}  \&
  {Tamajo}}{{Pavlovski} et~al.}{2018}]{pavlovski2018}
{Pavlovski} K.,  {Southworth} J.,   {Tamajo} E.,  2018, \mn@doi [\mnras]
  {10.1093/mnras/sty2516}, \href
  {https://ui.adsabs.harvard.edu/abs/2018MNRAS.481.3129P} {481, 3129}

\bibitem[\protect\citeauthoryear{{Pietrzy{\'n}ski} et~al.,}{{Pietrzy{\'n}ski}
  et~al.}{2019}]{pietrzynski2019}
{Pietrzy{\'n}ski} G.,  et~al., 2019, \mn@doi [\nat]
  {10.1038/s41586-019-0999-4}, \href
  {https://ui.adsabs.harvard.edu/abs/2019Natur.567..200P} {567, 200}

\bibitem[\protect\citeauthoryear{{Pollock}}{{Pollock}}{1987}]{pollock1987}
{Pollock} A.~M.~T.,  1987, \mn@doi [\apj] {10.1086/165539}, \href
  {https://ui.adsabs.harvard.edu/abs/1987ApJ...320..283P} {320, 283}

\bibitem[\protect\citeauthoryear{{Pollock}, {Crowther}, {Tehrani}, {Broos}  \&
  {Townsley}}{{Pollock} et~al.}{2018}]{pollock2018}
{Pollock} A.~M.~T.,  {Crowther} P.~A.,  {Tehrani} K.,  {Broos} P.~S.,
  {Townsley} L.~K.,  2018, \mn@doi [\mnras] {10.1093/mnras/stx2879}, \href
  {https://ui.adsabs.harvard.edu/abs/2018MNRAS.474.3228P} {474, 3228}

\bibitem[\protect\citeauthoryear{{Portegies Zwart}, {Pooley}  \&
  {Lewin}}{{Portegies Zwart} et~al.}{2002}]{portegies2002}
{Portegies Zwart} S.~F.,  {Pooley} D.,   {Lewin} W.~H.~G.,  2002, \mn@doi
  [\apj] {10.1086/340996}, \href
  {http://adsabs.harvard.edu/abs/2002ApJ...574..762P} {574, 762}

\bibitem[\protect\citeauthoryear{{Sana} et~al.,}{{Sana}
  et~al.}{2012}]{sana2012Sci}
{Sana} H.,  et~al., 2012, \mn@doi [Science] {10.1126/science.1223344}, \href
  {http://ads.ari.uni-heidelberg.de/abs/2012Sci...337..444S} {337, 444}

\bibitem[\protect\citeauthoryear{{Sana} et~al.,}{{Sana}
  et~al.}{2013}]{sana2013}
{Sana} H.,  et~al., 2013, \mn@doi [\aap] {10.1051/0004-6361/201219621}, \href
  {http://adsabs.harvard.edu/abs/2013A%26A...550A.107S} {550, A107}

\bibitem[\protect\citeauthoryear{{Schneider}, {Langer}, {de Koter}, {Brott},
  {Izzard}  \& {Lau}}{{Schneider} et~al.}{2014}]{schneider2014}
{Schneider} F.~R.~N.,  {Langer} N.,  {de Koter} A.,  {Brott} I.,  {Izzard}
  R.~G.,   {Lau} H.~H.~B.,  2014, \mn@doi [\aap] {10.1051/0004-6361/201424286},
  \href {http://adsabs.harvard.edu/abs/2014A%26A...570A..66S} {570, A66}

\bibitem[\protect\citeauthoryear{{Schnurr}, {Moffat}, {St-Louis}, {Morrell}  \&
  {Guerrero}}{{Schnurr} et~al.}{2008}]{schnurr2008}
{Schnurr} O.,  {Moffat} A.~F.~J.,  {St-Louis} N.,  {Morrell} N.~I.,
  {Guerrero} M.~A.,  2008, \mn@doi [\mnras] {10.1111/j.1365-2966.2008.13584.x},
  \href {http://adsabs.harvard.edu/abs/2008MNRAS.389..806S} {389, 806}

\bibitem[\protect\citeauthoryear{{Selman}, {Melnick}, {Bosch}  \&
  {Terlevich}}{{Selman} et~al.}{1999}]{selman1999a}
{Selman} F.,  {Melnick} J.,  {Bosch} G.,   {Terlevich} R.,  1999, \aap, \href
  {https://ui.adsabs.harvard.edu/abs/1999A&A...341...98S} {341, 98}

\bibitem[\protect\citeauthoryear{{Serenelli} et~al.,}{{Serenelli}
  et~al.}{2021}]{serenelli2021}
{Serenelli} A.,  et~al., 2021, \mn@doi [\aapr] {10.1007/s00159-021-00132-9},
  \href {https://ui.adsabs.harvard.edu/abs/2021A&ARv..29....4S} {29, 4}

\bibitem[\protect\citeauthoryear{{Shenar} et~al.,}{{Shenar}
  et~al.}{2017}]{shenar2017}
{Shenar} T.,  et~al., 2017, \mn@doi [\aap] {10.1051/0004-6361/201629621}, \href
  {https://ui.adsabs.harvard.edu/abs/2017A&A...598A..85S} {598, A85}

\bibitem[\protect\citeauthoryear{{Shenar} et~al.,}{{Shenar}
  et~al.}{2019}]{shenar2019}
{Shenar} T.,  et~al., 2019, \mn@doi [\aap] {10.1051/0004-6361/201935684}, \href
  {https://ui.adsabs.harvard.edu/abs/2019A&A...627A.151S} {627, A151}

\bibitem[\protect\citeauthoryear{{Shenar} et~al.,}{{Shenar}
  et~al.}{2021}]{shenar2021}
{Shenar} T.,  et~al., 2021, \mn@doi [\aap] {10.1051/0004-6361/202140693}, \href
  {https://ui.adsabs.harvard.edu/abs/2021A&A...650A.147S} {650, A147}

\bibitem[\protect\citeauthoryear{{Sim{\'o}n-D{\'{\i}}az} \&
  {Herrero}}{{Sim{\'o}n-D{\'{\i}}az} \& {Herrero}}{2014}]{simon-diaz2014}
{Sim{\'o}n-D{\'{\i}}az} S.,  {Herrero} A.,  2014, \mn@doi [\aap]
  {10.1051/0004-6361/201322758}, \href
  {http://adsabs.harvard.edu/abs/2014A%26A...562A.135S} {562, A135}

\bibitem[\protect\citeauthoryear{{Smith}, {Shara}  \& {Moffat}}{{Smith}
  et~al.}{1990}]{smith1990}
{Smith} L.~F.,  {Shara} M.~M.,   {Moffat} A. F.~J.,  1990, \mn@doi [\apj]
  {10.1086/168256}, \href
  {https://ui.adsabs.harvard.edu/abs/1990ApJ...348..471S} {348, 471}

\bibitem[\protect\citeauthoryear{{Sota}, {Ma{\'{\i}}z Apell{\'a}niz},
  {Walborn}, {Alfaro}, {Barb{\'a}}, {Morrell}, {Gamen}  \& {Arias}}{{Sota}
  et~al.}{2011}]{sota2011}
{Sota} A.,  {Ma{\'{\i}}z Apell{\'a}niz} J.,  {Walborn} N.~R.,  {Alfaro} E.~J.,
  {Barb{\'a}} R.~H.,  {Morrell} N.~I.,  {Gamen} R.~C.,   {Arias} J.~I.,  2011,
  \mn@doi [\apjs] {10.1088/0067-0049/193/2/24}, \href
  {http://adsabs.harvard.edu/abs/2011ApJS..193...24S} {193, 24}

\bibitem[\protect\citeauthoryear{{Southworth}, {Maxted}  \&
  {Smalley}}{{Southworth} et~al.}{2004}]{southworth2004}
{Southworth} J.,  {Maxted} P.~F.~L.,   {Smalley} B.,  2004, \mn@doi [\mnras]
  {10.1111/j.1365-2966.2004.07871.x}, \href
  {https://ui.adsabs.harvard.edu/abs/2004MNRAS.351.1277S} {351, 1277}

\bibitem[\protect\citeauthoryear{{Stevens}, {Blondin}  \& {Pollock}}{{Stevens}
  et~al.}{1992}]{stevens1992}
{Stevens} I.~R.,  {Blondin} J.~M.,   {Pollock} A.~M.~T.,  1992, \mn@doi [\apj]
  {10.1086/171013}, \href
  {https://ui.adsabs.harvard.edu/abs/1992ApJ...386..265S} {386, 265}

\bibitem[\protect\citeauthoryear{{Taylor} et~al.,}{{Taylor}
  et~al.}{2011}]{taylor2011}
{Taylor} W.~D.,  et~al., 2011, \mn@doi [\aap] {10.1051/0004-6361/201116785},
  \href {http://adsabs.harvard.edu/abs/2011A%26A...530L..10T} {530, L10}

\bibitem[\protect\citeauthoryear{{Tehrani}, {Crowther}, {Bestenlehner},
  {Littlefair}, {Pollock}, {Parker}  \& {Schnurr}}{{Tehrani}
  et~al.}{2019}]{tehrani2019}
{Tehrani} K.~A.,  {Crowther} P.~A.,  {Bestenlehner} J.~M.,  {Littlefair} S.~P.,
   {Pollock} A.~M.~T.,  {Parker} R.~J.,   {Schnurr} O.,  2019, \mn@doi [\mnras]
  {10.1093/mnras/stz147}, \href
  {https://ui.adsabs.harvard.edu/abs/2019MNRAS.484.2692T} {484, 2692}

\bibitem[\protect\citeauthoryear{{Tkachenko} et~al.,}{{Tkachenko}
  et~al.}{2014}]{tkachenko2014}
{Tkachenko} A.,  et~al., 2014, \mn@doi [\mnras] {10.1093/mnras/stt2421}, \href
  {https://ui.adsabs.harvard.edu/abs/2014MNRAS.438.3093T} {438, 3093}

\bibitem[\protect\citeauthoryear{{Townsley}, {Broos}, {Feigelson}, {Garmire}
  \& {Getman}}{{Townsley} et~al.}{2006}]{townsley2006}
{Townsley} L.~K.,  {Broos} P.~S.,  {Feigelson} E.~D.,  {Garmire} G.~P.,
  {Getman} K.~V.,  2006, \mn@doi [\aj] {10.1086/500535}, \href
  {https://ui.adsabs.harvard.edu/abs/2006AJ....131.2164T} {131, 2164}

\bibitem[\protect\citeauthoryear{Townsley, Broos, Garmire, Bouwman, Povich,
  Feigelson, Getman  \& Kuhn}{Townsley et~al.}{2014}]{townsley2014}
Townsley L.~K.,  Broos P.~S.,  Garmire G.~P.,  Bouwman J.,  Povich M.~S.,
  Feigelson E.~D.,  Getman K.~V.,   Kuhn M.~A.,  2014, \mn@doi [The
  Astrophysical Journal Supplement Series] {10.1088/0067-0049/213/1/1}, 213, 1

\bibitem[\protect\citeauthoryear{{Walborn} et~al.,}{{Walborn}
  et~al.}{2014}]{walborn2014}
{Walborn} N.~R.,  et~al., 2014, \mn@doi [\aap] {10.1051/0004-6361/201323082},
  \href {http://adsabs.harvard.edu/abs/2014A%26A...564A..40W} {564, A40}

\makeatother
\end{thebibliography}




\appendix

\section{Additional material}
\begin{table}
        \centering
        \caption{Measurements of $\varv \sin i$ and $\varv_{\rm mac}$ for each epoch and median spectrum.
        \label{t:line-broad}}
        \begin{tabular}{@{}l@{~~~}c@{~~~}c@{~~~}c@{}}
        \hline
      MJD & \ion{C}{iv}\,5801: $\varv \sin i$ & \ion{C}{iv}\,5801: $\varv_{\rm mac}$\\
		& [km/s] & [km/s] \\
		\hline
      59131.25 & $88.7  \pm 4.4$  & $116.3 \pm 4.4$ \\
      59132.25 & $98.9  \pm 4.9$  & $130.0 \pm 4.9$ \\
      59133.25 & $96.8  \pm 4.8$  & $ 93.6 \pm 4.8$ \\
      59134.21 & $93.8  \pm 4.7$  & $105.4 \pm 4.7$ \\
      59136.17 & $97.3  \pm 4.9$  & $144.4 \pm 4.9$ \\
      59137.20 & $89.2  \pm 4.5$  & $ 99.9 \pm 4.5$ \\
      59138.19 & $95.8  \pm 4.8$  & $ 88.0 \pm 4.8$ \\
      59139.24 & $120.8 \pm 6.0$  & $ 98.1 \pm 6.0$ \\
      59141.26 & $96.8  \pm 4.8$  & $ 84.4 \pm 4.8$ \\
      59142.21 & $140.1 \pm 7.0$  & $ 84.1 \pm 7.0$ \\
      59143.21 & $98.9  \pm 4.9$  & $130.2 \pm 4.9$ \\
      59144.17 & $116.2 \pm 5.8$  & $  8.5 \pm 5.8$ \\
      59146.19 & $59.1  \pm 3.0$  & $ 48.7 \pm 3.0$ \\
      59147.31 & $87.1  \pm 4.4$  & $ 83.8 \pm 4.4$ \\
      59148.18 & $80.0  \pm 4.0$  & $ 80.2 \pm 4.0$ \\
      \hline
      \hline
      Median spectrum &  \ion{N}{iv}\,4058: $\varv \sin i$ & \ion{N}{iv}\,4058: $\varv_{\rm mac}$\\
      	& [km/s] & [km/s] \\
      	\hline
      	Primary & $76.4 \pm 3.8$  & $95.6 \pm 3.8$ \\
      	\hline
      	&  \ion{C}{iv}\,5801: $\varv \sin i$ & \ion{C}{iv}\,5801: $\varv_{\rm mac}$\\
      	\hline
      	Primary & $74.9 \pm 3.5$  & $121.3 \pm 3.5$ \\
      	\hline
      	&  \ion{C}{iv}\,5812: $\varv \sin i$ & \ion{C}{iv}\,5812: $\varv_{\rm mac}$\\
      	\hline
      	Primary & $75.3 \pm 3.7$  & $112.0 \pm 3.7$ \\
      	Secondary & $78.0 \pm 3.9$  & $55.8 \pm 3.9$ \\
      	\hline
        \end{tabular}
\end{table}

\begin{figure*}
	\includegraphics[width=\textwidth]{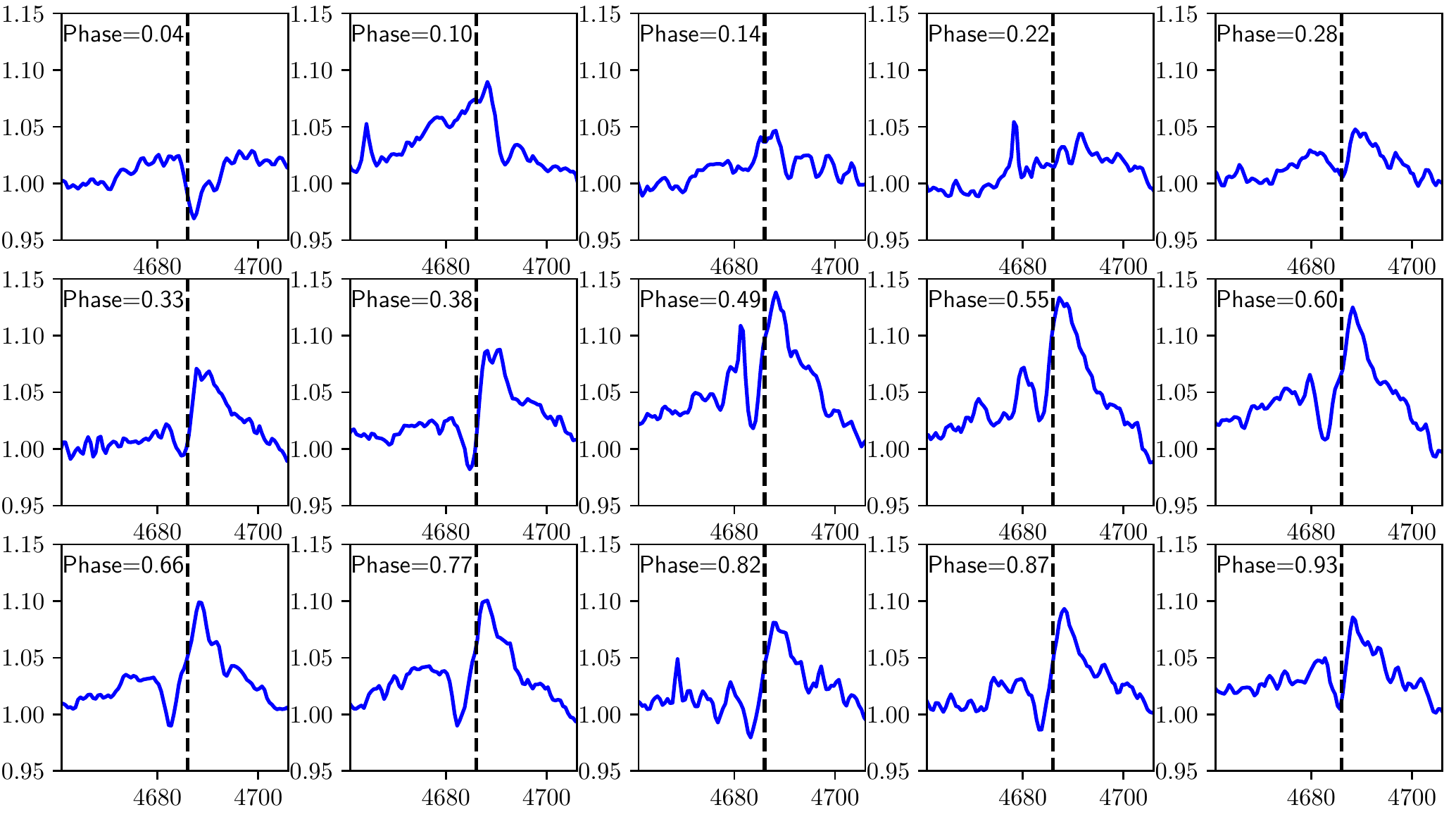}
    \caption{Line profile variations of \ion{He}{ii}\,4686 of all epochs as a function of phase.}
    \label{f:heii4684_ew}
\end{figure*}

\section{Detailed spectra disentangling}
\label{s:dsd}
\begin{figure*}
	\includegraphics[width=\textwidth]{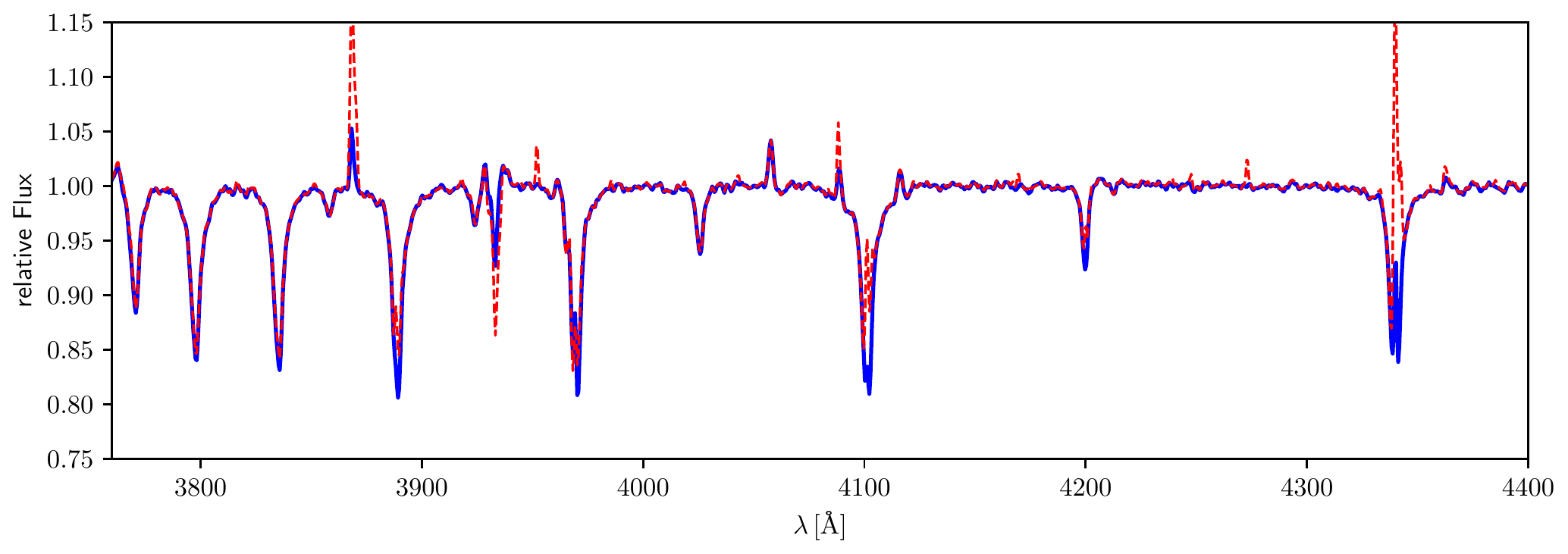}
	\caption{Blue solid line is the primary median spectrum while red dashed line is the averaged spectrum over all epochs (shift and add). Both spectra are similar, but the averaged spectrum (red dashed line) contains more artefacts.\label{f:dis_comp}}
\end{figure*}
\begin{figure*}
	\includegraphics[width=\textwidth]{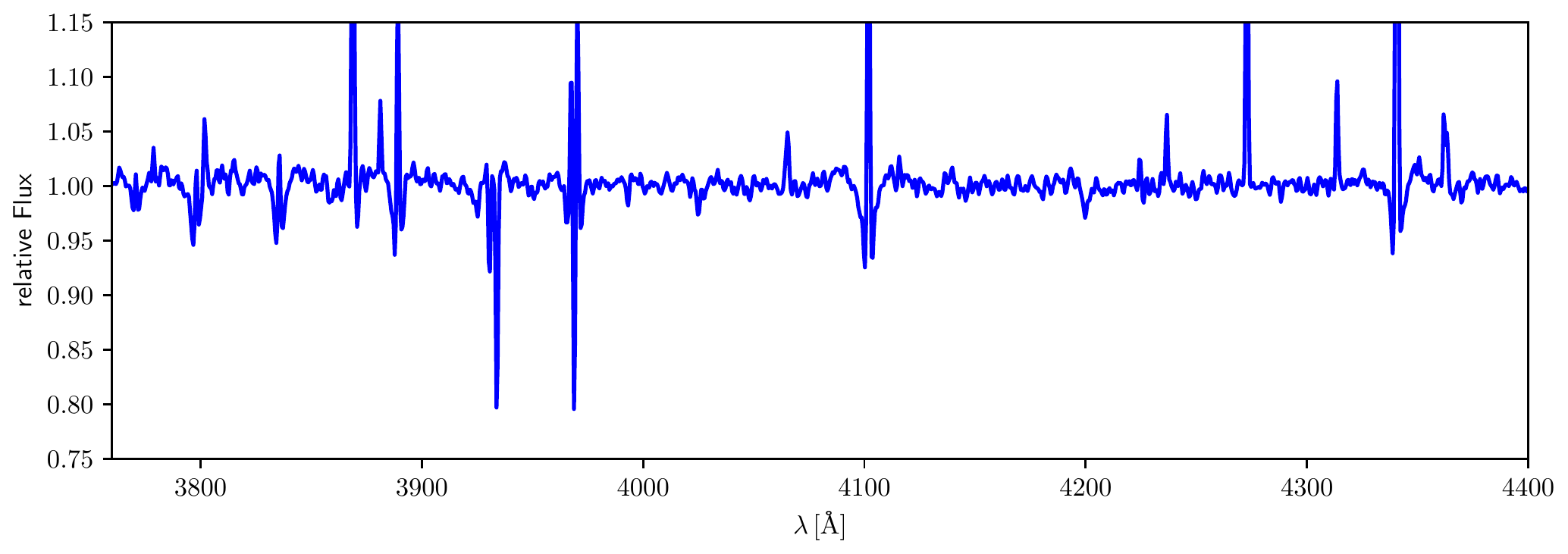}
	\caption{Random epoch divided by the primary median spectrum. The spectrum largely shows the spectral features of the secondary.\label{f:dis_med_sec}}
\end{figure*}
\begin{figure}
	\includegraphics[width=\columnwidth]{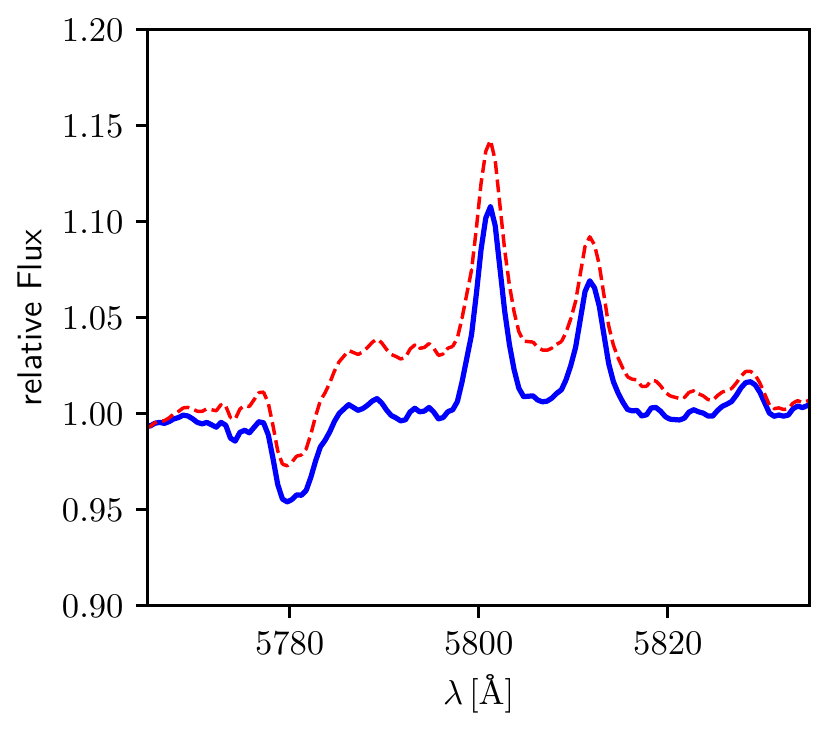}
	\caption{Blue solid is the primary median spectrum with removed broad \ion{C}{iv}\,5801-12 emission caused by the WC4 star Mk33Sb. Red dashed lined includes the broad \ion{C}{iv} emission.\label{f:dis_wc}}
\end{figure}

In the following disentangling work-flow we ignore any cross-contamination by nearby sources and only consider the fluxes of the primary and secondary. The SED of each components can be written as the product of relative flux $f$ and continuum SED $c$ ($\mathrm{SED_{1,2}} = f_{1,2} c_{1,2}$). The combined flux of the secondary and primary are recorded with VLT/UVES and after normalisation we obtained the following spectrum for each epoch i:
\begin{equation}
\frac{f^{\rm i}_1 c_1 + f^{\rm i}_2 c_2}{c_1 + c_2} = \mathrm{epoch^i}.
\end{equation}

In the first step we shifted the epochs to the rest-frame of the primary and calculate the median spectrum:
\begin{equation}
\mathrm{M}\left[ \frac{f^{\rm i}_1 c_1 + f^{\rm i}_2 c_2}{c_1 + c_2} \right] \approx \frac{f_1 c_1 +c_2}{c_1 + c_2}.\label{e:dis_pri}
\end{equation}
In the case of shift and add \citep[e.g.][]{gonzalez2006} the median is replaced with mean. A comparison is shown in Fig.~\ref{f:dis_comp}.

To obtain the spectrum of secondary we divide each epoch by the median primary spectrum ($\mathrm{Med_1}$). We could choose a subtraction instead of division, but this could add numerical noise to the spectrum because computers can only represent numbers with limited precision and 2 nearly equal subtracted numbers become indistinguishable. A division leads to the following equation and secondary median spectrum ($\mathrm{Med_2}$):
\begin{equation}
\mathrm{M}\left[\frac{f^{\rm i}_1 c_1 + f^{\rm i}_2 c_2}{c_1 + c_2}/\mathrm{Med_1}\right] = \mathrm{M} \left[ \frac{f^{\rm i}_1 c_1 + f^{\rm i}_2 c_2}{f^{\rm i}_1c_1+c_2}\right] \approx \frac{c_1 + f_2 c_2}{c_1 + c_2},
\end{equation}
which has the same form as the results of the primary in the first step (Eq.~\ref{e:dis_pri}). An example of an epoch spectrum after division by the primary median spectrum is given in Fig.~\ref{f:dis_med_sec}. The primary and secondary median spectra were iteratively used to divide each epoch before the median was calculated. The iteration was stopped, if no changes were observed to the previous iteration. The median spectra of both components are rescaled according to their flux ratio (Sect.~\ref{s:lum}). We divided the median primary spectrum by $c_1/(c_1 +c_2)$ or $c_2/(c_1 +c_2)$ for the secondary and than subtract $c_2/c_1$ (primary) or $c_1/c_2$ (secondary):
\begin{equation}
\mathrm{Med}_{1,2} \frac{c_1 + c_2}{c_{1,2}} -\frac{c_{2,1}}{c_{1,2}} = f_{1,2}.
\end{equation}

The broad \ion{C}{iv}\,5801-12 emission caused by the WC4 star Mk33Sb was removed in a similar way by just calculating the median without shifting the epochs (Fig.~\ref{f:dis_wc}). We do not know the actual relative flux of Mk33Sb to the combined Mk33Na system as the light of Mk33Sb leaked into the slit depending on seeing and airmass. In addition, we tried to minimise its contamination during the reduction process by using the 2D slit extraction (Sect.~\ref{s:opt_obs}). A rescaling might not make a noticeable difference.
   
The disentangled spectra for Mk\,33Na$_1$ and Mk\,33Na$_2$ are shown in Fig.~\ref{f:pri_dis} and \ref{f:sec_dis}.



\section{Spectroscopic fits of all epochs}

\begin{figure*}
	\includegraphics[width=\textwidth]{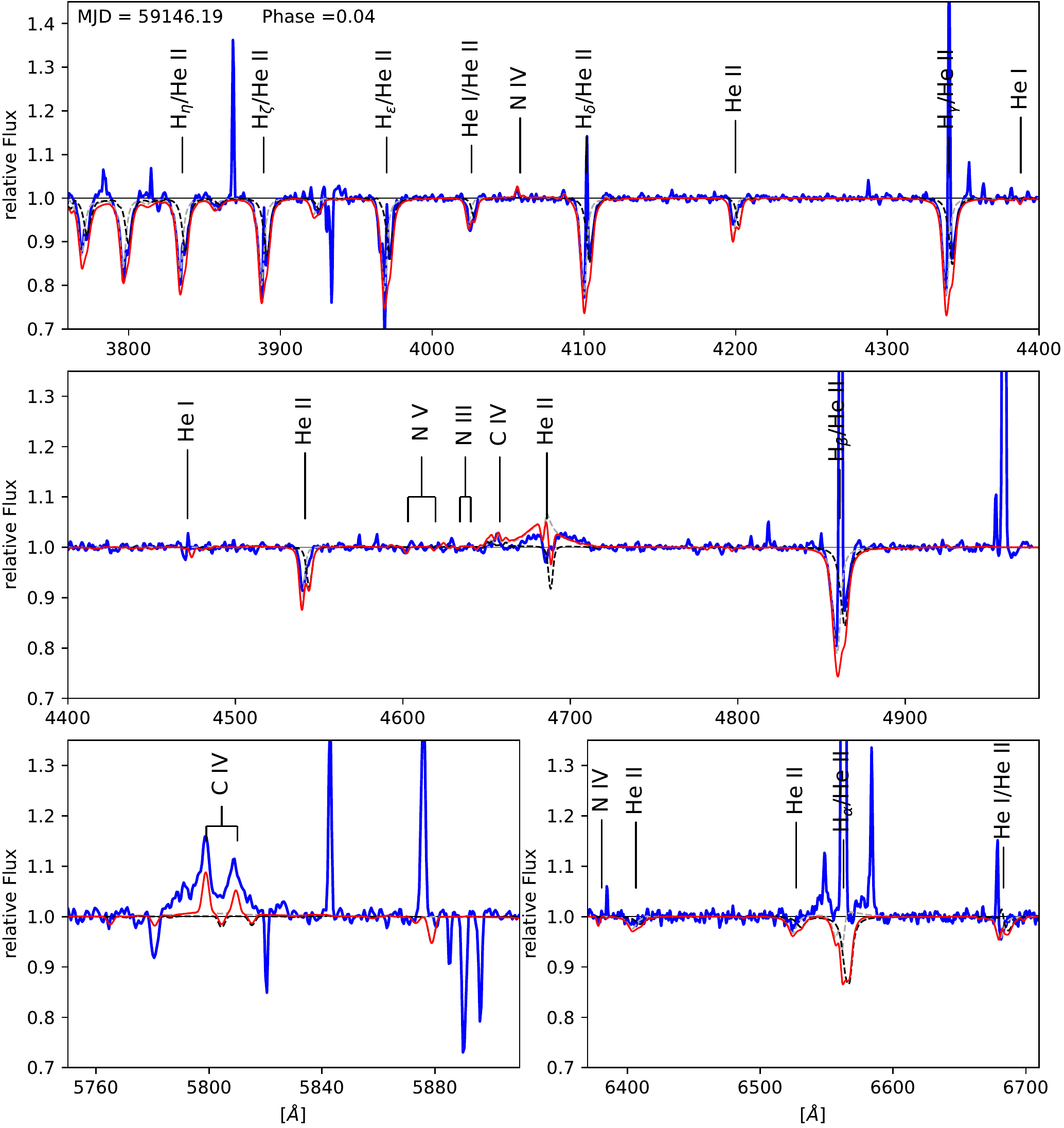}
    \caption{Spectroscopic fit of the original reduced and normalised data. Blue solid lines is the observations and red solid line is the combined model spectrum of the primary (dashed grey line) and secondary (black dashed line).}
\end{figure*}

\begin{figure*}	
	\includegraphics[width=\textwidth]{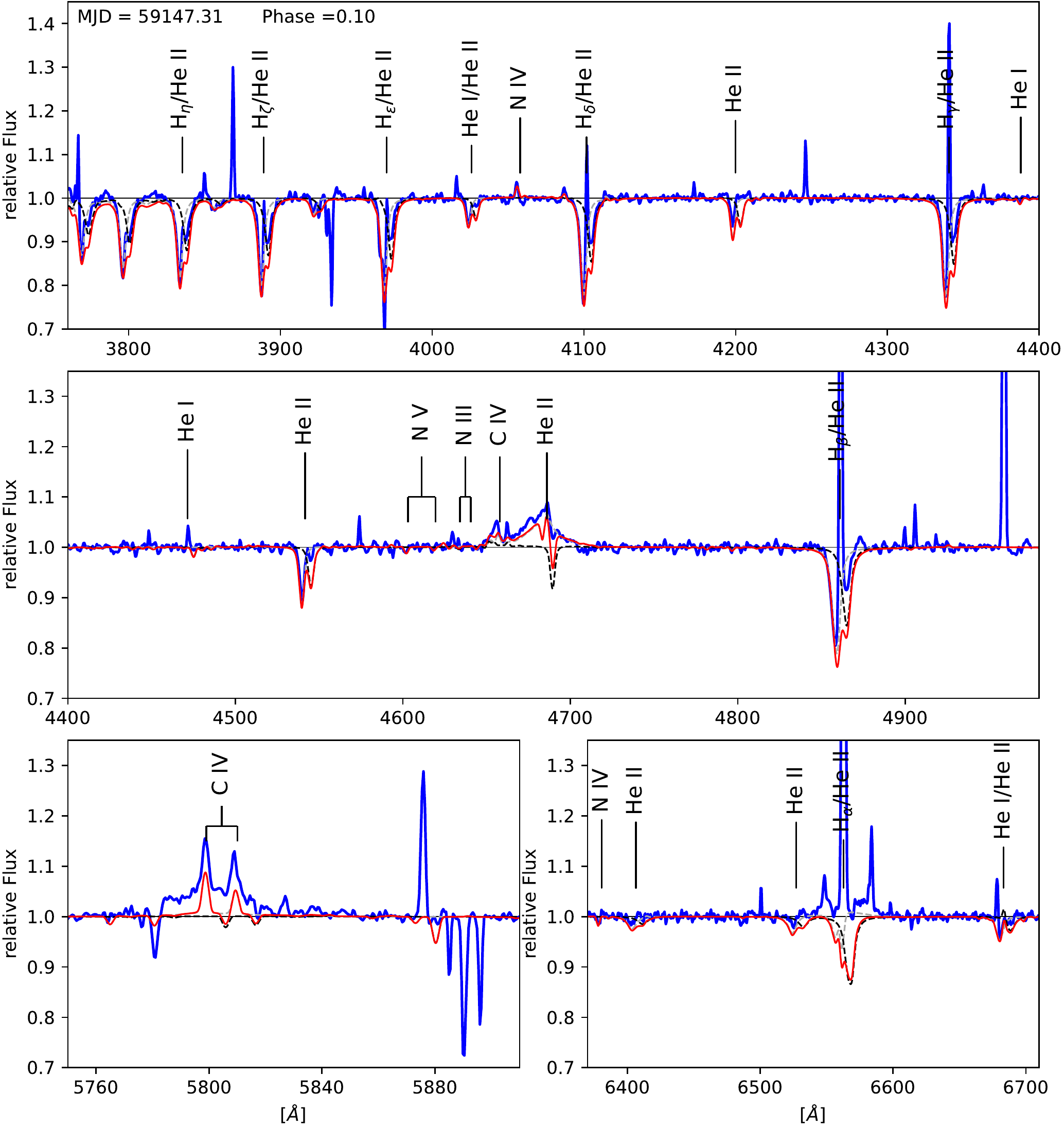}
	\caption{Spectroscopic fit of the original reduced and normalised data. Blue solid lines is the observations and red solid line is the combined model spectrum of the primary (dashed grey line) and secondary (black dashed line).\label{f:largest_sep}}
\end{figure*}

\begin{figure*}
	\includegraphics[width=\textwidth]{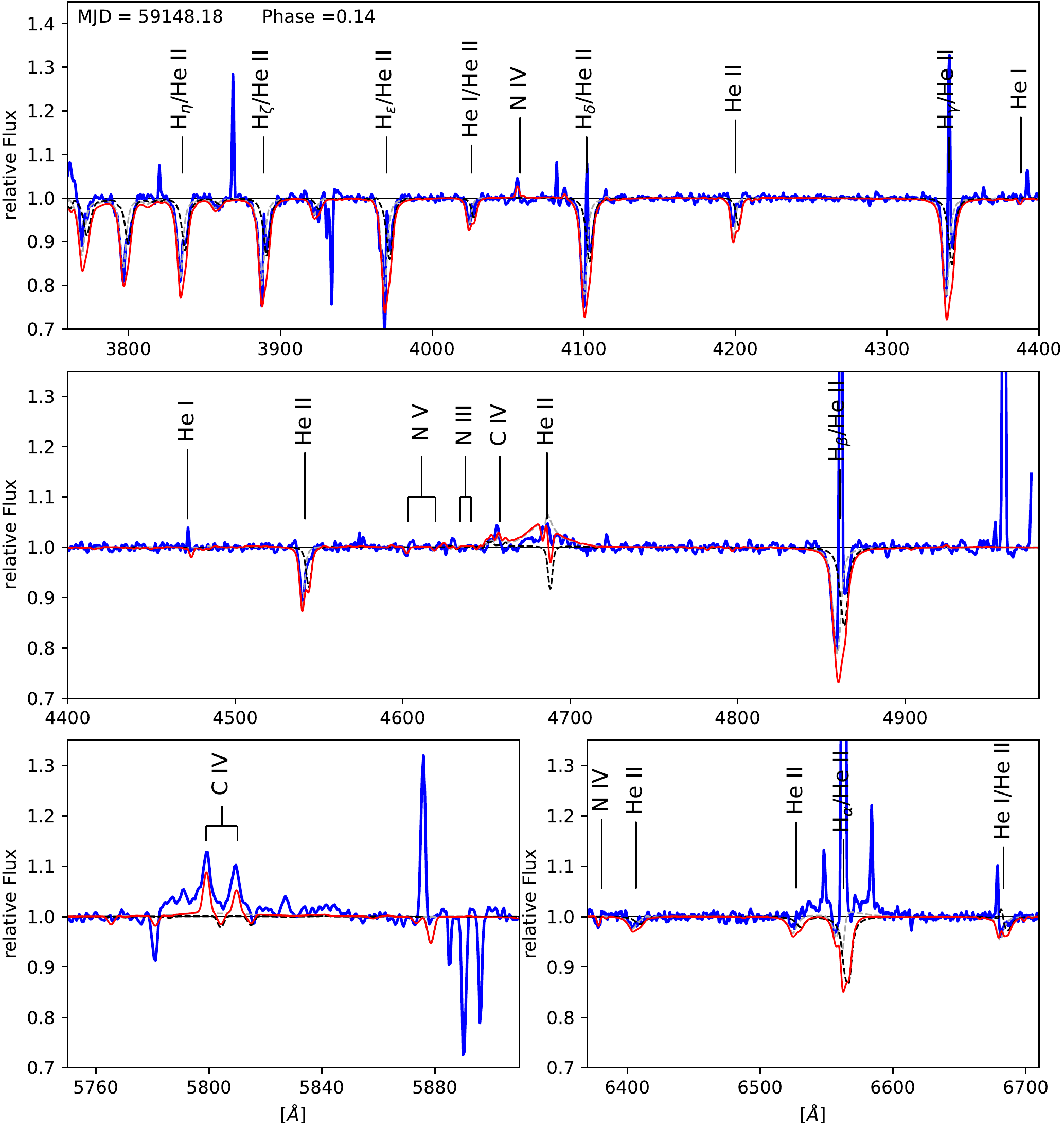}
	\caption{Spectroscopic fit of the original reduced and normalised data. Blue solid lines is the observations and red solid line is the combined model spectrum of the primary (dashed grey line) and secondary (black dashed line).}
\end{figure*}

\begin{figure*}
	\includegraphics[width=\textwidth]{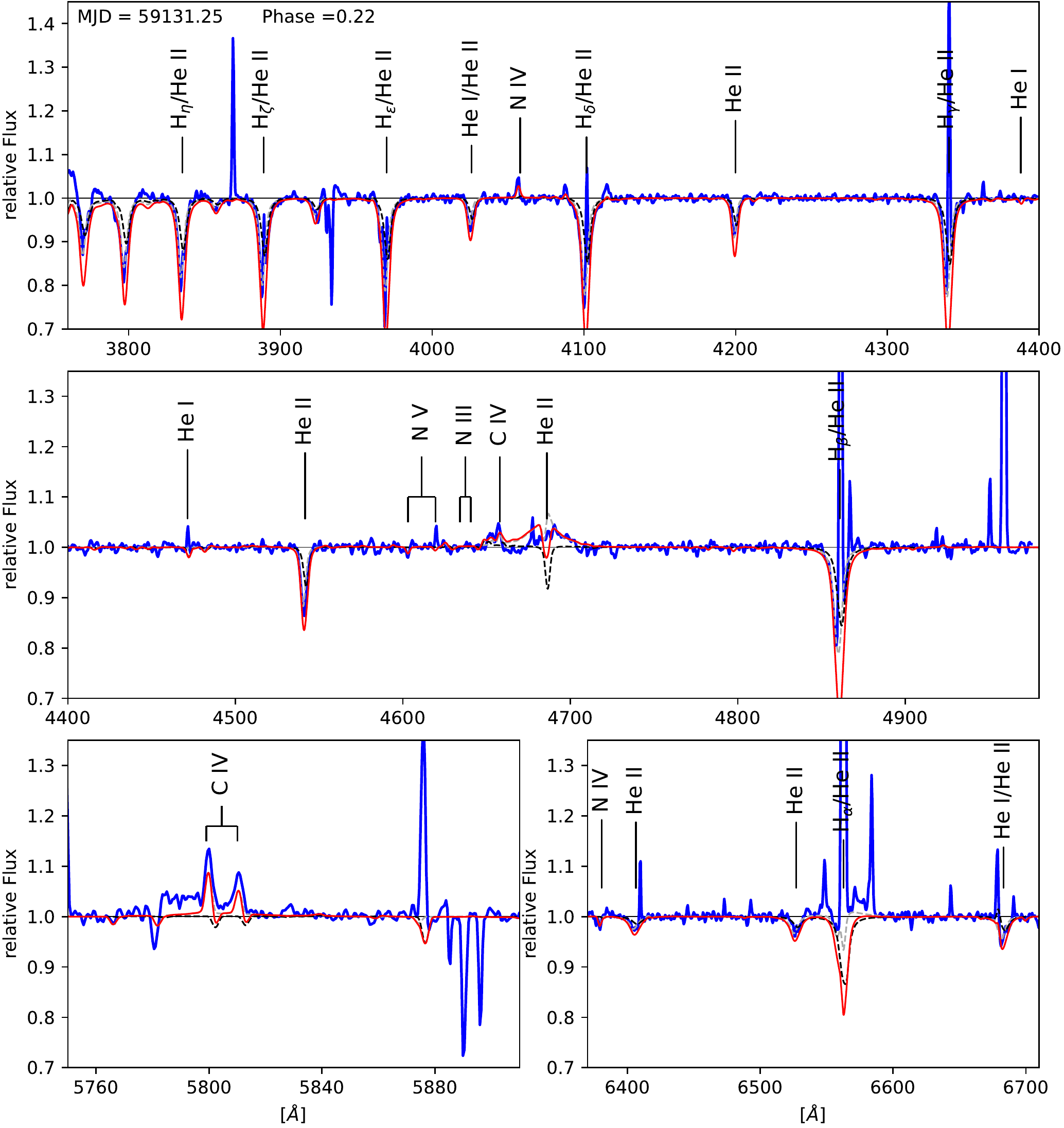}
	\caption{Spectroscopic fit of the original reduced and normalised data. Blue solid lines is the observations and red solid line is the combined model spectrum of the primary (dashed grey line) and secondary (black dashed line).}
\end{figure*}

\begin{figure*}
	\includegraphics[width=\textwidth]{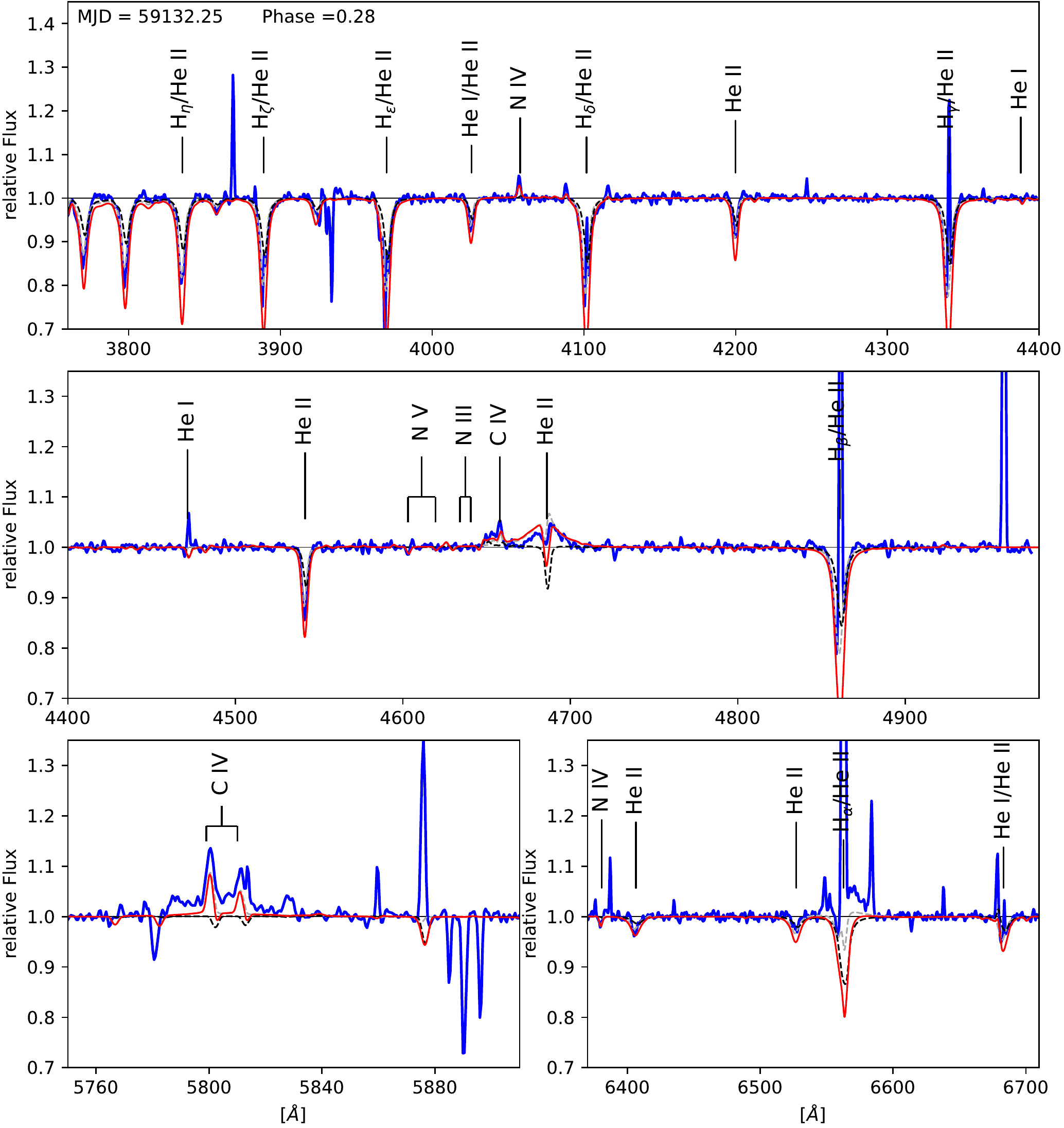}
	\caption{Spectroscopic fit of the original reduced and normalised data. Blue solid lines is the observations and red solid line is the combined model spectrum of the primary (dashed grey line) and secondary (black dashed line).}
\end{figure*}

\begin{figure*}
	\includegraphics[width=\textwidth]{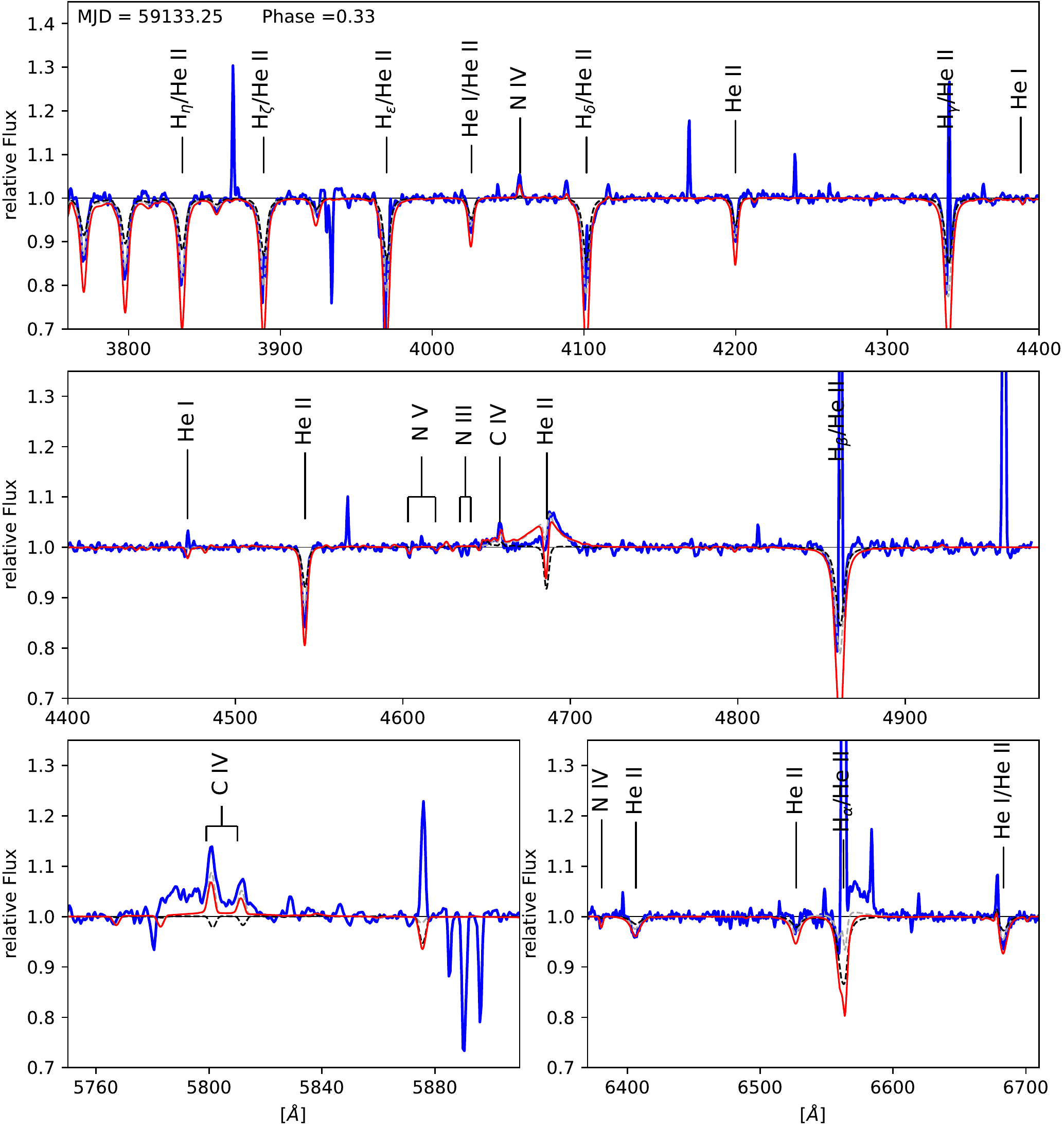}
	\caption{Spectroscopic fit of the original reduced and normalised data. Blue solid lines is the observations and red solid line is the combined model spectrum of the primary (dashed grey line) and secondary (black dashed line).}
\end{figure*}

\begin{figure*}
	\includegraphics[width=\textwidth]{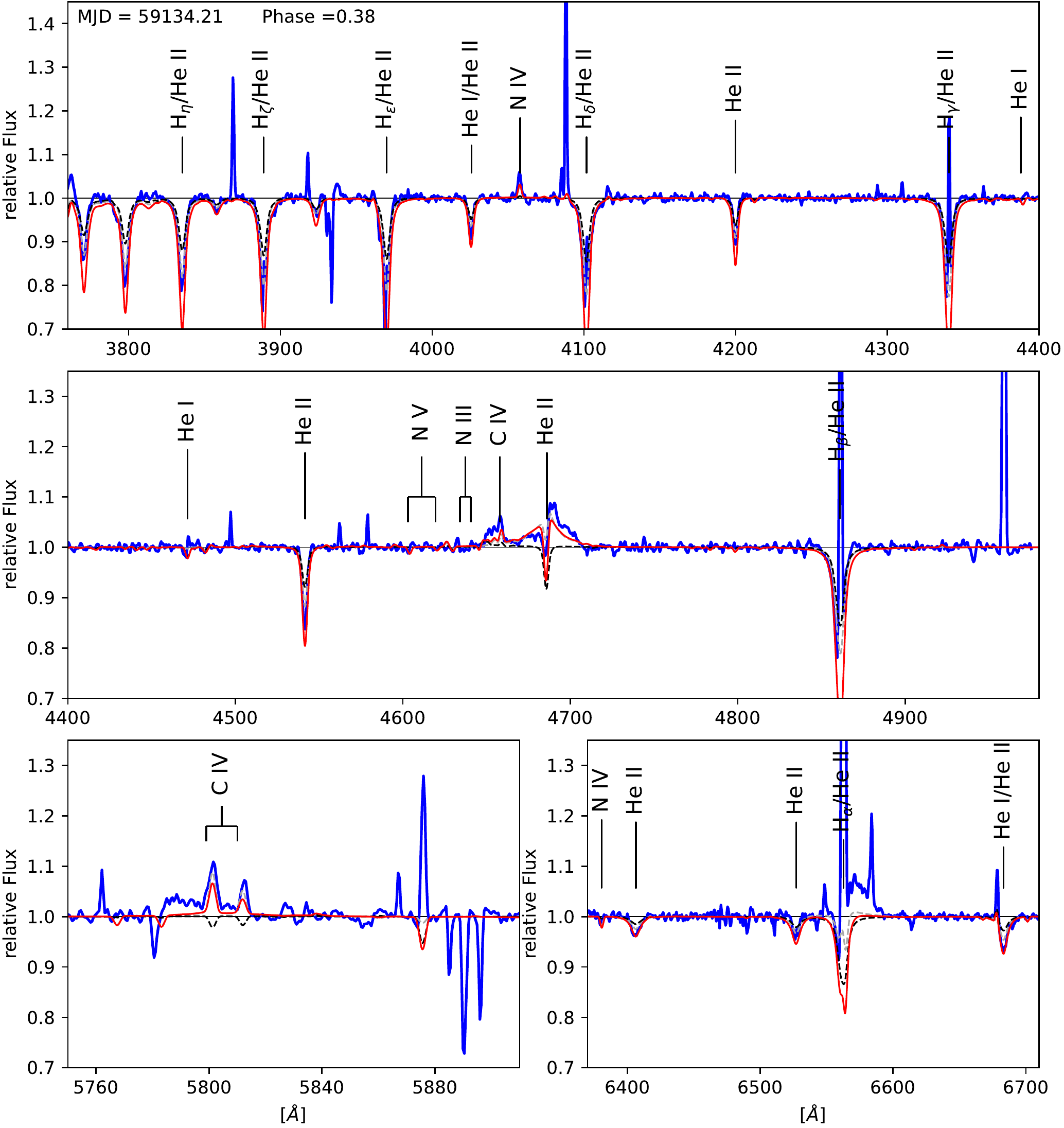}
	\caption{Spectroscopic fit of the original reduced and normalised data. Blue solid lines is the observations and red solid line is the combined model spectrum of the primary (dashed grey line) and secondary (black dashed line).}
\end{figure*}

\begin{figure*}
	\includegraphics[width=\textwidth]{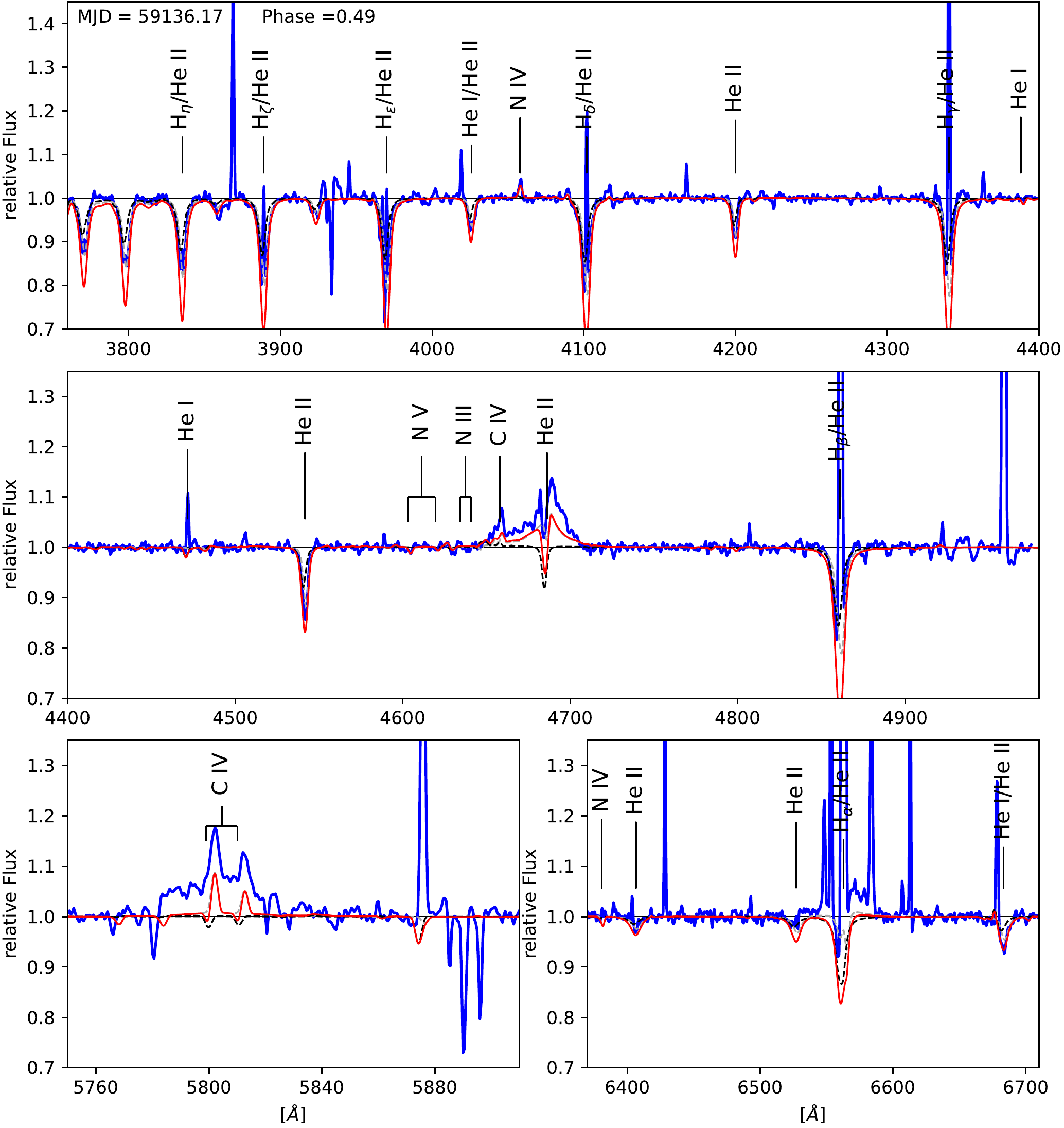}
	\caption{Spectroscopic fit of the original reduced and normalised data. Blue solid lines is the observations and red solid line is the combined model spectrum of the primary (dashed grey line) and secondary (black dashed line).}
\end{figure*}

\begin{figure*}
	\includegraphics[width=\textwidth]{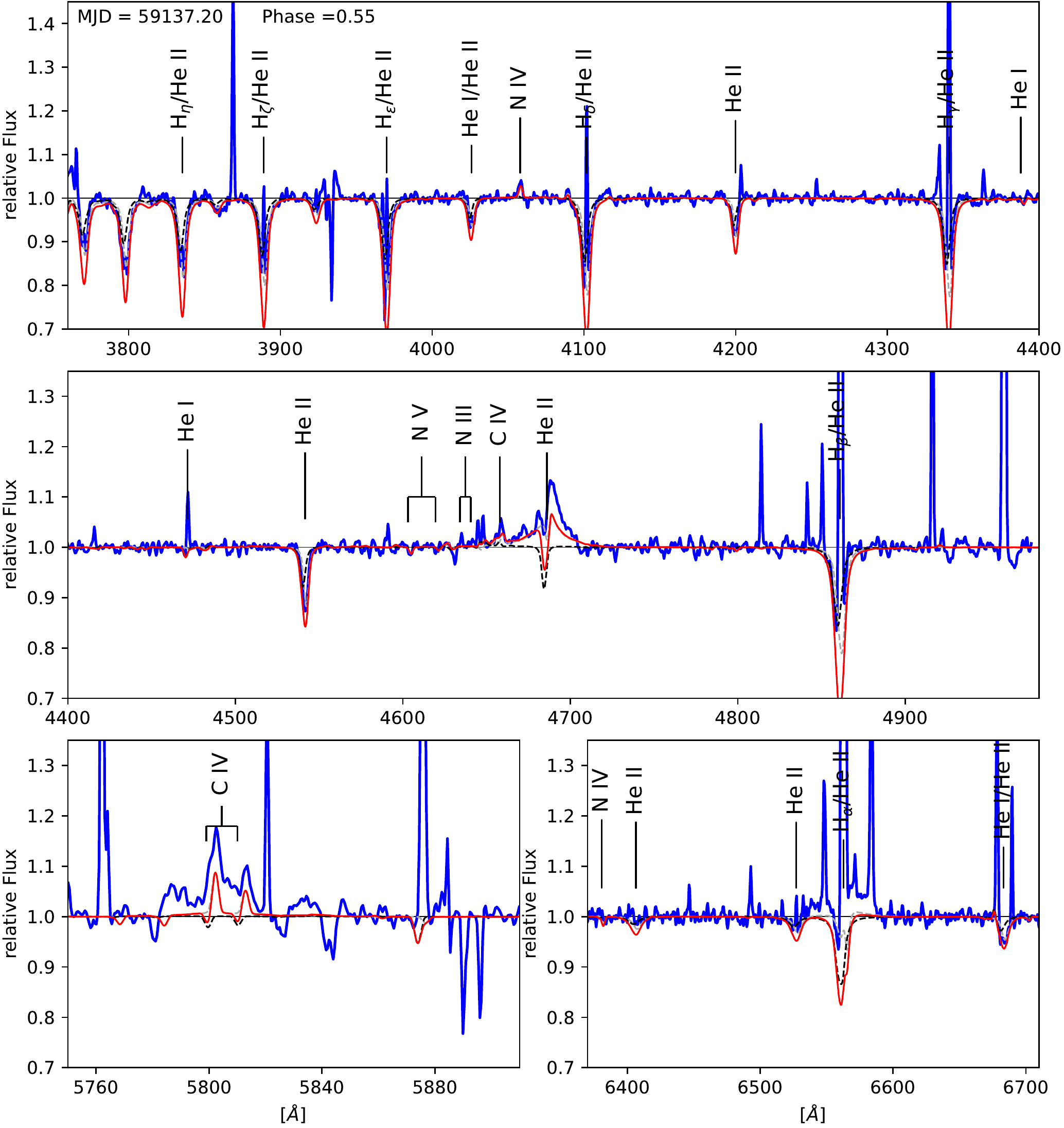}
	\caption{Spectroscopic fit of the original reduced and normalised data. Blue solid lines is the observations and red solid line is the combined model spectrum of the primary (dashed grey line) and secondary (black dashed line).}
\end{figure*}

\begin{figure*}
	\includegraphics[width=\textwidth]{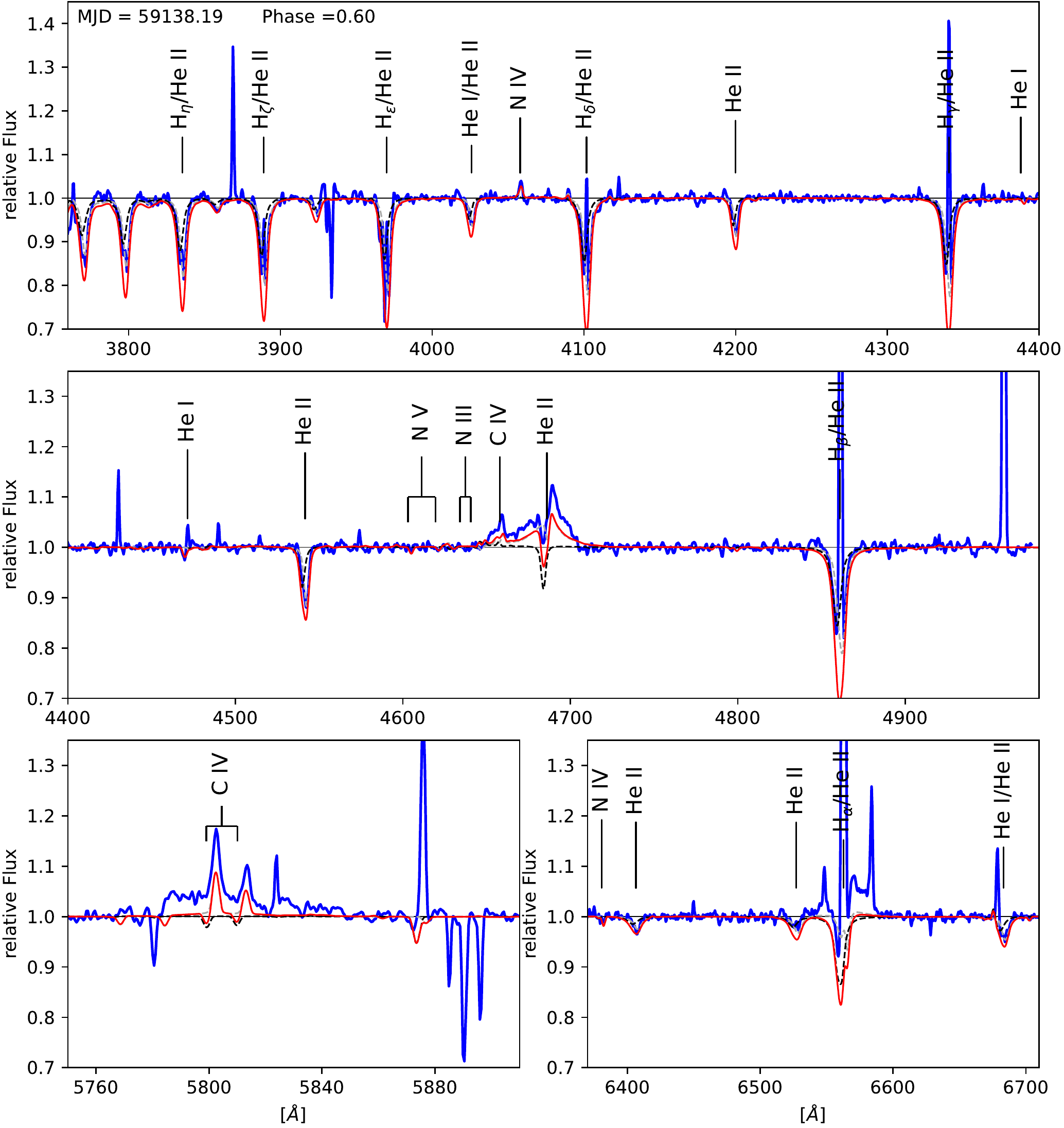}
	\caption{Spectroscopic fit of the original reduced and normalised data. Blue solid lines is the observations and red solid line is the combined model spectrum of the primary (dashed grey line) and secondary (black dashed line).}
\end{figure*}

\begin{figure*}
	\includegraphics[width=\textwidth]{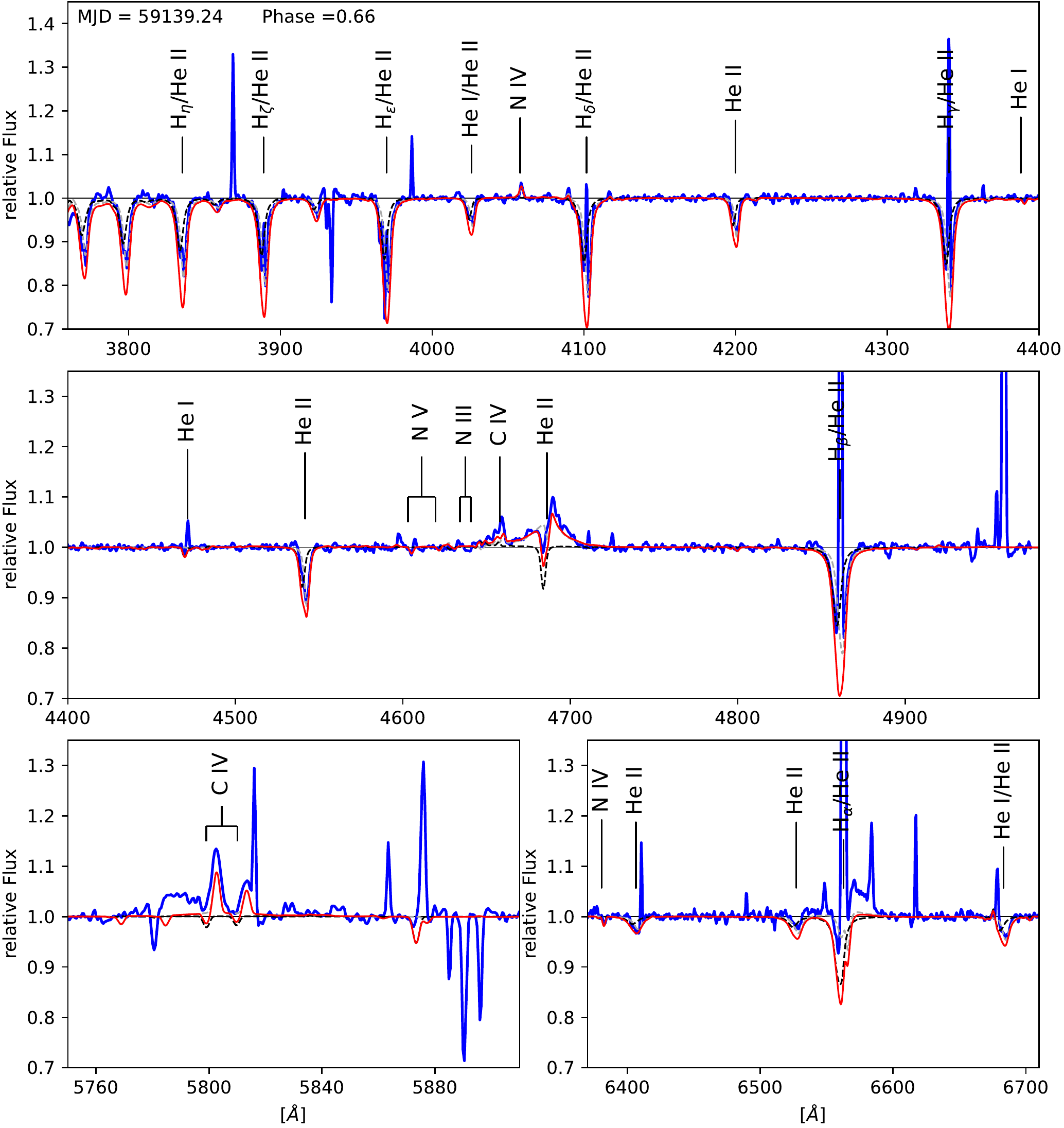}
	\caption{Spectroscopic fit of the original reduced and normalised data. Blue solid lines is the observations and red solid line is the combined model spectrum of the primary (dashed grey line) and secondary (black dashed line).}
\end{figure*}

\begin{figure*}
	\includegraphics[width=\textwidth]{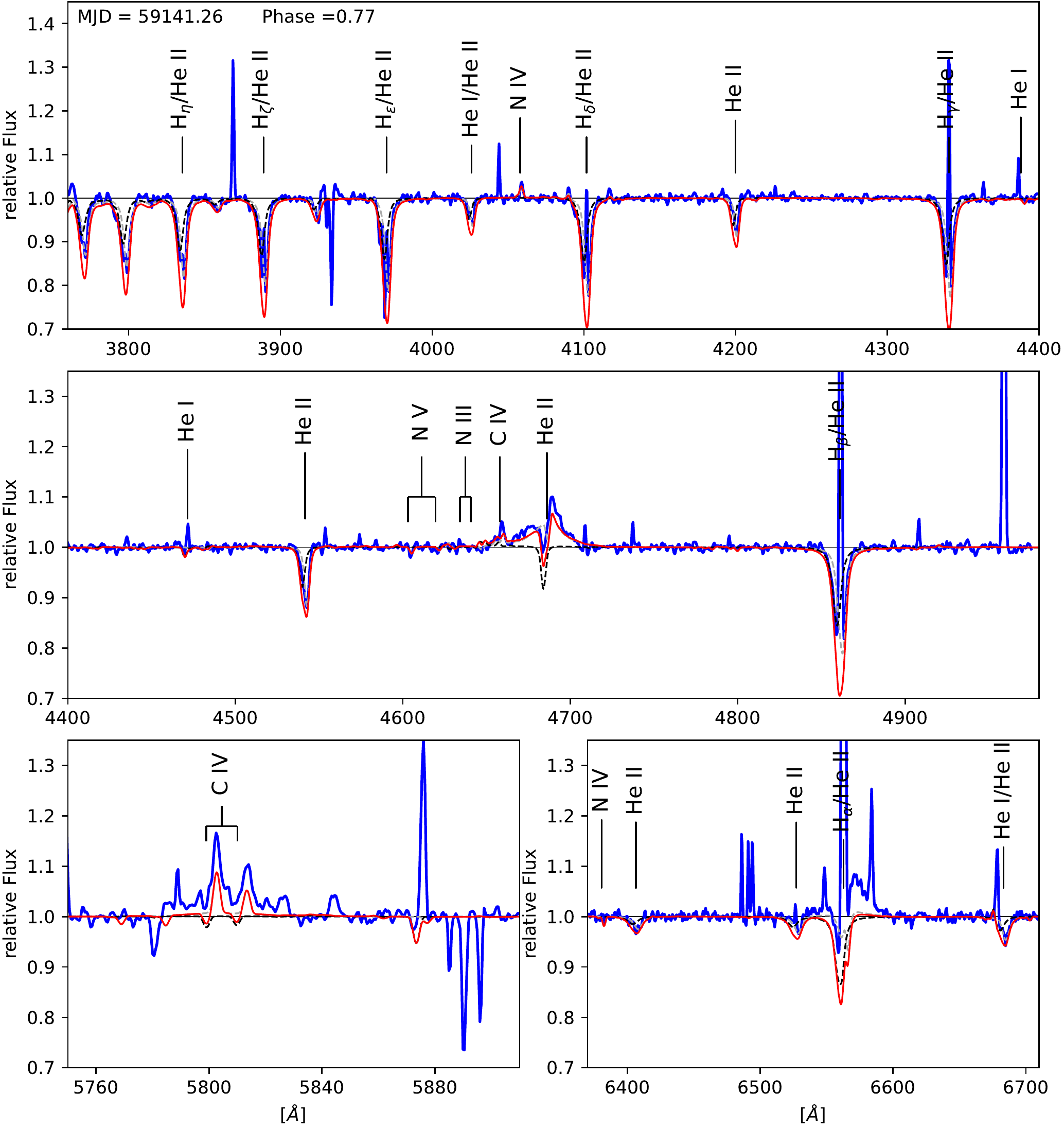}
	\caption{Spectroscopic fit of the original reduced and normalised data. Blue solid lines is the observations and red solid line is the combined model spectrum of the primary (dashed grey line) and secondary (black dashed line).}
\end{figure*}

\begin{figure*}
	\includegraphics[width=\textwidth]{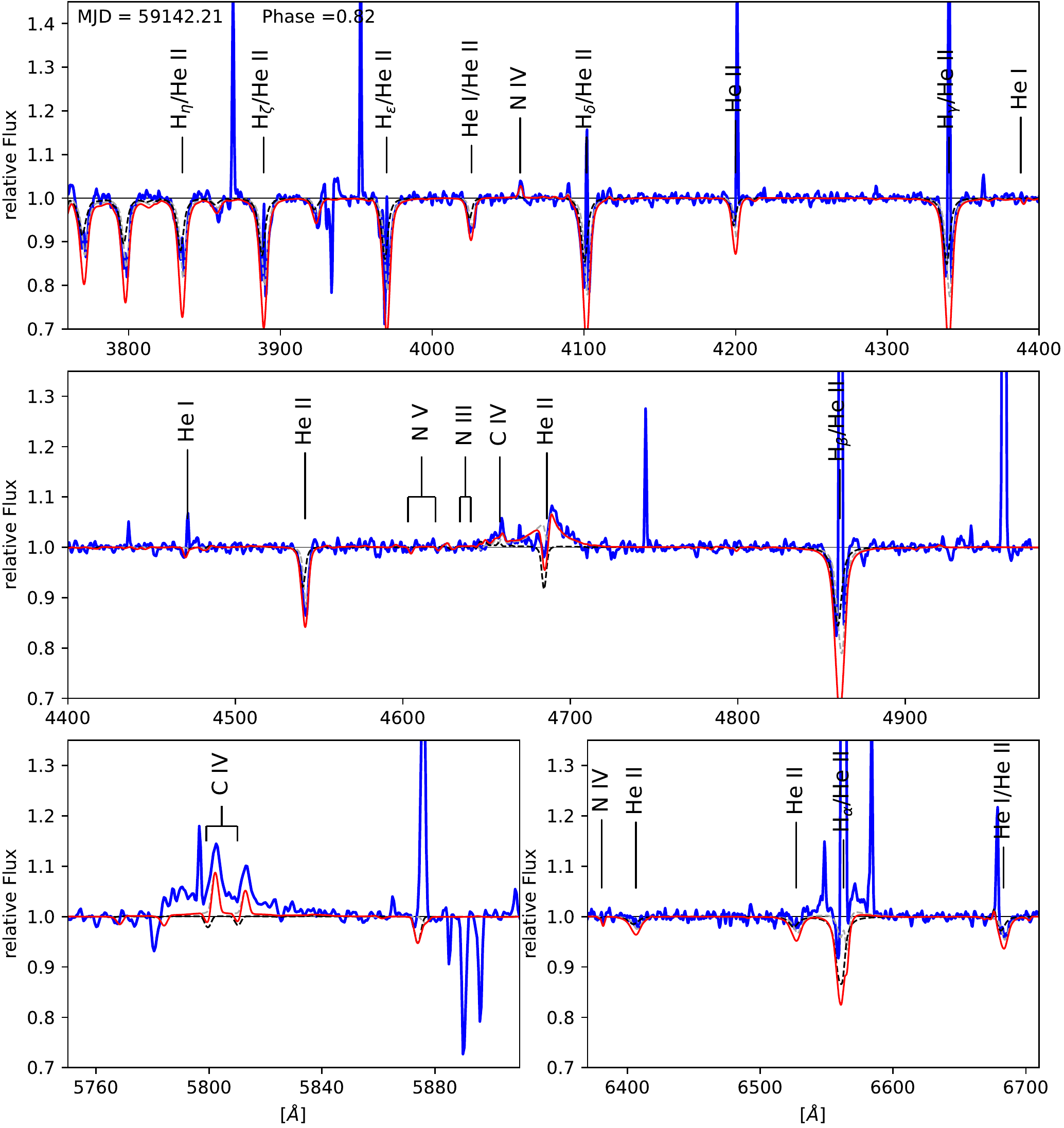}
	\caption{Spectroscopic fit of the original reduced and normalised data. Blue solid lines is the observations and red solid line is the combined model spectrum of the primary (dashed grey line) and secondary (black dashed line).}
\end{figure*}

\begin{figure*}
	\includegraphics[width=\textwidth]{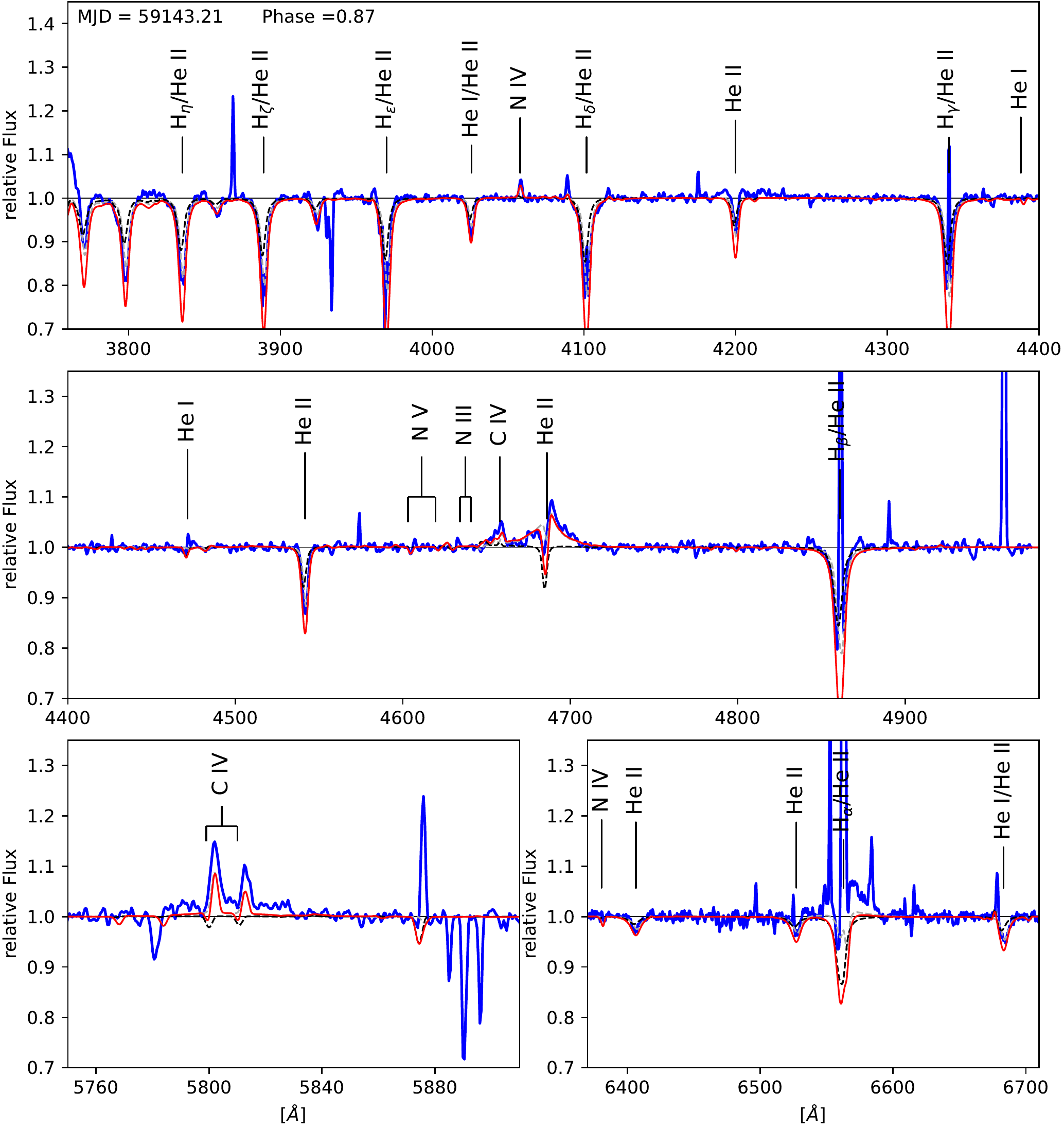}
	\caption{Spectroscopic fit of the original reduced and normalised data. Blue solid lines is the observations and red solid line is the combined model spectrum of the primary (dashed grey line) and secondary (black dashed line).}
\end{figure*}

\begin{figure*}
	\includegraphics[width=\textwidth]{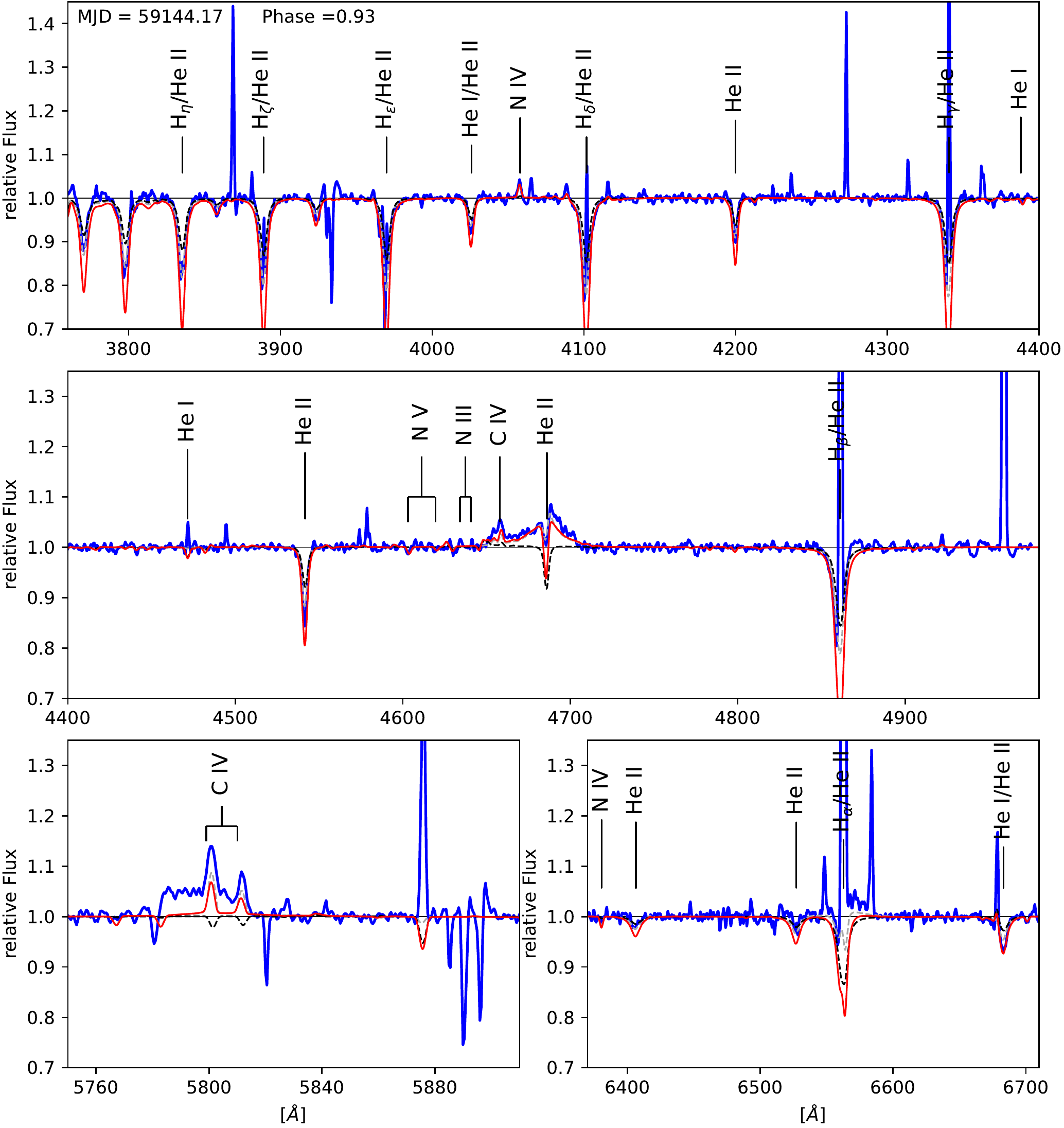}
	\caption{Spectroscopic fit of the original reduced and normalised data. Blue solid lines is the observations and red solid line is the combined model spectrum of the primary (dashed grey line) and secondary (black dashed line).}
\end{figure*}


%
%
%
%
%
%

\bsp	
\label{lastpage}
\end{document}